\documentclass[twocolumn]{aastex631}

\usepackage{graphicx}
\usepackage{natbib}
\usepackage{multirow}
\usepackage{textcomp}
\usepackage[caption=false]{subfig}
\usepackage{longtable}
\usepackage{ulem}
\usepackage{soul} 
\usepackage{enumitem}
\usepackage{chngcntr}
\usepackage{hyperref}
\usepackage{changepage}
\usepackage{amsmath}
\hypersetup{
    colorlinks=true,
    filecolor=magenta,      
    urlcolor=blue,
}
\usepackage{xcolor}
\usepackage{colortbl}
\newcommand{\greyline}{\arrayrulecolor{gray}\hline\arrayrulecolor{black}}

\newcommand{\HCfiveN}{HC$_{5}$N}
\newcommand{\HCsevenN}{HC$_{7}$N}

\newcommand{\am}{NH$_{3}$}
\newcommand{\msun}{M$_{\odot}$}

\newcommand{\Msuny}{M$_{\rm \odot}\, {\rm yr}^{-1}$}

\newcommand{\cyanobut}{HC$_5$N}

\newcommand{\til}{$\sim$}

\newcommand{\degree}{$^{\circ}$}
\newcommand{\kms}{km s$^{-1}$}
\newcommand{\kmsp}{km s$^{-1}$ pc$^{-1}$}

\newcommand{\jybe}{Jy beam$^{-1}$}
\newcommand{\hii}{H~{\scriptsize II}}
\newcommand{\cms}{cm$^{-2}$}

\defcitealias{Larson81}{Larson}

\begin{document}

\title{Discovery of a Giant Molecular Cloud at the Midpoint of the Galactic Bar Dust Lanes: M4.7--0.8}

\author[0000-0002-4013-6469]{Natalie O. Butterfield}
\affiliation{National Radio Astronomy Observatory, 520 Edgemont Road, Charlottesville, VA 22903, USA}
\email{nbutterf@nrao.edu}

\author[0000-0001-8708-5593]{Larry K. Morgan} 
\affil{Green Bank Observatory, 155 Observatory Rd, PO Box 2, Green Bank, WV 24944, USA}


\author[0000-0003-0410-4504]{Ashley T. Barnes}  
\affiliation{European Southern Observatory (ESO), Karl-Schwarzschild-Stra{\ss}e 2, 85748 Garching, Germany}

\author[0000-0001-6431-9633]{Adam Ginsburg}
\affiliation{Department of Astronomy, University of Florida, Bryant Space Science Center, Gainesville, FL, 32611, USA}

\author[0000-0002-1313-429X]{Savannah Gramze}
\affiliation{Department of Astronomy, University of Florida, Bryant Space Science Center, Gainesville, FL, 32611, USA}

\author[0000-0002-6753-2066]{Mark R. Morris}
\affil{Department of Physics and Astronomy, University of California, Los Angeles, Box 951547, Los Angeles, CA 90095 USA}

\author[0000-0001-6113-6241]{Mattia~C.~Sormani} 
\affil{Como Lake centre for AstroPhysics (CLAP), DiSAT, Universit{\`a} dell’Insubria, via Valleggio 11, 22100 Como, Italy}


\author[0000-0002-6073-9320]{Cara D. Battersby}
\affil{University of Connecticut, Department of Physics, 196A Auditorium Road, Unit 3046, Storrs, CT 06269, USA}

\author[0000-0003-3128-6542]{Charlie Burton}
\affil{Department of Physics, 4-183 CCIS, University of Alberta, Edmonton, Alberta T6G 2E1, Canada}

\author[0000-0002-7408-7589]{Allison H. Costa}
\affiliation{National Radio Astronomy Observatory, 520 Edgemont Road, Charlottesville, VA 22903, USA}

\author[0000-0001-8782-1992]{Elisabeth A. C. Mills}
\affiliation{Department of Physics and Astronomy, University of Kansas, 1251 Wescoe Hall Drive, Lawrence, KS 66045, USA}

\author[0000-0001-8224-1956]{J\"{u}rgen Ott}
\affil{National Radio Astronomy Observatory, 1011 Lopezville Road, Socorro, NM 87801, USA}

\author{Michael Rugel}
\affil{National Radio Astronomy Observatory, 1011 Lopezville Road, Socorro, NM 87801, USA}


\begin{abstract}
We present the detection of a previously unknown giant molecular cloud (GMC) located at the midpoint of the Galactic Bar Dust Lanes (M4.7--0.8), using spectral line observations taken with the Green Bank Telescope (GBT). This $\sim$60 pc long GMC is associated with accreting material that is transitioning from the quieter Galactic disk environment to the more extreme central molecular zone (CMZ) environment. Our 24 GHz single-dish radio observations targeted the NH$_3$ (1,1)$-$(4,4) and HC$_5$N (9$-$8), known dense gas tracers. The observations reveal the main features of the GMC, which we have dubbed the `Nexus' and `Filament', covering a 0$.\!\!^\circ$5$\times$0$.\!\!^\circ$25 area at 31$''$ angular resolution. In this publication we investigate the gas kinematics within the observed region and compare the distribution of molecular emission to previous infrared surveys to better understand the dust component. The observed gas tracers show centrally condensed cores corresponding to the positions of high dust column densities and low dust temperatures. We report the detection of a previously unknown NH$_3$ (3,3) maser, along with a 70$\mu$m source association, which supports the identification of this region as being actively star-forming. Gas emission in this region shows broad linewidths, comparable to values seen in CMZ clouds. The overall description of this cloud that we present is that of a highly dynamic region comprising dense gas and dust. This encapsulates a wide range of features associated with star formation, in addition to material transport related to the CMZ.

\end{abstract}


\section{introduction}

\begin{figure*}[t!]
\centering
\hspace{-2mm}
\subfloat{\includegraphics[width=0.494\textwidth]{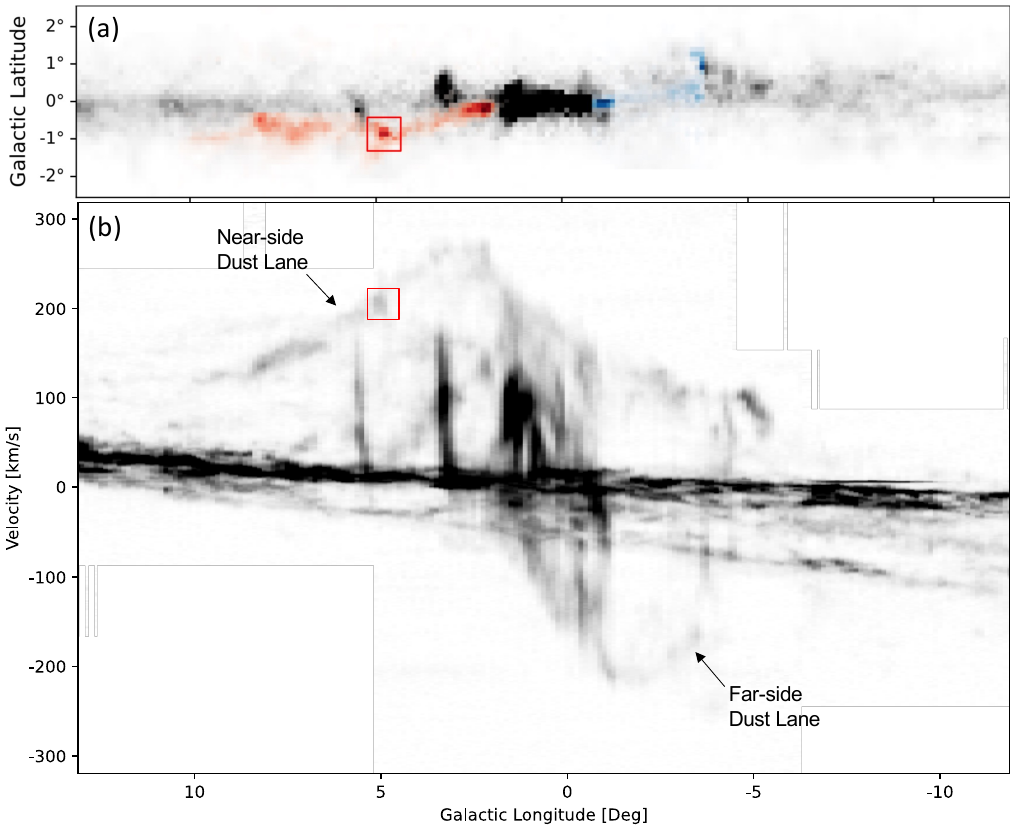}\label{introfig1}}
\subfloat{\label{introfig2}}
\subfloat{\includegraphics[width=0.487\textwidth]{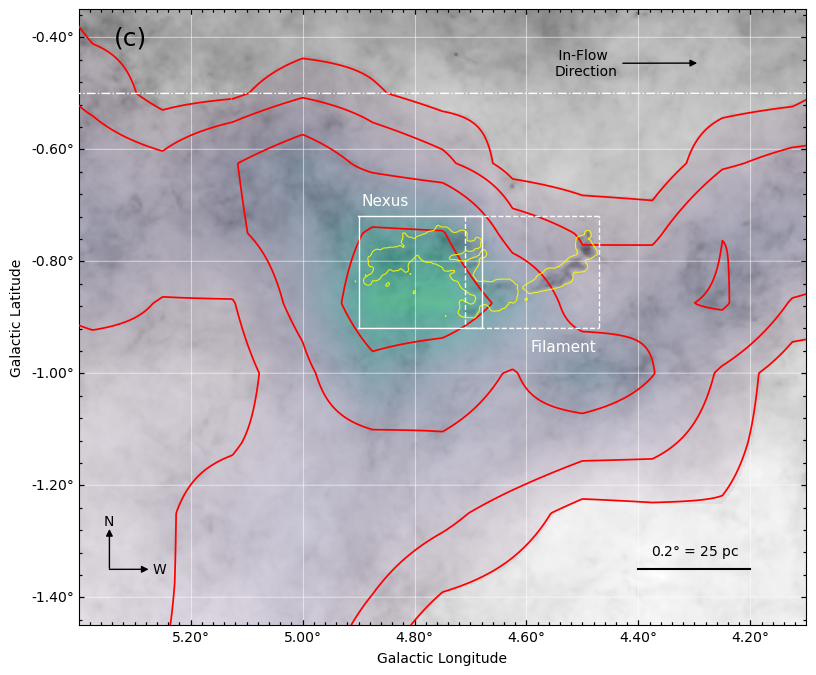}\label{introfig3}}
\caption{\textbf{(a)} Integrated $^{12}$CO (1--0) emission from the \cite{Dame01} survey, toward the inner \til25\degree\ of the Milky Way galaxy. Highlighted in red and blue are the Milky Way's Galactic Bar Dust Lanes, near-side and far-side respectively, from \cite{Marshall08} and \cite{sormani19a}. \textbf{(b)} Position-Velocity diagram of $^{12}$CO (1--0) for the emission shown in (a). Annotated in (b) are the Near-side and Far-side Dust Lane structures in position-velocity space. Red boxes in (a) and (b) show the location of the Midpoint (M4.7-0.8) of the Galactic Bar Dust Lanes in position-velocity space. \textbf{(c)} 250 \micron\ greyscale from Herschel \citep{Molinari16} with overlaid $^{12}$CO emission with red contours at 15, 25, 50, and 100 K \kms\ from \cite{sormani19a}, centered on the M4.7-0.8 cloud (red box in left panel). The image opacity of the $^{12}$CO emission is scaled by the intensity. The solid white box shows the field-of-view of the GBT observations targeting the main `Nexus' of the cloud (pilot study), and the dashed white box shows the field-of-view of the GBT observations targeting the `Filament' of the cloud (DDT). The light yellow contour in panel (c) shows the GBT \am\ (1,1) emission at 15$\sigma$ (150 mK, Figure \ref{morphfig}). The dot-dashed line shows the $b$=-0\fdg5 orientation - the lower Galactic latitude limit of many Galactic plane surveys (e.g., SEDGISIM, HOPS; \cite{Schuller21} and \cite{purcell12}, respectively). 
A compass for the cardinal directions in the Galactic coordinate system is shown in the bottom left corner of panel (c). 
}
\label{intro-fig}
\end{figure*}

Recent advancements in high spatial resolution observations at microwave, infrared, and optical wavelengths have greatly improved our understanding of bars in spiral galaxies \citep[e.g., SINGS, PHANGS, TIMER;][respectively]{sings,phangs, timer1}. One of the most intriguing features of bars is the presence of dark dust lanes that obscure the bright stellar emission. These dust lanes are essential in transporting material from the Disk toward the Galactic Center (GC), where it forms dense, ring-like structures that become sites of intense star formation \citep[e.g.,][]{Athanassoula92, Tress20b}. %

Star formation in the dust lanes of external galaxies is openly debated in the literature. Several studies have argued that star formation in galactic bars is suppressed due to the high shear forces in the region \citep{Emsellem15, Fraser20, Kim24, Neumann24b}. \cite{Neumann19} found that galactic bars fall into two main categories, those with significant star formation and those with little-to-no star formation in their bars. Most the star formation in these galactic bars tend to occur on the leading edge where shocks drive the formation of kpc-long filaments \citep{Athanassoula92, Renaud15, Neumann19, Fraser20}. These structures can have supersonic turbulence, which can help fragment these structures into separate clouds \citep[e.g.,][]{Renaud15}. Additionally, \cite{Maeda23} surveyed a sample of 18 barred galaxies and found that the star formation efficiency (SFE) is lower than values measured in the disk, with the SFE dipping at around R/R$_{bar}$=0.5 (see their Figure 8), where R$_{bar}$ is the radius of the bar and R is the position of the cloud within the bar. However, as highlighted by recent JWST observations by \citet{Schinnerer23}, dust lanes in external galaxies can also host numerous massive young star clusters, emphasizing their importance as key environments for studying star formation in highly dynamic regions.

The Milky Way has long been known to contain a bar \citep[e.g.,][]{deVauc64, Binney91}, making it likely to host dust lanes similar to those observed in other barred galaxies.
However, our position within the Galaxy makes identifying these regions particularly challenging. Recent advancements in simulations and observations have now enabled the identification of the Milky Way's bar dust lanes using both $^{12}$CO ($l$, $b$, v$_{los}$) datacubes \citep[e.g.,][]{fux99, Liszt06, Li16,sormani18a} and 3D dust extinction maps \citep{Marshall08}. Figure \ref{intro-fig} (a and b) shows the dust lanes of the Milky Way bar identified in $^{12}$CO datacubes \citep{Marshall08} - representing a direct connection between the Galactic Center and the Disk. With an estimated mass accretion rate of $\dot{M} \simeq 0.8 \pm 0.6 M_{\rm \odot}\, {\rm yr}^{-1}$ \citep{Hatchfield21}, the dust lanes supply mass and energy into the extreme environment of the Milky Way's Central Molecular Zone \citep[CMZ;][]{Henshaw23} -- providing fresh (and potentially chemically distinct) fuel for star formation. The dust lanes are also extremely interesting in their own right, being where extreme cloud collisions with relative velocities of more than 100 \kms\ \citep{sormani19b}, highly filamentary clouds \citep{Wallace22}, and the birth sites of peculiar massive star formation complexes \citep[e.g., Sgr E;][]{Anderson20} have been identified.

\subsection{M4.7-0.8: The Midpoint of the Near-side Galactic Bar Dust Lane}

The Midpoint of the Near-side Galactic Bar Dust Lane is observed to be one of the brightest regions observed in the $^{12}$CO dataset (Figure \ref{introfig1}, highlighted by the red box). 
Based on the geometrical model presented in \cite{sormani19a}, the Midpoint is located at R/R$_{bar}$\til0.5, at a distance of 7.0 kpc (we discuss this 3D geometry in more detail in Section \ref{sect:3D}). Gas located at the Midpoint is observed to have a line-of-sight velocity of \til200 \kms\ \citep[Figure \ref{introfig2};][]{fux99,Marshall08,sormani19a}. Based on previous work from \cite{sormani19a}, the $^{12}$CO at this location is estimated to have a mass of \til10$^6$ \msun\ and is suggested to accrete into the CMZ in 5--6 Myr, potentially forming a Sgr B2-like complex after CMZ accretion. This accretion is in the Westward direction for the Near-side Dust Lane (see Figure \ref{introfig3}), indicating the plane-of-sky velocity is also Westward. Figure \ref{introfig3} shows the $^{12}$CO contours overlaid on the Herschel 250 \micron\ dust continuum. The solid white box denotes the region we will be referring to as the ``Nexus" in this paper. The Nexus coincides with the brightest $^{12}$CO emission and appears correlated with the nexus of the dust continuum and $^{12}$CO emission. The dashed white box highlights the feature we will be referring to as the ``Filament". This narrow dust continuum feature appears to be an extension off of the Nexus of the cloud and shows a filamentary morphology.

In this paper we target these two fields (white boxes in Figure \ref{introfig3}) of the midpoint region of the Galactic Bar Dust Lanes using the Green Bank Telescope. We target multiple transitions of ammonia (\am) and cyanobutadiyne (\cyanobut) to investigate the higher resolution molecular gas morphology and kinematics of this region. We discuss the observations and data reduction in Section \ref{obs}. We present the detection of dense gas in these observations  in Section \ref{res} and analyze the morphology and kinematics of the cloud. We compare our dense gas observations to previous dust continuum studies in Section \ref{sec:dust}. Lastly, we investigate the 3D location, star formation, and other cloud properties in Section \ref{dis}. We summarize these conclusions in Section \ref{con}.

\begin{table}[bt!]
\caption{\textbf{List of Targeted Lines.}}
\vspace{-4mm}
\centering
\begin{tabular}{cccccc}
\hline\hline
\multicolumn{2}{c}{\textbf{--- \underline{Line Properties} ---}}  & \multicolumn{3}{c}{\textbf{--- \underline{Image Properties} ---}} \\
\textbf{Spectral} & \textbf{Rest} & \textbf{Beam}  &\textbf{RMS} & \textbf{Peak}  \\
\textbf{Line} & \textbf{Frequency} & \textbf{Size}   &\textbf{Noise} & \textbf{Intensity}  \\
& (GHz) & (\arcsec) & (mK) & (K) \\
\hline
\am\ (1,1) & 23.6944955 & 31.34 & 10 &  1.06  \\
\am\ (2,2) & 23.7226333 &  31.30  & 10 &  0.81  \\
\am\ (3,3) &  23.8701292 &  31.11 & 10 &  2.04  \\
\am\ (4,4) &  24.1394163 & 30.75 & 10 &  0.37  \\
~\am\ (5,5)\footnote{This transition was only observed in the Filament region and is therefore not used in the analysis of this paper beyond stating the detection. The \am\ (5,5) transition is shown in the Appendix (Section \ref{55 sec}). } &  24.5329887 & 30.27 &  10  & 0.22  \\
HC$_5$N (9-8) & 23.9639007 & 30.99  & 10  & 0.17  \\
\hline \hline
\end{tabular}
\label{lines}
\end{table}

\section{Observations and Data Reduction}
\label{obs}

The data presented in this paper were taken using the 100m Robert C. Byrd Green Bank Telescope (hereafter, GBT), located at the Green Bank Observatory in Green Bank, West Virginia. These single-dish radio observations were taken using the 18-26.5 GHz K-band focal plane array (KFPA) receiver (\til31\arcsec\ angular resolution; \til1 pc at the assumed distance to the Galactic Bar).

An initial study targeted a single 0\fdg25 $\times$ 0\fdg25 field centered at the location of strong CO emission associated with the midpoint of the near-side Dust Lane (see Figure \ref{introfig3}). Preliminary results indicated that a filamentary structure, visible in 250 \micron\ emission towards lower Galactic Longitudes, was also associated with \am\ emission detected in this preliminary field. Additional observations were therefore proposed and completed in order to extend the original field over the range of this filament.

Fully sampled maps were completed using the 7-pixel K-band Focal Plane Array (KFPA) on the GBT, centred at rest frequencies of 23.695, 23.723, 23.870, 24.139 and 24.533 GHz. These frequencies relate to the J = K = 1--5\footnote{The \am\ (5,5) transition was not observed in the Nexus region.} rotational inversion transitions of the \am\ molecule. A further rest frequency of 23.964 GHz was also observed, matching the \cyanobut\ (9--8) rotational transition. Observations took place over eight sessions from the 3rd of April, 2020 to the 19th of February, 2024.

\begin{figure*}[p]
\centering
\includegraphics[width=1.0\textwidth]{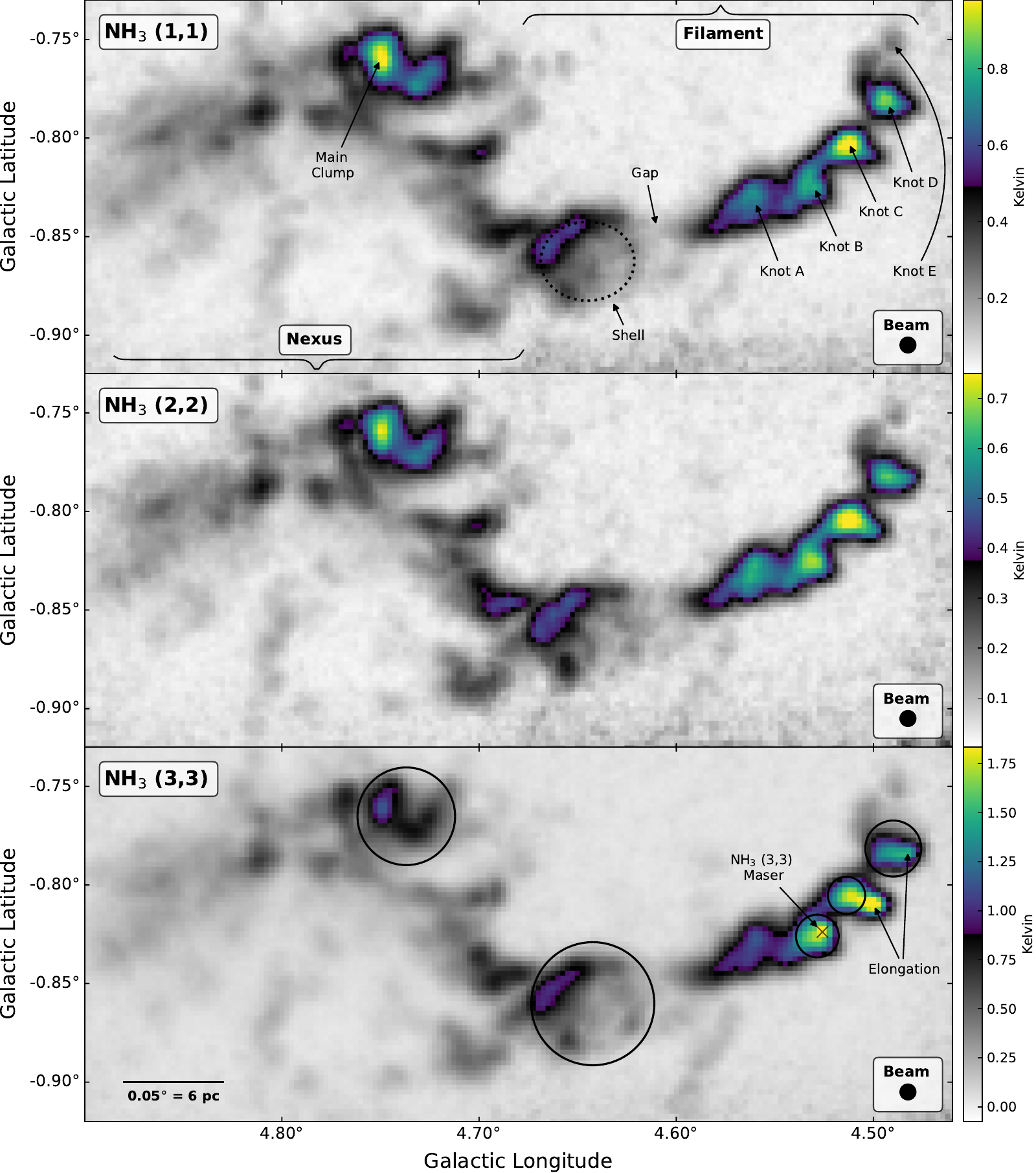}
\caption{Peak antenna temperature for the \am\ (1,1)--(3,3) transitions, top to bottom respectively, as indicated in the top left corner. Annotated on the top, \am\ (1,1), panel are several features and sources discussed in Section \ref{res}, including the ``Main Clump'' in the Nexus and the ``Gap" and ``Knots" A--E in the Filament. The black dashed ellipse illustrates the region we are annotating as the ``Shell''. The black `$\times$' symbol in the \am\ (3,3) bottom panel shows the location of the \am\ (3,3) maser, discussed in Section \ref{3maser}. The beam size is shown at the bottom right corner of each panel (see Table \ref{lines} for these values). A scale bar for the M4.7-0.8 cloud is shown in the bottom left of the \am\ (3,3) panel, assuming a distance of 7.0 kpc. The \am\ (3,3) panel also shows the apertures used to extract the spectra, shown in Figure \ref{spectra}, for regions: Main Clump, Shell, and Knots B-D (left to right, respectively). The knot elongation, observed in Knots C and D, in the higher J \am\ transitions, is annotated in the (3,3) panel.
}
\label{morphfig}
\end{figure*}

\subsection{Observational Setup}

The seven beams of the KFPA were used  for simultaneous, dual circular-polarization observations over the \til3\arcmin\ footprint of the receiver. Flux calibration was achieved via off-source `blank-sky' and switched noise-diode measurements, resulting in an estimated antenna temperature accuracy of \til20\%. Atmospheric opacity values were determined from NWS grid values using a combination of three local sites, depending upon the established weather conditions (i.e. wind direction and agreement between the sites) at the time of observation. The VEGAS spectrometer backend was used in mode 20 
(23.44 MHz (200 \kms) bandwidth per spectral window with 5.7 kHz (0.07 \kms) spectral resolution). Table \ref{lines} shows the line and imaging information (e.g., rest frequency, angular resolution) for the spectral lines presented in this study. More information on the imaging stage of the data reduction is discussed in Section \ref{data redux}.

\subsection{Mapping and Observing Strategies}

Maps were performed via an on-the-fly process with Nyquist-sampled rows in Galactic longitude. The `Off' positions were observed at intervals corresponding to the observation of four complete map rows. This allowed for sky emission compensation on a time scale of \til10 minutes, adequate for observations at K-band. Due to the high level of extended emission in directions toward the GC, large position offsets were necessary in order to ensure emission free `Off' measurements. These offsets were typically 5\degree\ and 3\degree\ in longitude and latitude, respectively. Pointing and focus calibrations were performed at the beginning of each observing session and approximately hourly thereafter.

\subsection{Data Reduction} 
\label{data redux}

Data were reduced primarily using the \texttt{GBTIDL} package \citep{gbtidl}.\footnote{The \texttt{GBTIDL} package can be found here: \hyperlink{https://gbtidl.nrao.edu/}{https://gbtidl.nrao.edu/}.} The `Off' position measurements were found to be stable over the duration of each individual session and so were averaged together in order to provide increased signal-to-noise in the final reduction. Each individual Nyquist-sampled integration was smoothed via a Gaussian convolution function. The resulting spectral resolution is 0.7 \kms. The `Off' spectra were further smoothed to a resolution of 0.84 \kms\ before being subtracted from the reference spectra and scaled via the system temperature determined from the `flickered' noise diode measurements. Atmospheric corrections (found as described above) were applied in order to place our data on the corrected antenna temperature (T$_A$*) scale.

An initial spectral baseline was fit to emission-free regions of each integration using orthogonal polynomials of order 7. The subsequent output was then gridded via the \texttt{gbtgridder}\footnote{The \texttt{gbtgridder} python program is available online at \href{https://github.com/GreenBankObservatory/gbtgridder}{https:// github.com/GreenBankObservatory/gbtgridder}.} routine using a Gaussian imaging kernel and a pixel width of 9\arcsec, approximately 1/3 of the observed beamsize. The FITS cubes resulting from each session were examined individually on the basis of rest frequency as well as on the individual KFPA beam/feed used. In some cases it was found that individual beams were unstable and/or contributing significantly higher levels of system noise. Where this was found to be detrimental to image quality, these beams were excised from the overall reduction. 

It was found that, over three observing sessions performed on the `filament' section of our overall region, there were significant disagreements in the positional metadata of our maps. This resulted in considerable differences in the antenna temperature measurements when compared pixel-to-pixel. It is believed that this problem arose as a result of poor elevation pointing calibration of the GBT, below elevations of \til20\degree\footnote{This issue was related to a deterioration in the telescope track foundation and has since been remedied (D.Frayer, priv comm).}. This was addressed in our data via careful examination of pointing offset data from peak scans and corrective offsets being applied to those data before gridding. The maps performed during the pilot phase of the study showed no evidence of this issue. Therefore we are confident in the astrometry of those maps as a reference and determine our pointing accuracy to be the same as our pixel width, 9\arcsec. Following the final combination of data from all individual sessions, a final baseline removal process was applied to each cube as a whole. This was done using a polynomial function fit to emission-free regions. The order of the polynomial fit was chosen to be as low as necessary to remove any obvious `ripple' structure on the scale of the spectral window (23.44 MHz). For the \am\ (1,1), (3,3), (4,4) and (5,5) lines, this was determined to be a third order polynomial. For the \am\ (2,2) and \cyanobut\ transitions, this was found to be second and fifth order respectively. No direct correlation was found between the level of sinusoidal variation in a transition's bandpass and any known hardware properties. System temperatures and measured RMS noise values were largely consistent across all observed frequencies. 

\begin{figure*}
\centering
\includegraphics[width=1.0\textwidth]{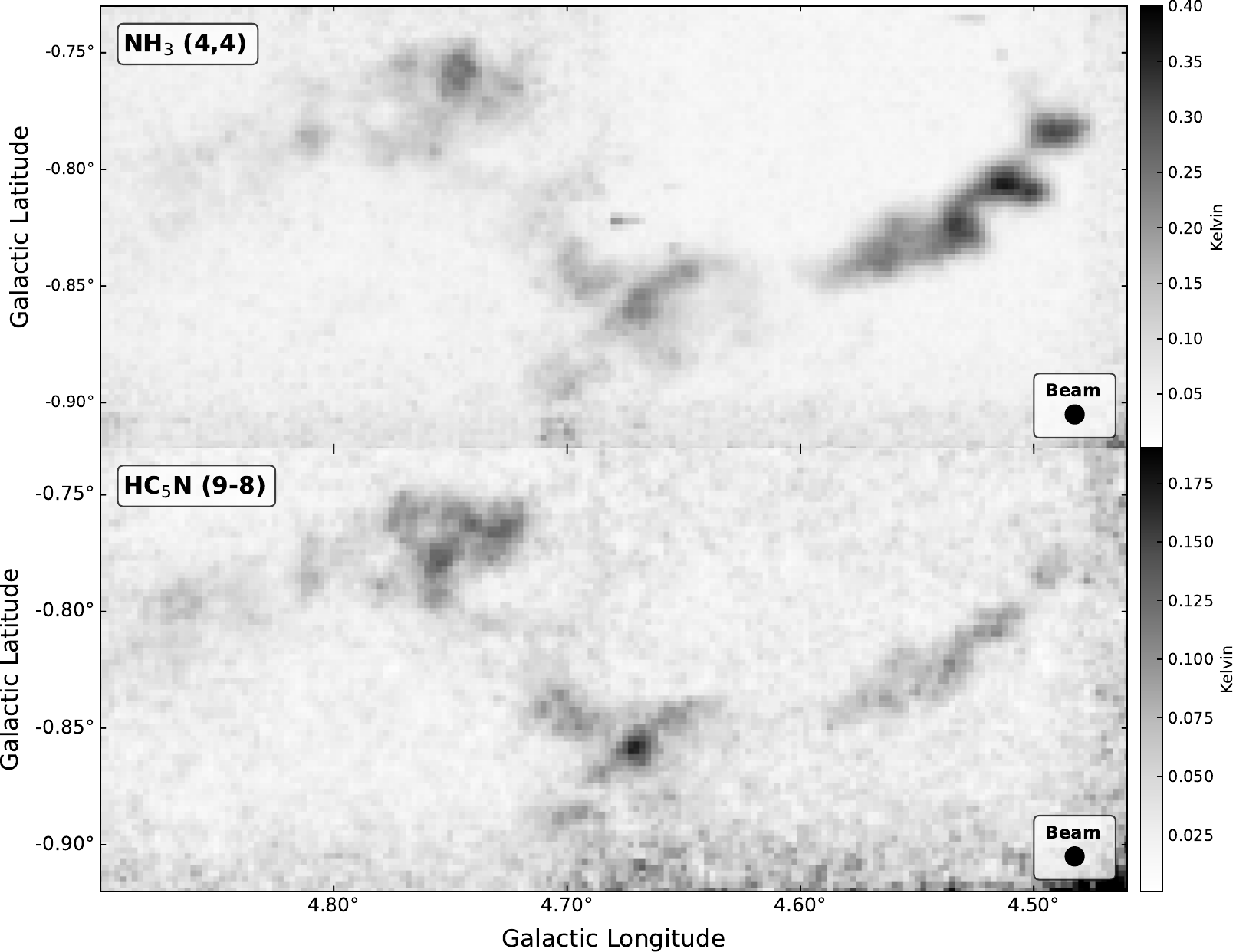}
\caption{Peak antenna temperature for the \am\ (4,4) and \cyanobut\ (9--8) transitions, top and bottom, respectively (as labeled in the top left corner). The beam is shown in the bottom left corner. The beam size is shown in bottom right corner (see Table \ref{lines} for these values). The noise floor in the Filament field is more prominent in these images, when compared with the \am\ (1,1)--(3,3) lines in Figure \ref{morphfig}, due to the relatively low SNR detection influencing the scaling of these images.  
}
\label{morphfig2}
\end{figure*}
\renewcommand{\thefigure}{\arabic{figure}}

\section{Detection of Dense Molecular Gas at the Midpoint}
\label{res}

We have detected dense gas, as traced by the \am\ and \cyanobut\ molecules, at the Midpoint of the Galactic Bar Dust Lanes: the M4.7-0.8 molecular cloud. In the following sections we discuss the morphological structure  (Section \ref{morph}) and the gas kinematics (Section \ref{kin}) of the molecular gas in this region. As we will show in Section \ref{kin}, M4.7-0.8 has a central velocity \til200 \kms, consistent with values for the Midpoint (Figure \ref{introfig2}). Therefore we will assume a distance of 7.0 kpc, with a conservative error estimate of 1 kpc, for all physical size calculations. 
This 1 kpc error estimate is similar to those used in \cite{Nilipour24}, who argue for a thick dust lane and use this value in their error estimate as well.

\subsection{Structure of the Dense Molecular Gas in M4.7-0.8}
\label{morph}

Figures \ref{morphfig} \& \ref{morphfig2} show the maximum emission in the \am\ (1,1)--(4,4) and the \cyanobut\ (9$-$8) transitions in the M4.7-0.8 molecular cloud.\footnote{The peak antenna temperature of the \am\ (5,5) line emission in the Filament is included in the Appendix (Figure \ref{fig:Am55}).} The \am\ (3,3) transition is the brightest out of all our detected spectral lines, with a peak emission of 2.04 K ($>$200$\sigma$; Table \ref{lines}). The \cyanobut\ (9--8) transition is the faintest of our detected spectral lines, with a peak of 0.17 K (17$\sigma$; Table \ref{lines}).

The dense molecular gas at the midpoint of the Galactic Bar Dust Lanes (M4.7-0.8) is observed to be extended in Galactic Longitude (\til0\fdg45, \til55 pc) and contains several bright clumps that are unresolved in these GBT observations (\til30\arcsec, \til1 pc; scale-bar is shown in Figure \ref{morphfig}, bottom). The large-scale structure is fairly consistent across all observed transitions, showing a similar morphology in all five spectral lines (Figures \ref{morphfig} and \ref{morphfig2}). However, when looking at the localized morphology in the cloud, we can observe interesting substructures. These features are annotated in the top \am\ (1,1) panel of Figure \ref{morphfig} and they will be discussed throughout this paper.

As mentioned previously in this paper, we highlight two main regions in the cloud: a `Nexus' (the region associated with the $^{12}$CO peak) and a `Filament' (the dust continuum filament extending out from the Nexus). The Nexus contains several faint, diffuse, extended structures (0\fdg05 -- 0\fdg15), with a main bright clump located at $l$=4\fdg74, $b$=-0\fdg76, labeled as ``Main Clump" in Figure \ref{morphfig}.

\begin{figure*}
\centering
\includegraphics[scale=0.27]{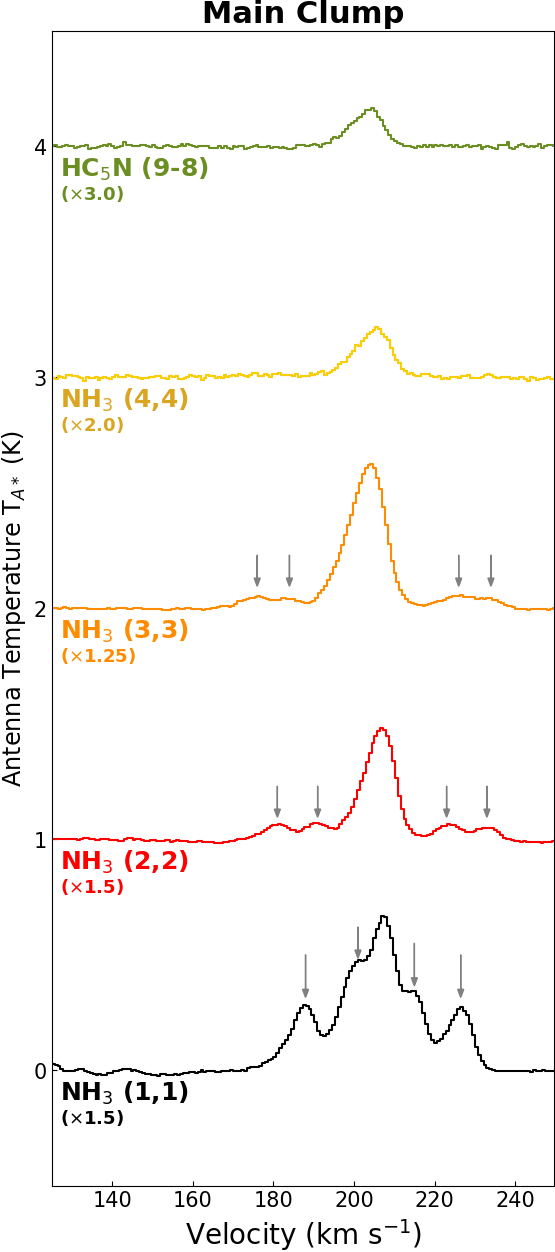}
\includegraphics[scale=0.27]{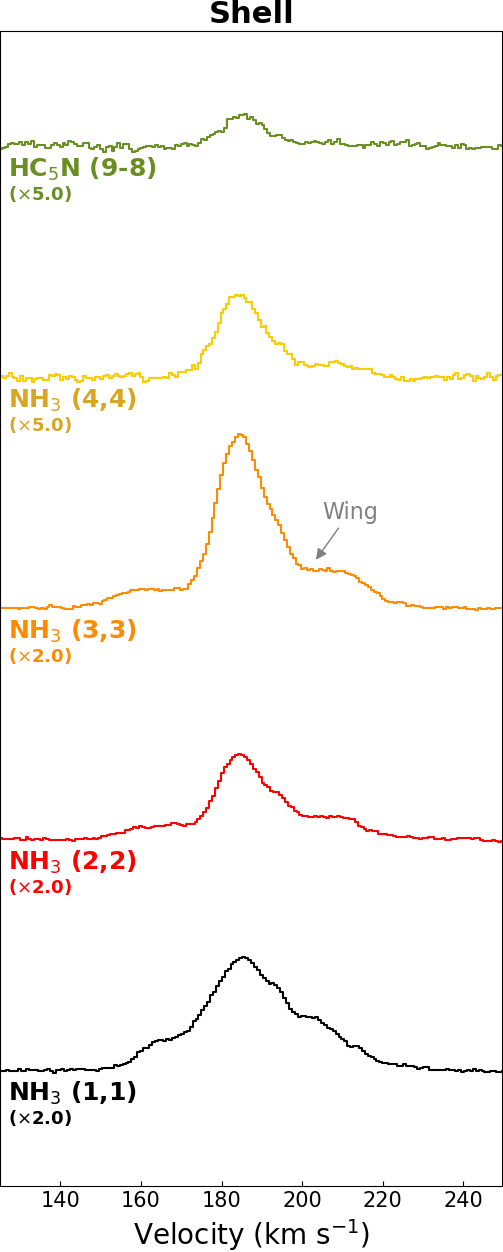}
\includegraphics[scale=0.27]{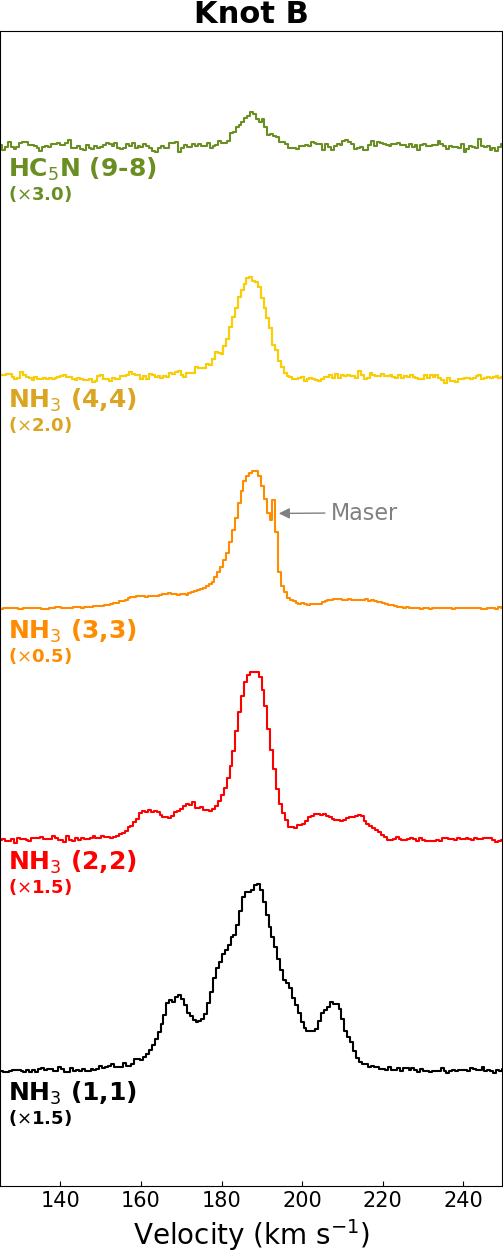}
\includegraphics[scale=0.27]{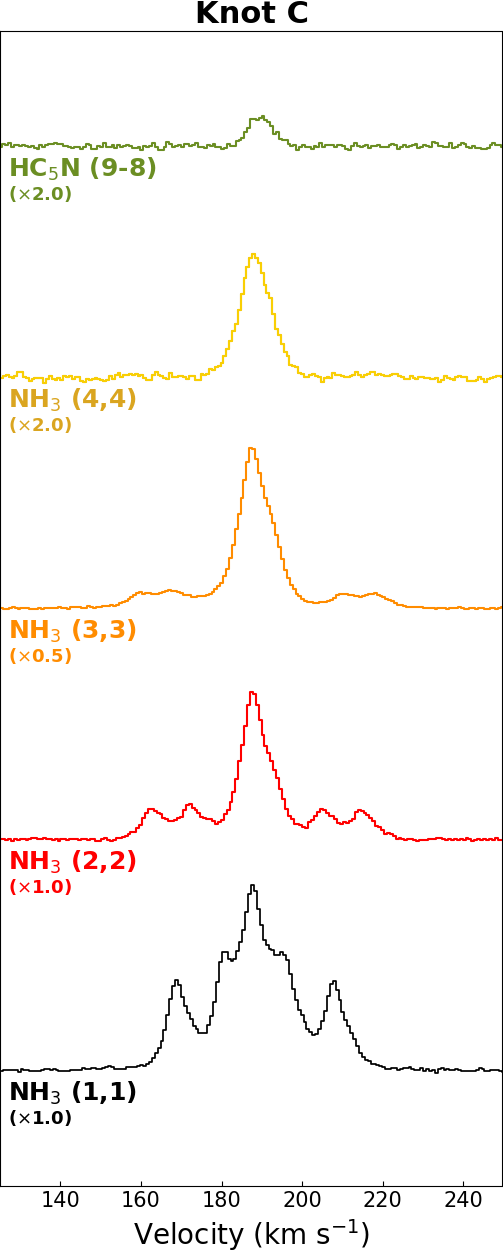}
\includegraphics[scale=0.27]{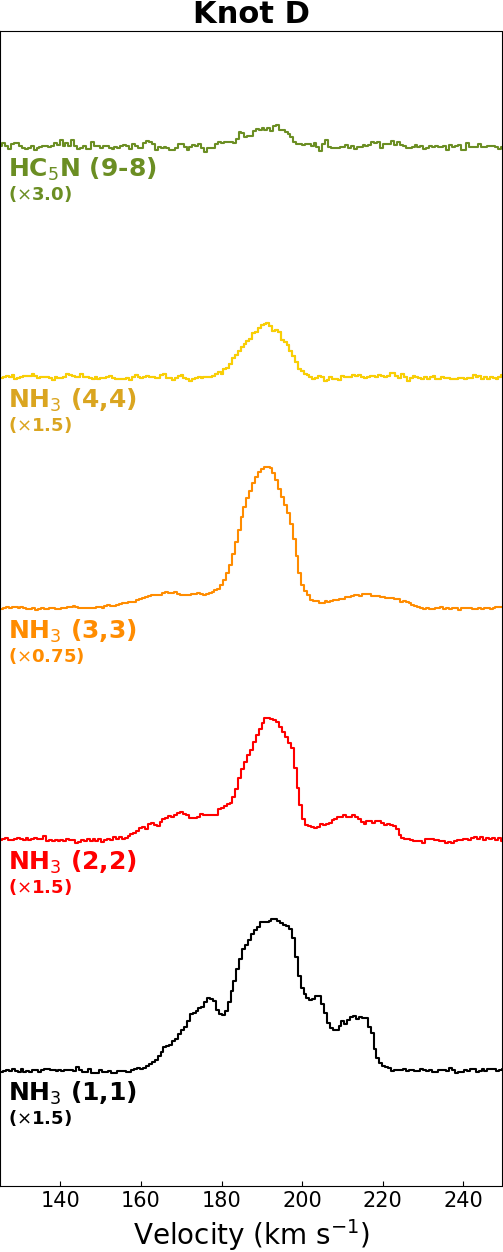}
\caption{Integrated spectra of the \am~(1,1)$-$(4,4) and \HCfiveN\ (9$-$8) lines in the M4.7-0.8 cloud for the five apertures shown in Figure \ref{morphfig}, with the label of the aperture shown on the top of each panel. Each spectrum is labeled by its corresponding transition on the left side of the plots. Several spectra have been scaled for visualization purposes. These spectra are denoted with the value used for this scaling below the transition name on the left side of the plot. The gray arrows in the first panel (Main Clump) show the location of the hyperfine satellite lines for the \am\ (1,1)--(3,3) transitions. The hyperfine lines are not noticeable in the (4,4) spectrum and are therefore not shown. The asymmetric `wing' in the spectral line profile, discussed in Section \ref{kin}, is annotated in Shell panel - where the offset in the secondary component is the most noticeable. The \am\ (3,3) maser, detected in Knot B, is annotated with a gray arrow in the middle panel.
}
\label{spectra}
\end{figure*}

The Filament is generally long (\til0\fdg3, \til40 pc) and narrow (\til0\fdg03, \til4 pc), resulting in a length-to-width ratio of roughly 10. While generally observed to be one continuous structure, we note a small discontinuity, labeled as the ``Gap" (Figure \ref{morphfig}; $l$=4\fdg61, $b$=-0\fdg85), where the emission is relatively faint when compared with adjacent sections of the Filament. This decrease in emission at this location in the structure is observed across all five transitions. On the Eastern side of the Gap, there is a shell-like structure, labeled as the ``Shell", which shows a brighter curved rim and a fainter cavity (dotted black line in Figure \ref{morphfig}). The Shell appears to be brightest on the Eastern rim with very low level emission on the Western rim.

On the Western-most region of the Filament, we observe five bright knots of emission in the \am\ transitions, labeled as Knots A--E in Figure \ref{morphfig}.  These sources are roughly evenly spaced, with a separation of \til2\arcmin\ (4.4 pc) between adjacent knots. While bright in the \am\ transitions, we do not observe evidence of these knots in the \cyanobut\ line above 125 mK (\til12$\sigma$) -- only in the large-scale structure of the Filament. Knot C is observed to be the brightest in every \am\ transition, including the \am\ (5,5) transition (Figure \ref{fig:Am55}). Knot E is observed to be the faintest in our observed transitions, and is only faintly detected in the J$\leq$4 transitions. Knot E is also observed to be smaller ($r$=20\arcsec; 0.7 pc) than Knots A--D and is unresolved in these observations.

Knots B, C and D exhibit some similar characteristics when comparing the \am\ (1,1) to \am\ (3,3) emission. Here, the \am\ (1,1) emission describes condensed circular emission peaks which are toward the northern edge of the Filament. In comparison, the \am\ (3,3) emission has a more extended morphology, with the \am\ (1,1) peaks occurring toward the eastern end of each respective prolate knot. This elongation in the higher J transitions is annotated in the bottom panel of Figure \ref{morphfig}. It should be noted that this is true for emission measured in the \am\ (4,4) and (5,5) transitions as well (Figures \ref{morphfig2} and \ref{fig:Am55}, respectively). We also note that this is true for emission measured both as the peak temperature as well as the integrated intensity of the respective \am\ transitions, as we will illustrate in the next section in our kinematic analysis.

\subsection{Kinematics of the Dense Molecular Gas in M4.7-0.8}
\label{kin}

In our study of the Near-side Dust Lane, we not only seek to determine the gas structure of the observed region, as discussed in Section \ref{morph}, but also aim to establish the bulk motions within that gas. Whether it is possible to confirm the transport of material along the Dust Lane with the presented data or to investigate any potential structural developments within the cloud as a single element, it is useful to examine the dynamic state of the observed gas. Here we present a kinematical analysis of the dense gas in the surveyed region using a through inspection of the observed spectra as a function of position.

The simplest way to investigate gas kinematics is by extracting spectra. We selected five regions to investigate: the Main Clump, the Shell, Knots B, C, and D. The apertures used to extract the spectra for each region are shown as black circles in Figure \ref{morphfig}.\footnote{A more detailed analysis on the clump structure in the cloud is underway and will be presented in a separate publication (Morgan et al., in prep)}. 
The spectra were then extracted and plotted in Figure \ref{spectra} using the python programs \texttt{SpectralCube}\footnote{The \texttt{SpectralCube} python program is available online at \href{https://github.com/radio-astro-tools/spectral-cube}{https://github.com/radio-astro-tools/spectral-cube}.} and \texttt{pyspeckit}\footnote{The \texttt{pyspeckit} python program is available online at \href{https://github.com/pyspeckit/pyspeckit}{https://github.com/pyspeckit/pyspeckit}.} \citep{pyspeckit}.

\begin{figure}
\centering
\includegraphics[width=0.47\textwidth]{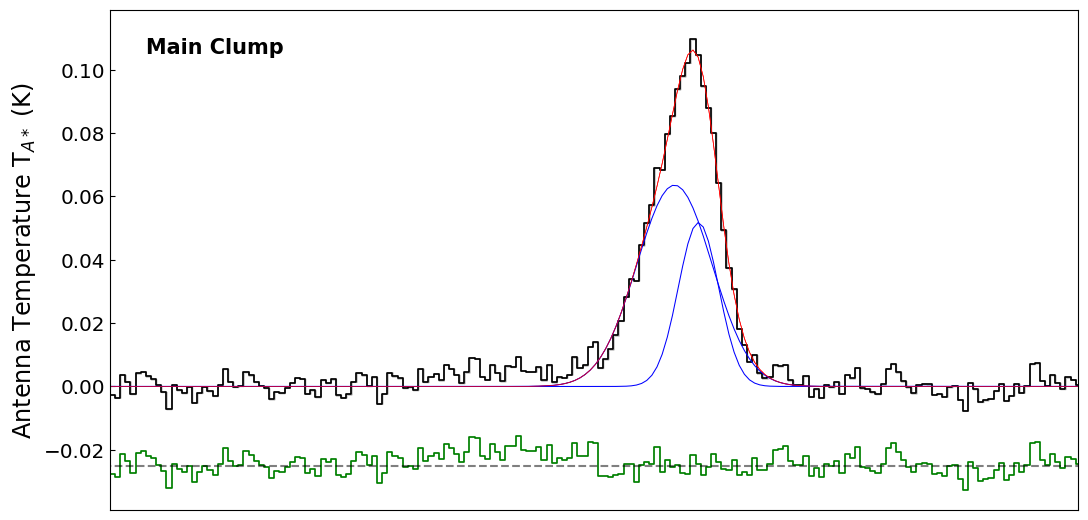}\\
\includegraphics[width=0.47\textwidth]{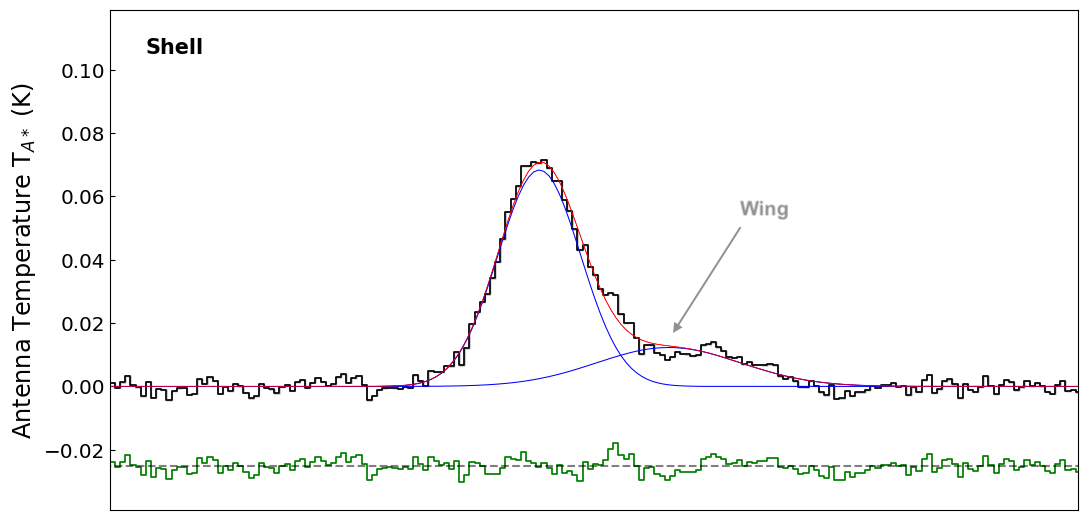}\\
\includegraphics[width=0.47\textwidth]{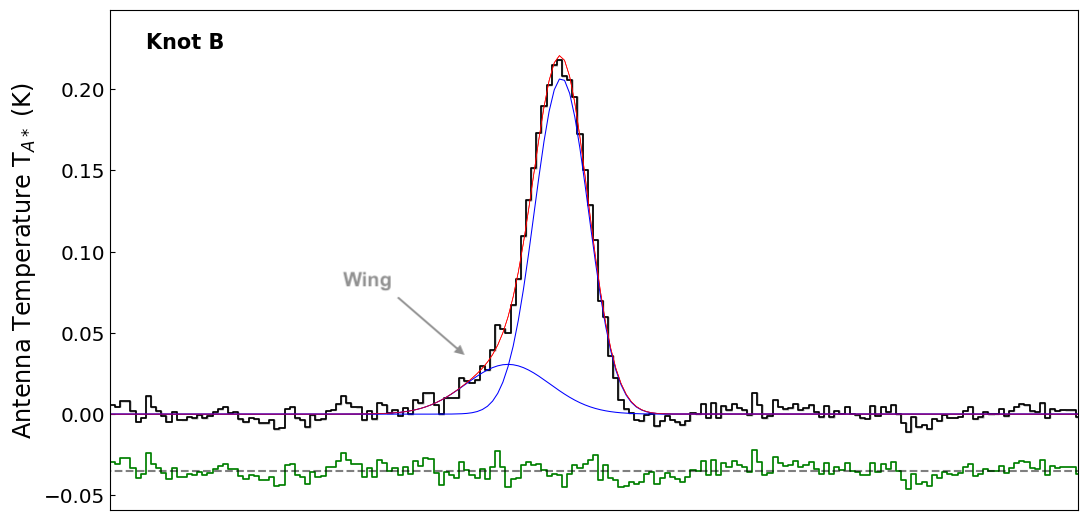}\\
\includegraphics[width=0.47\textwidth]{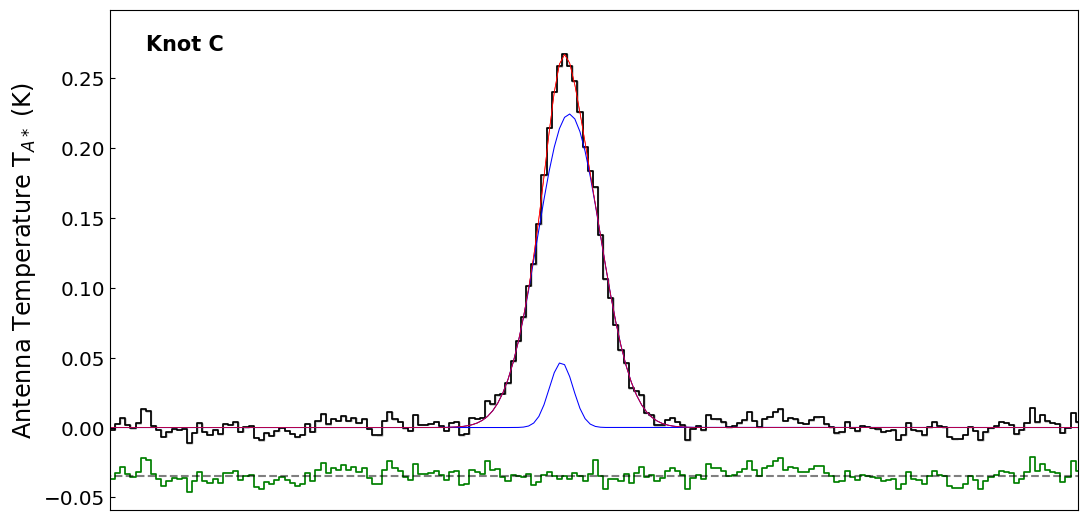}\\
\includegraphics[width=0.47\textwidth]{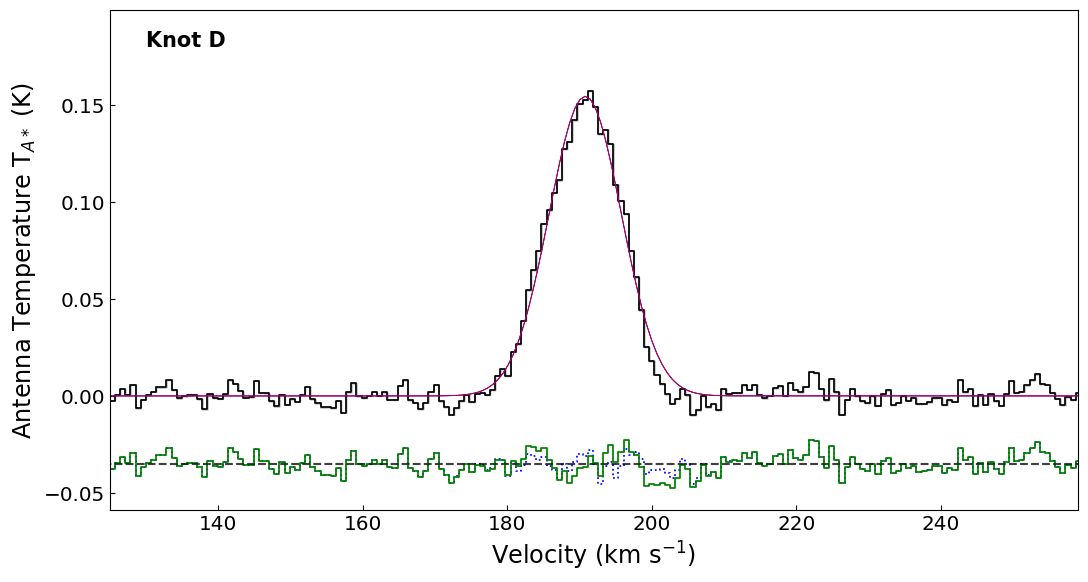}\\
\caption{Fitted spectra of the \am\ (4,4) line. Red line shows the best Gaussian fit to the extracted spectra (black line). The green line shows the residuals from the best fit. The blue profiles show the individual Gaussian components. The fit parameters are presented in Table \ref{spec-table}.}
\label{spectra-fit}
\end{figure}

\begin{table}[bt!]
\caption{\textbf{Kinematics of the \am~(4,4) transition.} }
\vspace{-4mm}
\centering
\begin{tabular}{lcc}
\hline\hline
\textbf{Parameter}\footnote{$A_c$ is the amplitude of the fit, $v_c$ is the central velocity of the component and $\sigma$ is the velocity dispersion. The $r$ parameter on the same line as each of the region names defines the radii value of that region.} ~ ~~~~~ ~	~~~ & ~ ~~ ~~~~ ~	&  \textbf{Value}  \\
\hline
\multicolumn{2}{l}{\textbf{Main Clump}} & \textbf{$r \sim$ 89\farcs1 (3.46 pc)} \\ [0.05cm]
\hline
~ ~ ~$A_{c}$  	&  ~~~ ~~~	&  64 $\pm$ 7 mK 	  \\ 
~ ~ ~$v_{c}$  	&  ~~~ ~~~	&  203.1 $\pm$ 0.4 \kms 	  \\
~ ~ ~$\sigma$ 	& 	& 5.2 $\pm$ 0.1 \kms	 \\
\greyline
~ ~ ~$A_{c}$  	&  ~~~ ~~~	&  52 $\pm$ 8 mK 	  \\
~ ~ ~$v_{c}$  	&  ~~~ ~~~	&  206.5 $\pm$ 0.2 \kms 	  \\
~ ~ ~$\sigma$ 	& 	& 2.8 $\pm$ 0.3 \kms	 \\
\hline
\multicolumn{2}{l}{\textbf{Shell}}  & \textbf{$r \sim$ 112\farcs4 (4.36 pc)}  \\ [0.05cm]
\hline
~ ~ ~$A_{c}$  	&  ~~~ ~~~	&  68 $\pm$ 7 mK 	  \\
~ ~ ~$v_{c}$ 	&	&  184.4 $\pm$ 0.1 \kms	   \\
~ ~ ~$\sigma$ 	& 	&  5.7 $\pm$ 0.1 \kms 	 \\
\greyline
~ ~ ~$A_{c}$  	&  ~~~ ~~~	&  12 $\pm$ 1 mK 	  \\
~ ~ ~$v_{c}$ 	&	&  202.3 $\pm$ 1.5 \kms	   \\
~ ~ ~$\sigma$ 	& 	&  10.0 $\pm$ 1.2 \kms 	 \\
\hline
\multicolumn{2}{l}{\textbf{Knot B}}  & \textbf{$r \sim$ 39\farcs0 (1.51 pc)}  \\ [0.05cm]
\hline
~ ~ ~$A_{c}$  	&  ~~~ ~~~	&  207 $\pm$ 26 mK 	  \\ 
~ ~ ~$v_{c}$ 	&	&  187.5 $\pm$ 0.2 \kms 	 \\
~ ~ ~$\sigma$ 	& 	&  3.7 $\pm$ 0.2 \kms 	  \\
\greyline
~ ~ ~$A_{c}$  	&  ~~~ ~~~	&  31 $\pm$ 12 mK 	  \\ 
~ ~ ~$v_{c}$ 	&	&  180.1 $\pm$ 4.5 \kms 	 \\
~ ~ ~$\sigma$ 	& 	&  5.8 $\pm$ 2.1 \kms 	  \\
\hline
\multicolumn{2}{l}{\textbf{Knot C}}  & \textbf{$r \sim$ 34\farcs1 (1.32 pc)}  \\ [0.05cm]
\hline
~ ~ ~$A_{c}$  	&  ~~~ ~~~	&  224 $\pm$ 8 mK 	  \\ 
~ ~ ~$v_{c}$ 	&	&  188.6 $\pm$ 0.1 \kms 	 \\
~ ~ ~$\sigma$ 	& 	&  4.4 $\pm$ 0.1 \kms 	  \\
\greyline
~ ~ ~$A_{c}$  	&  ~~~ ~~~	&  47 $\pm$ 8 mK 	  \\ 
~ ~ ~$v_{c}$ 	&	&  187.5 $\pm$ 0.2 \kms 	 \\
~ ~ ~$\sigma$ 	& 	&  1.7 $\pm$ 0.3 \kms 	  \\
\hline
\multicolumn{2}{l}{\textbf{Knot D}}  & \textbf{$r \sim$ 51\farcs1 (1.98 pc)} \\ [0.05cm]
\hline
~ ~ ~$A_{c}$  	&  ~~~ ~~~	&  154 $\pm$ 2 mK 	  \\ 
~ ~ ~$v_{c}$ 	&	&  190.8 $\pm$ 0.1 \kms 	 \\
~ ~ ~$\sigma$ 	& 	&  5.0 $\pm$ 0.1 \kms 	  \\
\hline\hline
\end{tabular}
\label{spec-table}
\end{table}

The \am\ molecule has a known hyperfine structure containing 4 primary satellite lines that are offset from the main component \citep[e.g.,][gray arrows in the left-most panel of Figure \ref{spectra}]{Ho83}. These satellite lines are clearly detected in the (1,1)--(3,3) transitions. However, drawing upon the (1,1) transitions as an example, we can see that the inner satellite lines are blended with the main component in all of the extracted spectra. It can be seen that the outer satellite lines are also not clearly separated, particularly in the cases of the Shell and Knot D. We attribute this blending to the particularly large linewidths seen in this region and discuss this further in Section \ref{sec:mom-map}.
The \am\ (3,3) line in the Knot B region shows a narrow-line component around 195 \kms\ that is not observed in any of the other regions (Figure \ref{spectra}). This feature appears to be contained to the area marked with the black `$\times$' in Figure \ref{morphfig}. This narrow-line velocity feature could be associated with \am\ (3,3) maser emission and is discussed in detail in Section \ref{3maser}. Several of the profiles show `wings' in the main profile component, as annotated in the Shell panel of Figure \ref{spectra}, suggesting a secondary component offset in velocity. This secondary component is most visible in the Shell profile, but is also present in the Main Clump and Knot B (Figures \ref{spectra} \& \ref{spectra-fit}).  

The hyperfine structure of \am\ molecular emission has been used in many studies to extract valuable information on the physical properties (primarily temperature and optical depth/column density) of the regions under examination \citep[e.g., GAS, KEYSTONE, RAMPS;][respectively]{gas1,keystone1,ramps1}. However, due to the line blending seen here, such analysis is particularly complicated and may even be impossible (or so degenerate as to be inconclusive). In this publication we are only focusing on the gas morphology and kinematics and will therefore not be using the hyperfine lines in our analysis. A follow up publication will investigate the hyperfine lines in more detail. Thus, we have chosen to use the (4,4) transition as the primary line from which to extract the basic velocity parameters (e.g., $v_{c}$, $\sigma$) of our observations. 

Figure \ref{spectra-fit} shows the \am\ (4,4) line profile for these regions (black histogram), fit with 1--2 Gaussian components (blue profiles) using the python program \texttt{pyspeckit}. The parameters for these fit components are shown in Table \ref{spec-table}.  Most of the spectra are best fit with two Gaussian components with the exception of Knot D, for which a single component is optimal. The solid green line, offset at -0.025 K in all panels, shows the residual of the best-fit line profile, shown as the red line in Figure \ref{spectra-fit}. A secondary residual (blue dotted line) is shown for Knot D, for which there is slight improvement in the fit if a secondary component is included. This improvement in the residuals is best seen around 200 \kms. However, as this improvement is at the 1$\sigma$ level of the noise we will examine this source using the single Gaussian fit.

Most of the measured velocity dispersion values are between 2--5 \kms\ (Table \ref{spec-table}). The largest velocity dispersion measured in our spectra is associated with the Shell (10.0 $\pm$ 1.2 \kms). Knot D is best fit with a broadest single component spectrum (5.0 $\pm$ 0.1 \kms). Knot C is the brightest of the 5 regions shown here.
The offset in the secondary velocity component is the most extreme in the Shell and Knot B, with velocity offsets $>$ 5 \kms. These lower intensity velocity components, discussed previously as `wings', are annotated in these two panels of Figure \ref{spectra-fit}

Due to the hyperfine structure of the \am\ molecule, analysis of the kinematics can be quite complex because of the satellite lines. These hyperfine lines become much more prominent at the lower transitions (e.g., Figure \ref{spectra}). Therefore, we will be using the \am\ (3,3) line for the kinematic analysis for the following arguments. First, the emission in the (3,3) line is relatively bright when compared with the (1,1) line (see Figure \ref{morphfig}). Second, hyperfine satellite lines in the (3,3) transition are relatively faint compared to the main component. These strong hyperfine lines in the (1,1) and (2,2) transitions are visible in Figure \ref{spectra}. Although the hyperfine lines in the (3,3) transition are visible in Figure \ref{spectra}, the ratio of the hyperfine lines to the main component is lower than the (1,1) and (2,2) transitions.  Third, the larger velocity offset between the satellite lines and the main component allows for better isolation of the main component in velocity space. We note in latter sections where the analysis could be influenced by the lower level hyperfine emission. And lastly, while the \cyanobut\ (9-8) transition would be an excellent candidate to use for kinematic analysis, due to the lack of hyperfine structure in the line profile, the relatively low SNR makes fitting the velocity components across the structure unfeasible. 
Although the \am\ (4,4) line was used for the spectral line analysis, discussed earlier in the section and shown in Figure \ref{spectra-fit}, the relatively faint nature of this transition on a pixel-by-pixel analysis (e.g., see the maximum intensity emission shown in Figure \ref{morphfig2}), makes this transition unfeasible to use in the `moment' analysis in the following section.

\begin{figure*}
\centering
\includegraphics[width=1.0\textwidth]{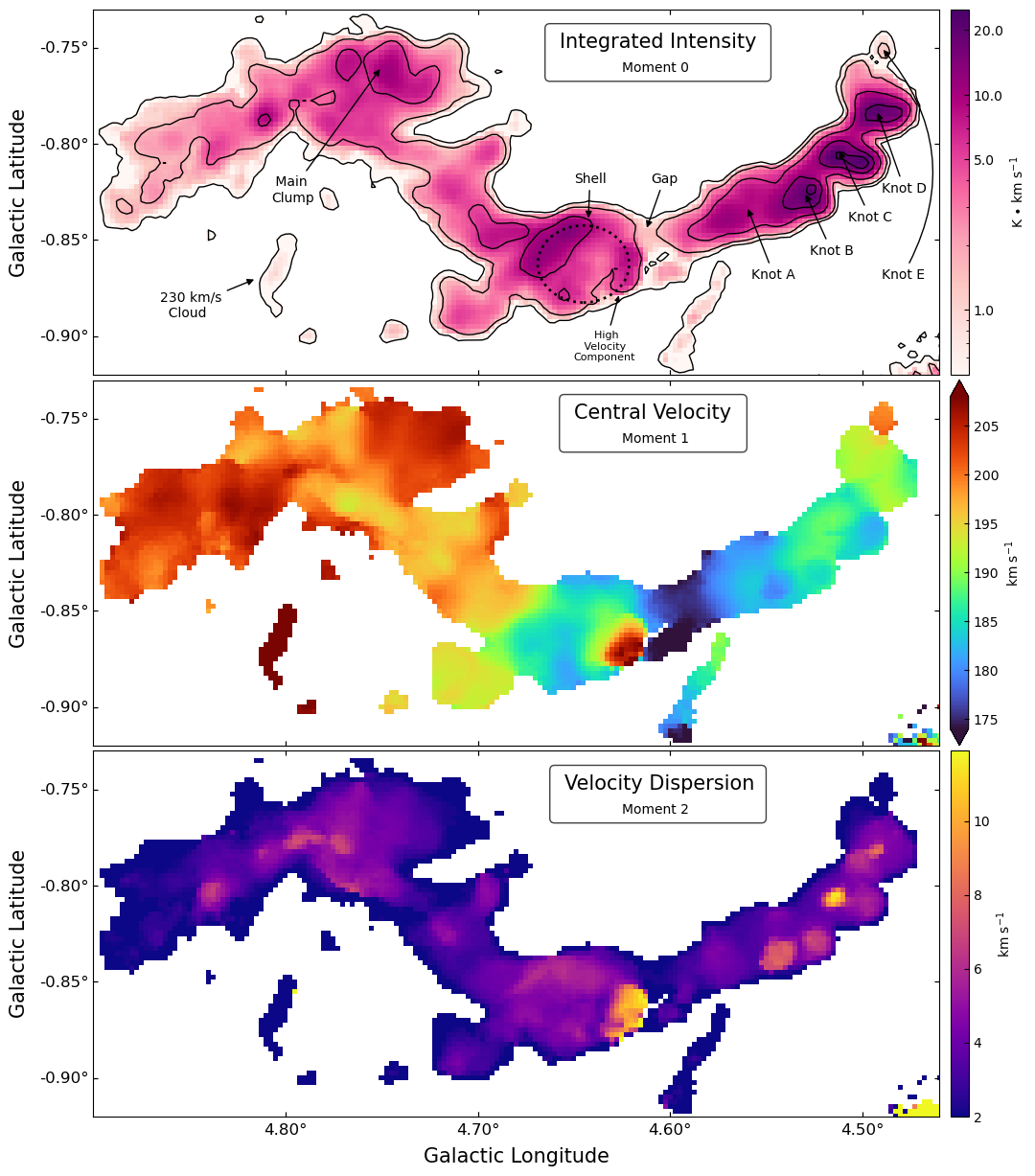}
\caption{Integrated Intensity (Moment 0; top panel), Central Velocity (Moment 1; middle panel) and Velocity Dispersion (Moment 2; bottom panel) for the \am\ (3,3) transition in the M4.7-0.8 cloud. Contours on the peak intensity (top panel) show the 15, 25, 60, 120, 175 $\times$ 10 mK (rms noise). All three panels use a lower limit cut-off of 150 mk (15$\sigma$; lowest level contour in the top panel). Annotated in the top panel are the features discussed in Section \ref{res}. 
}
\label{kin-fig}
\end{figure*}

\subsubsection{Central Velocity \& Velocity Dispersion}
\label{sec:mom-map}

In this section, we use the \am\ (3,3) data to investigate the gas kinematics in the cloud by creating a moment 0 ($M_0$, integrated intensity), moment 1 ($M_1$, central velocity) and moment 2 ($M_2$, velocity dispersion) maps. These moment maps are defined by the following equations:

\begin{equation}
M_0 = \int{T_A^*}(v)dv,
\label{M0}
\end{equation}

\begin{equation}
M_1 = \frac{\int{T_A^*}(v)v~dv}{\int{T_A^*}(v)dv},
\label{M1}
\end{equation}

and 

\begin{equation}
M_2 = \sqrt{\frac{\int T_A^*(v)(v-M_1)^2dv}{\int{T_A^*}(v)dv}} .
\label{M2}
\end{equation}

Where $T_A^*$ is the antenna corrected temperature of the data cube in Kelvin and $v$ is the velocity in \kms. 
These types of visualizations can illustrate velocity gradients across the source and can reveal regions with high velocity dispersion. However, care is needed when viewing these images - especially in cases where hyperfine lines are strong or multiple components are present. Figure \ref{kin-fig} shows these three moment images, discussed above, for the \am\ (3,3) line.

All three moment maps shown in Figure \ref{kin-fig}, M$_0$--M$_2$, were created in \texttt{CASA} \citep{CASA}, using the \texttt{immoments} command. We used a robust lower limit cut off of 150 mK (15$\sigma$) for all three moments to avoid fitting noise components in our integration. We used a velocity range of 146.4--257.8 \kms\ when making these images.

The \am\ (3,3) integrated emission (Moment 0; Figure \ref{kin-fig}, top) generally traces a distribution similar to that of the \am\ (3,3) peak intensity (Figure \ref{morphfig}, bottom). As noted previously in Section \ref{morph}, the elongation observed in the peak emission in Knots B--D is also observed to be present in the integrated emission as well. 
The \am\ (3,3) central velocity (Moment 1; Figure \ref{kin-fig}; middle panel) shows that velocities in the cloud generally range between 180$-$210 \kms. This velocity range is consistent with values expected for the Midpoint of the dust lanes (\til200 \kms; see Figure \ref{introfig2}). 

In general, the Nexus has slightly higher velocities, around 195$-$210 \kms, when compared to most of the emission in the Filament (180$-$195 \kms). This subtle velocity distribution (\til15 \kms) across the broad \til60 pc structure results in a large-scale velocity gradient across the M4.7-0.8 cloud of \til0.25 \kmsp, from East to West.

Within the Nexus region of the M4.7-0.8 cloud, there are no broad-scale velocity gradients. There are a few localized gradients of a few \kms\ associated with emission across adjacent cloud clumps. South of the bulk emission in the Nexus is an elongated vertical clump that is labeled as the 230 \kms\ cloud in the top panel of Figure \ref{kin-fig}. This feature is over-saturated in the M$_1$ map (Figure \ref{kin-fig}, middle panel) and is a distinct component along our line-of-sight. This feature is also observed to be discrete component in position-velocity space (see Figure \ref{Sine-pv} and Section \ref{pv-section} in the Appendix).

The Filament shows a more complex velocity structure compared with the Nexus. Towards the `Gap' in the Filament (annotated in Figure \ref{kin-fig}, top) we observe the lowest velocity across the entire cloud, with velocity values extending down to 175 \kms. There is a generally increasing velocity gradient away from the `Gap' location in both directions along the Filament. Adjacent to the `Gap', near the western rim of the `Shell', we observe the highest velocity associated with the Filament (210--215 \kms), which we have labeled as the `High Velocity Component'. In the Moment 1 map, in the region between the High Velocity Component and the Gap, we observed a sharp jump in the central velocity across a few pixels. This type of discontinuity in the velocity distribution typically occurs when two components are located along the same line of sight. In this case, the central velocity in the fitting algorithm is shifted by bright emission in one component compared with the other (see Equation \ref{M1}) across a few pixels. This higher velocity emission appears to be constrained to the western rim of the Shell suggesting this feature is likely a secondary component along the line-of-sight. Furthermore, the high velocity component is observed to be a distinct feature in the position-velocity diagram (Figure \ref{Sine-pv}) and in channel maps (Figure \ref{channels}) - both of which are discussed in Appendix Section \ref{sec:B}.
This high-velocity component is likely the secondary profile observed in the Shell spectrum (10 \kms; Figure \ref{spectra-fit}, Table \ref{spec-table}). 
As the spectrum shown in Figure \ref{spectra-fit} is from the \am\ (4,4), and therefore from the same \am\ molecule shown in Figure \ref{kin-fig}, they are likely tracing the same velocity components.

The general velocity gradient in the Filament is clear, with an increasing velocity from Knot A (\til180 \kms) to Knot E (200 \kms).\footnote{Figure \ref{pillars} shows a scaled version of the moment 1 map for the knots associated with the Filament.} Knots B and C show a velocity gradient across the elongated knot substructure. In these two sources, we also observe a velocity gradient from the northeast to the southwest. In the NE, the emission is higher, around 190 \kms, whereas in the SW the emission lower, around 183--185 \kms. This results in a velocity gradient across the substructure of 2.5 \kmsp. Knot E is observed to have the highest velocity and appears distinct from its neighbor, Knot D.

The \am\ (3,3) velocity dispersion (Moment 2; Figure \ref{kin-fig}, bottom) in the M4.7-0.8 cloud is generally around 2--10 \kms. The periphery of the cloud is observed to have lower velocity dispersion ($\leq$2 \kms), whereas the interior of the cloud is observed to have higher velocity dispersion (\til6 \kms). There are a few clumps where the velocity dispersion exceeds 10 \kms. The high velocity component is observed to be one of these regions. This high velocity dispersion is consistent with the spectral fitting results for the secondary component in the Shell (Table \ref{spec-table}).

However, many of these other clumps with large velocity dispersion values have prominent hyperfine lines. Due to the approach in which \texttt{CASA} calculates the velocity dispersion, by using a simple, single Gaussian fit, satellite lines can result in a broadening of the velocity dispersion in regions with relatively bright hyperfine lines. While the hyperfine lines are suppressed in the (3,3) transition when compared with the (1,1) line (see Figure \ref{spectra}), regions where the hyperfine line are quite strong can result in high apparent velocity dispersion values in Moment 2 maps. Knot C is one of these regions which shows relatively high velocity dispersion in the Moment 2 map, but an examination of the (3,3) profile for this region reveals relatively strong hyperfine lines which are likely causing the high dispersion values in this plot. Here we take the presented dispersion values, in Figure \ref{kin-fig}, as illustrative and reserve a full analysis of the individual hyperfine component velocity dispersions for a future publication.

\subsubsection{Detection of a New \am\ (3,3) Maser}
\label{3maser}

\begin{figure}
\centering
\includegraphics[width=0.48\textwidth]{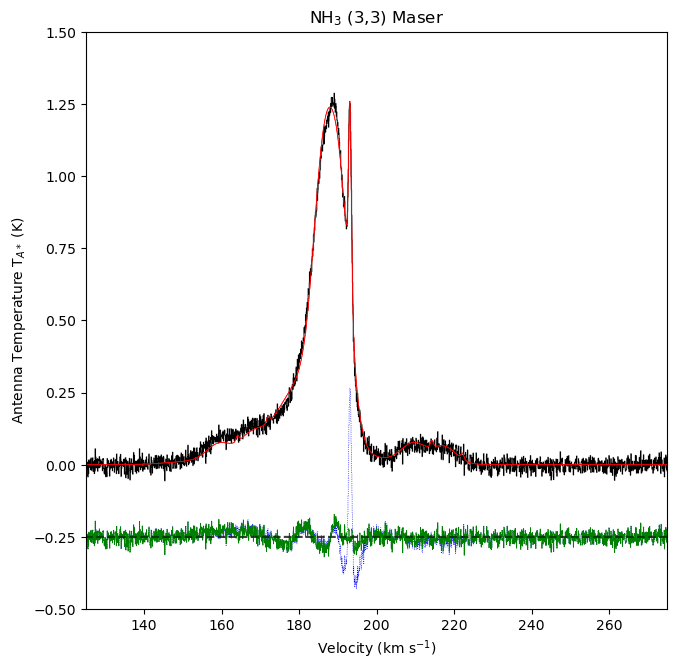}
\caption{Spectrum showing the GBT \am\ (3,3) maser, using the native resolution dataset ($\Delta$V=0.07 \kms). The black line shows the raw data extracted from the Knot B region indicated in Figure \ref{morphfig}. The red line shows the sum of the three components fitted with the ammonia wrapper in \texttt{pyspeckit}. The values for these three components are shown in Table \ref{maser-table}. The solid green line shows the residuals from fitting all three components from Table \ref{maser-table}. The dotted blue line shows the residuals of the fit excluding the narrow line component ($v_c$=193.16 \kms) - illustrating the maser line in the residuals.}
\label{am-maser}
\end{figure}

A narrow-line velocity feature is observed in the \am\ (3,3) spectral cube at $l$=4\fdg53, $b$=-0\fdg82 (see black `$\times$' in Figure \ref{morphfig}), and is clearly visible in Figure \ref{spectra} (Knot B, \am~(3,3) panel), and Appendix Figures \ref{Sine-pv} \& \ref{channels}. The location of this narrow line emission feature suggests it is associated with the north-western region of Knot B. Bright, narrow-line profiles in the \am\ (3,3) are often indicative of maser emission \citep[e.g.,][]{Mangum94,Kraemer95}, supporting our identification of this feature as a new \am\ (3,3) maser detection.

For this analysis, we will be using an un-smoothed version of the dataset. This dataset was sampled at the native resolution of 0.07 \kms, with no additional smoothing (see Section \ref{data redux}). This higher resolution channel width is necessary to fully resolve and sample the narrow line component associated with the maser emission. To better understand the nature of the source, we fit the \am\ (3,3) spectrum, shown in Figure \ref{spectra}, using the python program \texttt{pyspeckit} \citep{pyspeckit}. This package allows us to account for the prominent hyperfine components of the \am\ (3,3) transition when performing our fit. The raw \am\ (3,3) spectrum for Knot B is shown in Figure \ref{am-maser} as a black histogram. We fit this profile with a two-component fit, using similar central velocity and velocity dispersion values previously measured in the \am\ (4,4) data (see Table \ref{spec-table}). We note that despite the higher spectral resolution of this data, compared with the spectrum used in Figure \ref{spectra-fit}, we still recover the same components. The residuals from this two-component fit are shown as a dashed blue line at -0.25 K, in which the narrow line feature is clearly detected. When adding an additional narrow-line component to our fitting algorithm, we can see that the residuals are greatly improved (solid green line at -0.25 K in Figure \ref{am-maser}). The parameters of this 3-component fit, with the combined fit shown as a red line in Figure \ref{am-maser}, are presented in Table \ref{maser-table}. The introduction of a fourth component, with a central velocity of 181 \kms\ and a broad velocity dispersion of \til8 \kms, can slightly improve the residuals in the fit around 180 \kms\ (solid blue line in Figure \ref{am-maser}). However, without additional evidence to support this fourth component, it is unclear whether this is an additional velocity component or the result of noise in the brightest emission channels. Higher resolution observations are necessary to spatially disentangle these multiple components and determine whether a fourth component is indeed present in this region.

\begin{table}[bt!]
\caption{\textbf{Kinematics of the \am~(3,3) transition in Knot B}}
\vspace{-4mm}
\centering
\begin{tabular}{lcc}
\hline\hline
\textbf{Parameter}\footnote{$v_c$ is the central velocity of the component and $\sigma$ is the velocity dispersion.} ~ ~~~~~ ~	 & ~ ~~ ~~~~ ~	&  \textbf{Value}  \\
\hline
\multicolumn{3}{l}{\textbf{Main Component}} \\ [0.05cm]
\hline
~ ~ ~$v_{c}$  	&  ~~~ ~~~	&  188.42 $\pm$ 0.02 \kms 	  \\
~ ~ ~$\sigma$ 	& 	& 3.05 $\pm$ 0.03 \kms	 \\
\hline
\multicolumn{3}{l}{\textbf{Secondary Component}} \\ [0.05cm]
\hline
~ ~ ~$v_{c}$ 	&	&  181.88 $\pm$ 0.21 \kms	   \\
~ ~ ~$\sigma$ 	& 	&  7.49 $\pm$ 0.09 \kms 	 \\
\hline
\multicolumn{3}{l}{\textbf{Maser Component}} \\ [0.05cm]
\hline
~ ~ ~$v_{c}$ 	&	&  193.16 $\pm$ 0.01 \kms 	 \\
~ ~ ~$\sigma$ 	& 	&  0.37 $\pm$ 0.01 \kms 	  \\
\hline\hline
\end{tabular}
\label{maser-table}
\end{table}

\begin{figure*}
\centering
\includegraphics[width=1.0\textwidth]{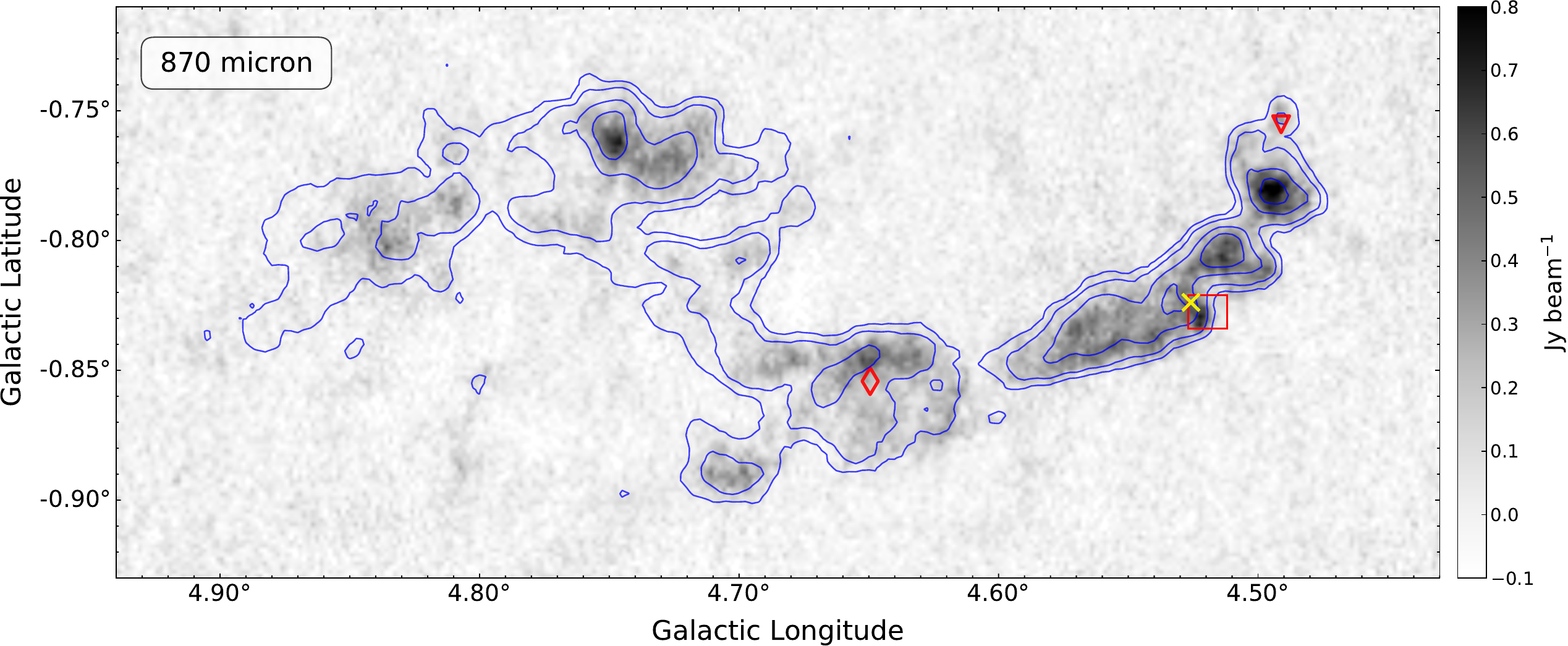}
\caption{Sub-mm dust emission in the (Midpoint) Cloud at 870 \micron\ from ATLASGAL \citep[18\farcs2 resolution, 0.04 \jybe\ sensitivity;][]{Schuller09}. Overlaid are the \am\ (1,1) peak emission at 0.15, 0.25, 0.5, and 0.75 K level contours (Figure \ref{morphfig}). The yellow $\times$ symbol marks the location of the \am\ (3,3) maser (Section \ref{3maser}). The red diamond marks the location of the radio point source detected in the VLASS survey \citep[][see Figure \ref{vlass-ps}]{VLASS}. The red triangle marks the location of the 70 \micron\ point source from \cite{Elia17}.
The red box shows the FOV region for the source targeted in Figure \ref{hii reg}.
}
\label{atlasgal}
\end{figure*}

\begin{figure*}
\centering
\includegraphics[width=1.0\textwidth]{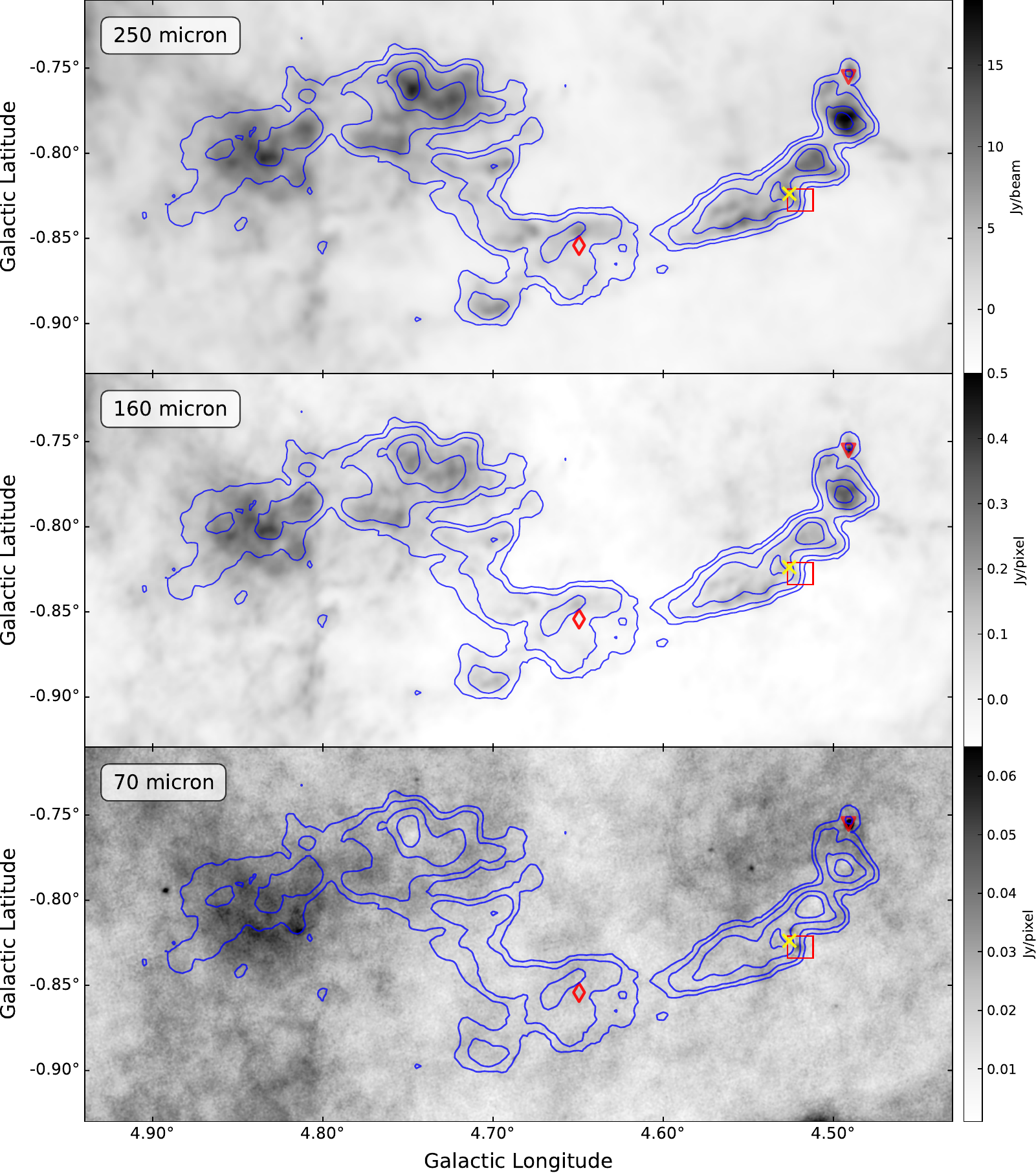}
\caption{Far-Infrared Herschel dust emission in the Midpoint Cloud at 250 \micron, 160 \micron, and 70 \micron\ \citep{Molinari16}. The Herschel bands have a resolution of 6, 3.2 and 3.2 arcsec/pixel (top to bottom respectively). The Herschel bands have a sensitivity of 0.2 \jybe\ (250 \micron), 10 mJy/pixel (160 \micron), and 4 mJy/pixel (70 \micron). Overlaid are the \am\ (1,1) 0.15, 0.25, 0.5, and 0.75 K level contours from Figure \ref{morphfig}. The yellow $\times$ symbol marks the location of the \am\ (3,3) maser, discussed in Section \ref{3maser}. The red diamond marks the location of the radio point source detected in the VLASS survey \citep[][see Figure \ref{vlass-ps}]{VLASS}. The red triangle marks the location of the 70 \micron\ point source from \cite{Elia17}. The red box shows the FOV region for the source targeted in Figure \ref{hii reg}. A 3-color image of these three datasets is shown in the Appendix (Figure \ref{introfigr}). 
}
\label{herschel}
\end{figure*}

\begin{figure*}
\centering
\includegraphics[width=1.0\textwidth]{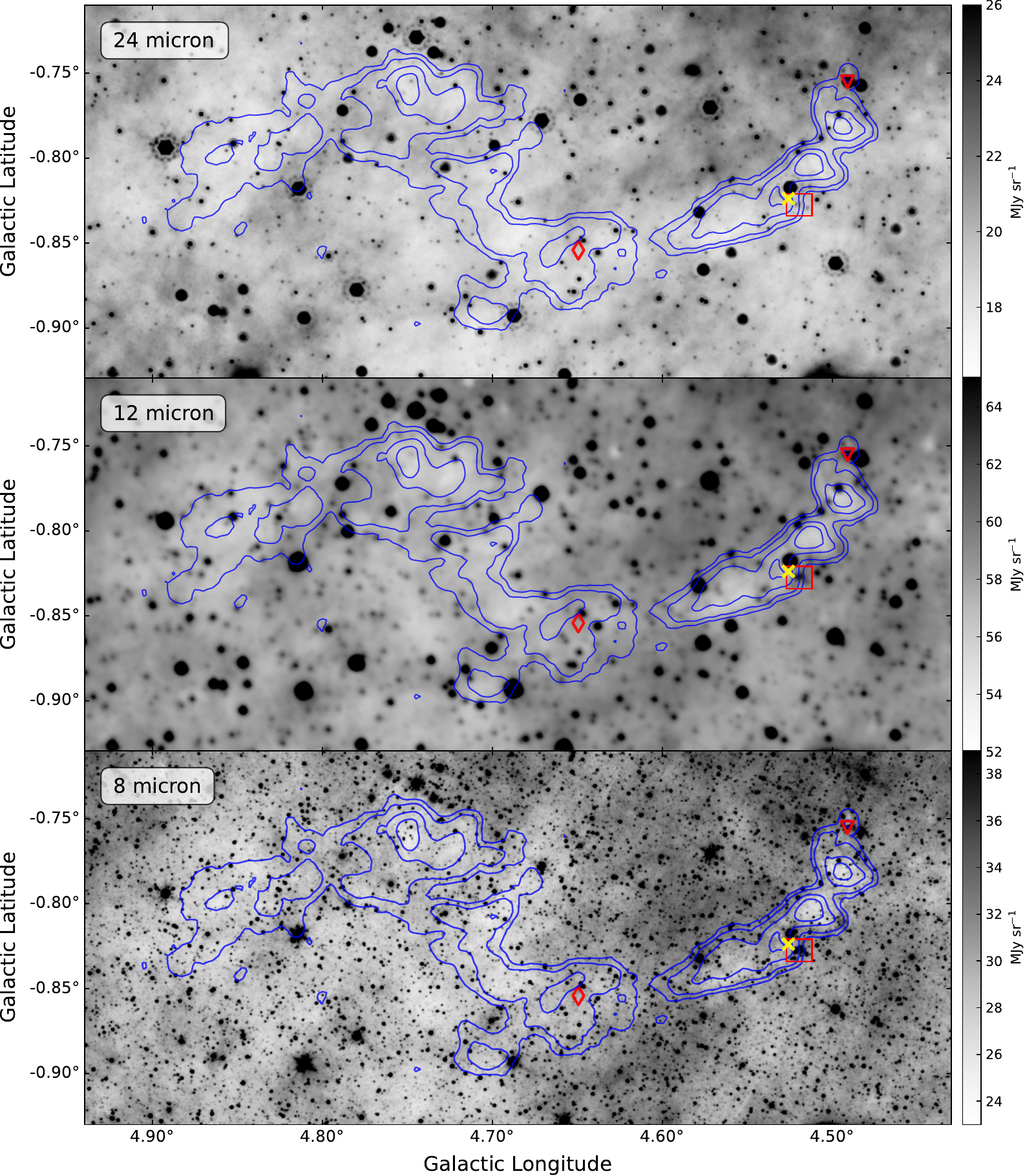}
\caption{Mid-Infrared Dust emission in the Midpoint Cloud at 24 \micron\ \citep[MIPSGAL, 6\arcsec\ resolution, 0.6 m\jybe;][]{mipsgal}, 12 \micron\ \citep[WISE, 6\farcs5, 0.2 m\jybe;][]{wise}, 8 \micron\ \citep[IRAC-4, 1\farcs2, {0.4 m\jybe};][]{Churchwell09}. Overlaid are the \am\ (1,1) at 0.15, 0.25, 0.5, and 0.75 K level contours from Figure \ref{morphfig}. The yellow $\times$ symbol marks the location of the \am\ (3,3) maser, discussed in Section \ref{3maser}. The red diamond marks the location of the radio point source detected in the VLASS survey \citep[][see Figure \ref{vlass-ps}]{VLASS}. The red triangle marks the location of the 70 \micron\ point source from \cite{Elia17}. The red box shows the FOV region for the source targeted in Figure \ref{hii reg}. 
}
\label{mid-ir}
\end{figure*}

\section{Dust Emission in M4.7-0.8} 
\label{sec:dust}

In this section, we compare our \am\ radio observations (Section \ref{res}) to previous infrared and sub-mm survey datasets as a way of investigating the dust continuum emission associated with the dense molecular gas.
Figure \ref{atlasgal} shows the sub-mm (870 \micron) continuum from the Apex ATLASGAL survey \citep{Schuller09}. The blue \am\ (1,1) contours clearly trace a similar structure in the sub-mm emission indicating the 870 \micron\ continuum emission is correlated with the \til200 \kms\ gas. 
The Main Clump in the Nexus and Knot D in the Filament are relatively bright at 870 \micron. We also observe a brighter feature located toward the lower right of the \am\ (3,3) maser (yellow $\times$), as highlighted by the red box in Figure \ref{atlasgal} (this annotated region will also be highlighted in Figures \ref{herschel}--\ref{mid-ir}). We will discuss a multi-wavelength analysis of this source in Section \ref{knotB}.

Figure \ref{herschel} shows the far-IR emission in 
three Herschel datasets from \cite{Molinari16}, at: 250 \micron\ (top), 160 \micron\ (middle), and 70 \micron\ (bottom).  
The 250 and 160 \micron\ emission show a somewhat similar emission distribution as the 870 \micron\ emission (Figure \ref{atlasgal}), where we observe similar features (e.g., Main Clump in the Nexus, bright knots in the Filament). The 70 \micron\ emission appears quite different from the 160 and 250 \micron\ emission, with many of the extended emission features not detected at 70 \micron. 

The cloud appears to have two main dust components: one associated with the left side of the Nexus, and one associated with the Filament and the right side of the Nexus (see Figure \ref{introfigr}). 
The left side of the Nexus is observed to be relatively faint in the 870 \micron\ data (Figure \ref{atlasgal}) but increases in brightness towards the shorter IR wavelengths (Figure \ref{herschel}).

The right side of the Nexus and the region associated with the Filament are relatively bright in the 870 and 250 \micron\ emission, but become relatively faint in the 160 \micron\ emission, when compared with the left side of the Nexus. These emission regions are not observed in the 70 \micron\ emission and in the case for the Main Clump and Knots C and D they are observed to be somewhat extincted (Figure \ref{herschel}). 
The Gap in the Filament is also observed in the Far-IR and sub-mm emission as well. 
As the Filament is not present in the 70 \micron\ data, we do not detect the Gap in the 70 \micron\ emission.

The 70 \micron\ emission also appears to have several compact sources throughout the structure. The most notable of these is the compact source observed with Knot E (red triangle in Figure \ref{atlasgal}-\ref{mid-ir}). This feature is discussed in more detail in Section \ref{knotE}. We also note a compact 70 \micron\ source that coincides with compact emission observed in the 870 \micron\ emission. This source is located in Knot B, near the \am\ (3,3) maser (yellow `$\times$' in Figures \ref{atlasgal} and \ref{herschel}). This compact feature does not appear to be detected in the 160 and 250 \micron\ datasets. We will discuss this feature in more detail, later in this section.

\begin{figure*}
\centering
\includegraphics[width=1.0\textwidth]{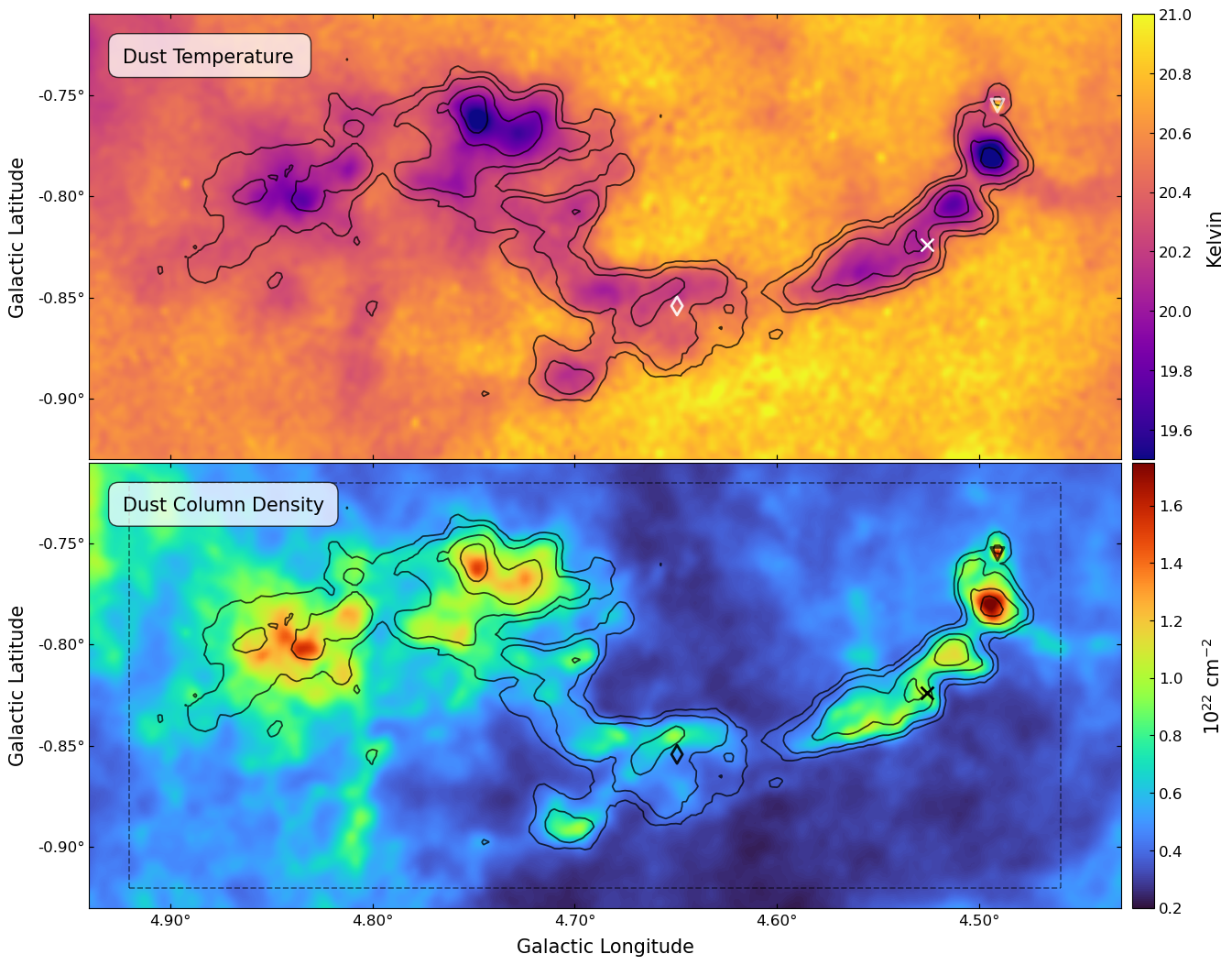}
\caption{Dust temperature (top) and column density (bottom) of the Midpoint cloud from the \cite{Marsh17} PPMAP dataset. Both images have a resolution of 12\arcsec\ and a sensitivity of \til0.1 K (top) and $6 \times 10^{19}$ \cms\ (bottom). Contours show the maximum intensity \am\ (1,1) emission at 0.15, 0.25, 0.5, and 0.75 K (see Figure \ref{morphfig}). The $\times$ symbol shows the locations of the \am\ (3,3) maser from Section \ref{3maser}. The triangle marks the location of the 70 \micron\ point source discussed in Section \ref{knotE}. The diamond shows the location of the radio point source discussed in Section \ref{feedback}. The grey dashed box, in the bottom panel, illustrates the region used to calculate the cloud mass - utilizing the column density data shown here.
}
\label{dust-temp}
\end{figure*}

Figure \ref{mid-ir} shows the mid-IR wavelength emission at 24 \micron\ (top; MIPSGAL), 12 \micron\ (middle; WISE), and 8 \micron\ (bottom; GLIMPSE). The mid-IR emission reveals the cloud to be an infrared dark cloud at short wavelengths due to the relatively low emission from within the cloud compared to the mid-IR background. Although this cloud is observed to be IR-dark, we can still observe hints of the same general structure as traced by the \am\ contours. This general structure is the clearest towards the Filament region, where we observe a sharp contrast between the Filament region and the background emission. The Nexus region shows a much more extended IR-dark region, indicating the dense gas, as traced by the \am, is concentrated within this larger extinction feature. Furthermore, this extended region around the Nexus correlates with the bulk $^{12}$CO emission shown in Figure \ref{introfig3}. We will discuss this extended component in the Nexus in more detail in Section \ref{sect:411}.
This lower-intensity mid-IR emission is strongest towards Knots C and D and the Main Clump of the cloud, indicating higher extinction in these regions. These are the same regions where we also observed 70 \micron\ extinction (Figure \ref{herschel}, bottom). While this lower-level emission is the most extreme in these regions, all of the knots in the Filament show mid-IR extinction, including the most compact knot, Knot E. This higher extinction in these sources is likely due to the higher column densities in this region, which we discuss further in Section \ref{dust-prop}.

The compact region observed to the lower right of the \am\ (3,3) maser location (yellow $\times$), is also detected in the 8 \micron\ emission, and faintly observed in the 12 and 24 \micron\ emission. This source is discussed in more detail in Section \ref{knotB}.

\subsection{Dust properties in M4.7-0.8}
\label{dust-prop}

Figure \ref{dust-temp} shows the dust temperature (top) and N(H$_2$) column density (bottom) from \cite{Marsh17} for M4.7-0.8,\footnote{These data are publicly available \hyperlink{http://www.astro.cardiff.ac.uk/research/ViaLactea/}{here}. M4.7-0.8 is located in field l004.} based on the Herschel 70, 160, 250, 350, and 500 \micron\ measurements. 
The dust temperature in the cloud is observed to be between 19--21 K (Figure \ref{dust-temp}, top). The morphology of the cooler dust component ($T_{dust}$ $<$ 20.5 K) appears to trace the M4.7-0.8 cloud (black contours; see Figure \ref{morphfig}). The dust in the region surrounding the \am\ emission in the Nexus (\til20.25 K) is slightly cooler than the dust in the region around the Filament (\til20.75 K). This could be an indication that the environment around the two regions is different. The extended emission around the Nexus, discussed in Section \ref{sec:dust}, is also relatively cool and shows similar dust temperature values as the bulk M4.7-0.8 cloud. The region around the Filament is observed to be relatively warm compared with the colder interiors.

The colder dust temperatures in the cloud (i.e., $<$ 20 K) are strongly correlated with the bright \am\ (1,1) regions, located in the inner-most regions of the clumps and knots. The coldest of these is Knot D, with a measured dust temperature of 19.2 K. 
Knot E is observed to be hotter than all other \am\ peaks, with a temperature of 21 K. We note a slight positional offset between this hot dust core and the \am\ (1,1) peak of \til10\arcsec. This offset is smaller than our GBT beam size (35\arcsec) and therefore could be caused by a pointing offset rather than a physical offset of the source.  
The emission surrounding Knot E appears to be slightly cooler (20.1 K) than the background emission (20.8 K), indicating this hot dust core could be embedded in a cooler envelope. We investigate this source, and the implication of an embedded core, in more detail in Section \ref{knotE}.

The N(H$_2$) column density in the cloud is around 0.5--2.0 $\times$ 10$^{22}$ cm$^{-2}$ (Figure \ref{dust-temp}, bottom). The highest column densities in the region trace the \am\ emission morphology of M4.7-0.8. All of the knots located in the Filament show higher column densities in their interiors than the surrounding regions, with the highest column densities in M4.7-0.8 (1.82$\times$10$^{22}$ cm$^{-2}$) associated with Knot D. Knot E also shows relatively high column densities (1.65$\times$10$^{22}$ cm$^{-2}$), indicating that in addition to being quite compact ($<$10\arcsec; 0.4 pc), it is quite dense as well. We also note that these column density values are consistent with those in GMCs (see Section \ref{GMC}).

The Nexus region of the cloud also shows moderately high column densities, with values around 1.0$\times$10$^{22}$ cm$^{-2}$. There are two main high column density clumps in this region, the Main Clump (as annotated in Figure \ref{morphfig}) and a secondary clump located in the Eastern region of the cloud ($l$=4\fdg84, $b$=-0\fdg80). This Eastern source does not appear to stand out in the \am\ emission nor the \cyanobut\ emission. However, this region does appear to be relatively bright in the 250 \micron\ emission and in the Herschel short wavelength datasets (70 and 160 \micron). This region of the cloud also appears to be quite large, extending well past the \am\ (1,1) contours and appearing to correlate with the IR-dark region around the Nexus, observed in the Spitzer 8 \micron\ data (Figure \ref{herschel}, bottom).

The Gap region in the Filament also exhibits relatively low column density, with only a faint bridge connecting the eastern and western sides of the Filament. The low column density regime appears to be associated with a large cavity that extends towards the south, below the filament, and carves a lower density region north above the Gap. This lower column density at the location of the Gap could be an indication that feature is produced by nearby stellar feedback that has disrupted the Filament \citep[e.g.,][]{Zhang16,Chen23}.

\section{Discussion}
\label{dis}

In the following sections we discuss the physical properties of the M4.7-0.8 cloud and the implications of our results from the previous section.

\subsection{3D location of the Midpoint} 
\label{sect:3D}

\begin{figure}
\centering
\hspace{-7mm}
\includegraphics[width=.48\textwidth]{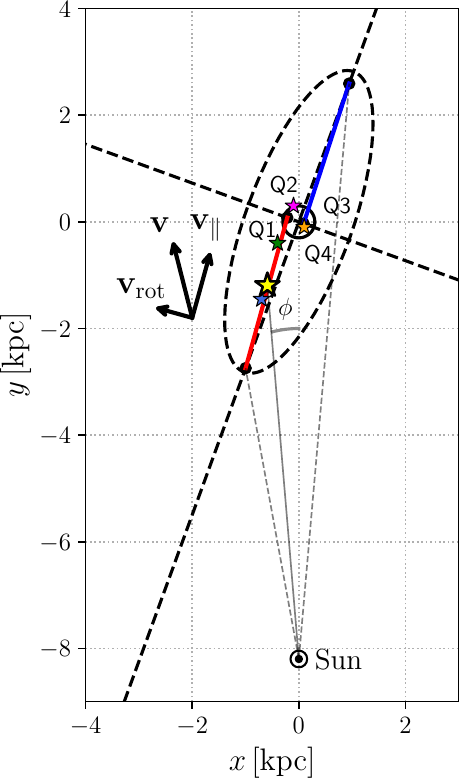}
\caption{Top-down geometry of the Milky Way Galactic Bar Dust Lanes from \citet[their Figure 2]{sormani19a}. Here, the red and blue lines show the near-side and far-side dust lanes, respectively, matching the orientation shown in Figure \ref{introfig1}. The yellow star shows the location of the Midpoint cloud in this diagram at (-0.59, -1.19). Labeled are four quadrants of the Galactic Bar: Q1-Q4. \textit{Quadrant 1} (Q1) corresponds to gas and dust on the near side of the Galactic bar dust lanes that is accreting into the Galactic Center. \textit{Quadrant 2} (Q2) corresponds to \textbf{matter} from Q1 that has `overshot' the GC. \textit{Quadrant 3} (Q3) corresponds to the far-side Galactic bar dust lanes that is accreting into the GC. Lastly, \textit{Quadrant} 4 (Q4) corresponds to gas and dust that has `overshot' the GC from the far-side dust lane. $\phi$ shows an angle of 4\fdg8, corresponding to the Galactic Longitude of the cloud. The solid gray line shows the distance to the Midpoint cloud at 7.035 kpc (yellow star; centered at $x$=-0.59, $y$=-1.19), while the dashed gray lines show the angular extent of the dust lanes (Figure \ref{introfig1}). The additional stars show the relative locations of G5 (blue), Bania's clump (green), the Helix stream (pink), and Sgr E (orange). 
}
\label{3D}
\end{figure}

Figure \ref{3D} shows the top-down geometry of the Milky Way Galactic Bar Dust Lanes from \citet[their Figure 2]{sormani19a}. 
The orientation and extent of the Galactic Bar Dust Lanes is marked with a dashed ellipse, with red and blue lines illustrating the near and far side dust lanes, respectively. The Galactic Bar Dust Lanes can roughly be divided into four main quadrants, labeled as Q1--Q4 in Figure \ref{3D}. Quadrants 1 and 3 correspond to the Near-side and Far-side Dust Lanes, respectively. Quadrants 2 and 4, correspond to gas and dust that have `overshot' the GC \citep[i.e., `spray',][see their Figure 1]{Kim24}, from Quadrants 1 and 3 respectively, and are continuing to traverse the Bar. The gas and dust in these overshot regions can then interact with the approaching material on the opposing side to produce strong shocks from cloud-cloud collisions \citep{sormani18b, sormani21, Gramze23}. These shocks are visible as vertical features in position-velocity space \citep[see Figure \ref{introfig2} and][]{sormani21}.  

The Midpoint cloud is located at a Galactic longitude of roughly $l$\til5\degree\ (angle $\phi$ in Figure \ref{3D}). Based on the geometry shown in Figure \ref{3D}, this suggests the cloud is located at (-0.59, -1.19), as marked by the yellow star. At this location, the cloud is calculated to be at a distance 7.035 kpc from the Sun. We also note the significantly low Galactic latitude of the cloud ($b$=-0\fdg85), which suggests the cloud is below the midplane of the Milky Way galaxy. Using the geometry presented in \citet[their Figure 2]{Goodman14}, we calculate M4.7-0.8 is located at 118.5 pc below the Sun, or 93.5 pc below the midplane (assuming the cloud has a distance of 7.0 kpc and the Sun has a positive $z$-height of 25 pc).

The cloud is also likely to be on the leading edge of the Dust Lanes. 
Orbital motions within the bar can create a shearing effect on the interstellar medium \citep[ISM,][]{Athanassoula92,Emsellem15,Renaud15}. This strong shear can inhibit the formation of dense structures, such as clouds, which leads to low star formation within the bar, even in regions with relatively high densities. However, at the bar's leading edge, supersonic turbulence, along with reduced shear and weaker tidal forces compared to the inner regions, promotes the formation and persistence of high-density structures \citep{Renaud15}. The leading edges of the bar can also produce converging flows and large-scale shocks that promote the accumulation of dense gas into kiloparsec-scale structures \citep[e.g.,][]{Athanassoula92}.

\subsubsection{Relative Location to Other Dust Lane Clouds} 
\label{sect:411}

The connection between M4.7-0.8 and the Galactic Bar dust lanes, as indicated by the line-of-sight velocity, implies this cloud is part of a much larger structure of gas and dust. Part of this larger structure, in the region surrounding M4.7-0.8, is shown in Figure \ref{introfig3} - as highlighted by the $^{12}$CO contours. 
The Nexus region of the Midpoint is located at a concentration of CO-rich gas and dust, with filamentary extensions towards the NE, S, and W directions (see Figure \ref{introfig3}). The western filamentary extension, which we have labeled as the Filament, is the narrowest, and brightest, of these three in the 250 \micron\ dust continuum. The southern and north-eastern extensions appear to follow a larger, \til20-30 pc wide, diffuse feature - as traced by $^{12}$CO emission. While the Filament may be too narrow to be resolved by the \cite{Dame01} dataset, southwest of the Filament is a large-scale structure in the dust continuum that appears to be traced by $^{12}$CO emission. The orientation of the Filament is similar to that of this western bridge, although follow-up spectral line observations of this region are necessary to confirm this possible association.\footnote{A follow up program to study the entirety of the Dust Lanes using the GBT has been approved, for 200 hours, under the ``Bar Ammonia Radiation in Lanes Filaments and YSOs Survey (\hyperlink{https://greenbankobservatory.org/science/gbt-surveys/barflys/}{BARFLYS})" large program, and will be carried out over the next 3 years.} Our observations of the Nexus reveal that the region contains numerous diffuse structures distributed throughout the region (Figure \ref{morphfig}). Additionally, most of the dense gas in this region is located along the northern edge of the $^{12}$CO emission (see Figures \ref{introfig3}). However, we note that this offset could be the result of the low-resolution of the $^{12}$CO (0\fdg125 pixel resolution). Follow-up observations of $^{12}$CO are necessary to confirm whether there is an offset between the $^{12}$CO and the dense gas, as traced by \am.

Slightly `upstream' from the Midpoint is the G5 complex \citep[M5.4-0.4;][]{rod06,Gramze23, Nilipour24}. G5 is suggested to be the result of a cloud-cloud collision \citep[][Figure 15]{Gramze23} from `spray' in Q4 that has overshot the CMZ and is colliding with accreting material in Q1 (see Figure \ref{3D}). This results in the broad linewidths (i.e., `vertical feature') observed in Figure \ref{introfig2} at $l$=5\fdg5 \citep{rod06,sormani21}. The 3D location of G5 is shown as a light blue star in Figure \ref{3D}. 
The location of the Midpoint `downstream' from G5, in the direction of accretion, suggests M4.7-0.8 maybe the result of a prior interaction. If the locations of the `spray' are relatively constant, over a few Myr, the accretion time to traverse between the G5 and M4.7-0.8 locations, then M4.7-0.8 could be the result of a G5-like collision that has relaxed into a filamentary structure. 

Further `downstream' from the Midpoint, on the Near-side Galactic Bar Dust Lane, is Bania's clump ($l$=3\degree; green star) and the Q1/Q2 transition (i.e, the $x_1$ (Bar) and $x_2$ (CMZ) intersection). The intersection of the CMZ and Bar are suggested to undergo strong shocks from cloud-cloud collisions \citep{Anderson20}. Sgr E (orange star; Figure \ref{3D}) is also suggested to be at the intersection between the $x_1$ and $x_2$ orbits and is argued to be undergoing intense star formation \citep{Anderson20}
However, only \til30\% (0.8 \Msuny) of the mass is suggested to accrete into the CMZ \citep{Hatchfield21}, indicating that a majority of the material overshoots the GC. The Helix stream is thought to be one of these regions \citep[][pink star in Figure \ref{3D}]{Veena24}. 
The Helix stream is argued to be associated with material that has overshot the CMZ from the near-side Dust lane and is continuing to traverse the Bar \citep{Veena24}. The central velocities of this structure are quite high (100--200 \kms), indicating that they are not likely associated with the $x_2$ orbit.

\subsection{Is M4.7-0.8 a GMC?} 
\label{GMC}

The leading edge of the Galactic Bar dust lanes tends to accumulate dense gas and promote the formation of over-dense structures \citep{Athanassoula92, Renaud15}. These over-dense structures have the potential to form GMCs within the larger Dust Lane. 
The physical extent of the M4.7-0.8 cloud (\til60 pc) suggests this structure is likely a GMC within the leading edge of the Near-side Dust Lane. GMCs are typically long (\til50--100 pc), dense ($\gtrsim$100 cm$^{-3}$), massive (10$^4$--10$^6$ \msun) clouds, often observed to be stellar nurseries \citep[see][for a review of GMCs]{Chevance23}. 

The observed length of the M4.7-0.8 cloud is consistent with many other GMCs observed in the Milky Way and other local GMCs in nearby galaxies \citep[e.g.,][]{Chevance23}. These local GMCs tend to be filamentary in nature and have been argued to be the `bones' of the Milky Way spiral arms \citep[e.g.,][]{Goodman14,Zucker15}. Many of these GMCs also tend to have localized regions of higher densities (i.e., knots) where the gas has been concentrated by its self gravity. The M4.7-0.8 cloud also shows these types of localized knots that are observed in the Filament (Knots A--E).

In general, the N(H$_2$) column densities in M4.7-0.8 (\til10$^{22}$ cm$^{-2}$; Figure \ref{dust-temp}, bottom) are similar to column density values measured in Disk GMCs \citep[e.g.,][]{Marsh17}. However, when comparing our measured column density to values found in CMZ clouds, we find that CMZ clouds are an order of magnitude higher than those observed in M4.7-0.8 \citep[\til10$^{23}$ cm$^{-2}$;][]{CMZoom}. This suggests that the column density in the cloud will either likely increase during accretion into the CMZ or will coalesce with other clouds in the CMZ to result in the higher values observed in the inner region.

Using the column density data from \cite{Marsh17}, shown in Figure \ref{dust-temp} (bottom), we can calculate a total mass for the M4.7-0.8 GMC. We used the \texttt{PPMAP-MASS jupyter} notebook which calculates a cloud mass from the \cite{Marsh17} column density dataset \citep{Gramze23}.\footnote{The \texttt{jupyter} notebook used to calculate the GMC mass from the \cite{Marsh17} column density dataset can be found \hyperlink{https://github.com/SpacialTree/lactea-filament/blob/main/lactea-filament/notebooks/ppmap_mass.ipynb}{here}.} This notebook first calculates an average column density over a specified region which, in our case, uses the grey dashed box in Figure \ref{dust-temp} (bottom). Using this average column density and area, with a set distance of 7.0 kpc (as calculated in Section \ref{sect:3D}), we estimate a cloud mass of 1.6 $\times$ 10$^5$ \msun. We note, however, that this cloud mass is lower than previous mass estimates calculated using the \cite{Dame01} CO dataset \citep{sormani19a}.

\cite{sormani19a} calculate a mass estimate of this region of roughly 10$^5$--10$^6$ \msun\ and used an $X_{CO}$ factor of 2$\times$10$^{20}$($N$(H$_2$) cm$^{-2}$)/($W$($l,b$) K \kms) for their mass estimates (see their Section 2.2 and Figure 3). While this $X_{CO}$ factor, used in \cite{sormani19a}, represents the Galactic average, it relatively high for CMZ clouds \citep{Bolatto13}. 
Extragalactic studies of the CO-to-H$_2$ conversion factor illustrate that this value tends to be lower in the inner region of galaxies when compared to the disk \citep[e.g.,][]{Teng23, Chang24}.

In the \cite{Gramze23} mass calculation of the G5 GMC, they calculate a mass of 2 $\times$ $10^5$ \msun\ - using an $X_{CO}$ factor of 1.5 $\times$ 10$^{19}$ \cms. However, in their PPMAP calculation, they are measuring a mass of 2 $\times$ 10$^4$ \msun. This order of magnitude increase in the $X_{CO}$ calculation, when compared with the PPMAP estimate, is similar to what we are observing in M4.7-0.8. This could be an indication that the $X_{CO}$ factor could be different from values used in the Disk and CMZ. Despite the range of mass estimates for this cloud (10$^5$--10$^6$ \msun), which could be constrained in later publication, the estimated cloud mass falls well within the range for GMCs \citep[10$^4$--10$^6$ \msun; e.g.,][]{Chevance23} - supporting our argument that M4.7-0.8 is a GMC.

Based on the large-scale velocity gradient of the near-side of the Galactic Bar Dust Lanes, shown in Figure \ref{introfig2}, we would expect an increasing velocity gradient from E to W across the cloud of $\sim$+0.25 \kmsp. However, we instead observe a decreasing velocity gradient of -0.25 \kmsp\ (see Section \ref{sec:mom-map}). We note that this is a average across the entire length of the GMC - where as local velocity gradients in the Filament could be an order of magnitude higher (see Figure \ref{Sine-pv}). This difference in the averaged velocity gradient across the cloud and the large-scale velocity gradient of the dust lanes could be an indication of some internal dynamics affecting the velocity structure of the GMC. 

Velocity gradients across GMCs is not uncommon - as observed in \cite{ros03} who investigated the kinematics of 45 GMCs in M33 at 20 pc resolution (which would correspond to roughly 3 pixels/beams across M4.7-0.8). They observed, both positive and negative, velocity gradients across the GMCs in their sample that tend to be around 0.1 \kmsp\ (see their Figure 7) - indicating that our measure velocity gradient across the entire GMC is slightly more extreme than their average. In 40\% of their sample GMCs, \cite{ros03} saw counter-rotation when compared with the disk rotation of M33. If M4.7-0.8 is the result of a previous cloud-cloud collision, as hypothesized in our previous section, then the resulting material could `decelerate', resulting in this apparent counter-rotation, as has been observed in previous simulations \citep[][see their Figure 16 and Section 5]{Tress20b}. Follow up CO observations at $<$30\arcsec\ resolution, is needed to determine the kinematics of the extended region to ascertain whether the Filament is showing counter-rotation relative to the bulk emission in the GMC. 
For this publication we will limit our analysis to investigating the cloud kinematics and turbulence by using the spectral fitting information from Table \ref{spec-table} in the following section.

\subsubsection{Elevated Turbulence in M4.7-0.8 When Compared to GMCs in the Disk}
\label{linewdith}

\begin{figure}[]
\centering
\includegraphics[width=0.48\textwidth]{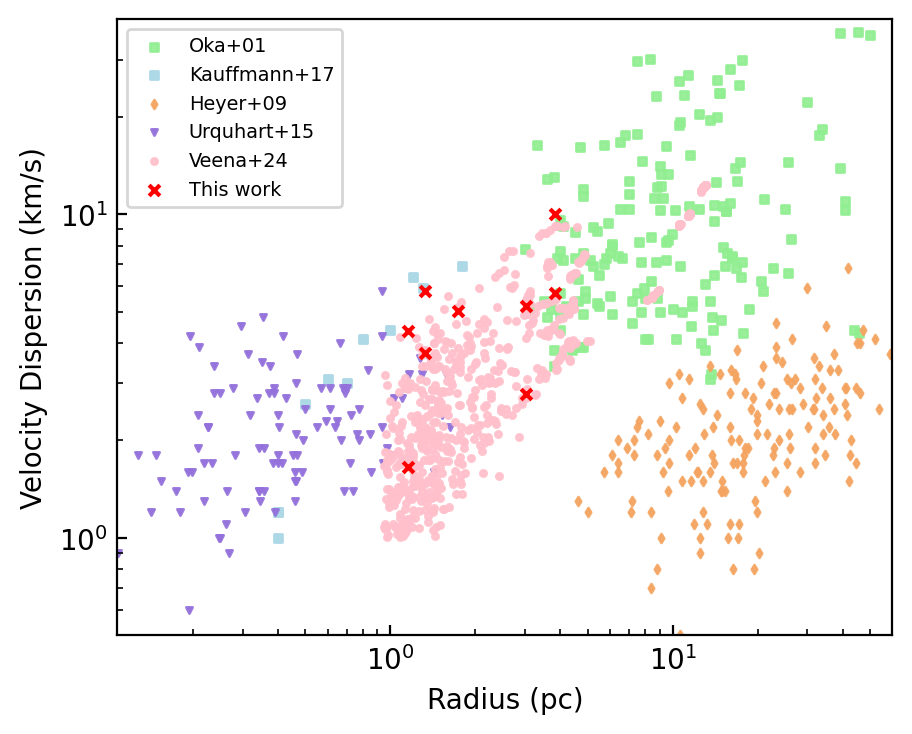}
\caption{Larson linewidth-size relationship for the 5 regions shown in Figure \ref{morphfig} (marked as red $\times$ symbols here) compared with other survey datasets as labeled in the top left legend. The velocity dispersion is measured from the spectral line fits shown in Figure \ref{spectra-fit}. The values for the velocity dispersion and the radius are reported in Table \ref{spec-table}. The green and blue squares show the GC data points from \cite{Oka01b} and \cite{Kauffmann17} respectively. The orange diamonds show the Disk clouds from \cite{Heyer09}. The pink circles show the measurements from \cite{Veena24}, which is suggested to be down stream from the Midpoint cloud. The purple triangles show the measured \am\ (1,1) linewidths of high mass star forming regions in the Disk \citep{Urquhart15}. 
}
\label{larson}
\end{figure}

The CMZ is observed to have broader linewidths than the Galactic disk on similar size scales \citep[e.g.,][]{Shetty12}. These broad lines are typically an order of magnitude greater than values measured in the disk. This increase is often attributed to turbulence, however other possible scenarios include tidal shearing and feedback \citep[e.g.,][see their Table 2]{Henshaw23}. One method of understanding this turbulence is by plotting the measured linewidth against the physical radius of the region used to extract the line, also known as a Larson relationship \citep{Larson81}, defined as:
\begin{equation}
\left(\frac{\sigma}{\textrm{km~s}^{-1}}\right) = \alpha  \left(\frac{R}{\textrm{pc}}\right)^{\beta}
\end{equation} 
(see their Equation 1). Here $\sigma$ is the velocity dispersion in \kms\ and R is the radius of the cloud.

In Figure \ref{larson}, we plot the \citetalias{Larson81} relationship for the cloud using the 5 spectra shown in Table \ref{spec-table} (shown as red `$\times$' symbols) against other clouds in the CMZ \citep[squares;][green and blue data points, respectively]{Oka01a, Kauffmann17} and Galactic Disk GMCs \citep[orange diamonds;][]{Heyer09}. Data for these other clouds was pulled from \cite{Urquhart15}, who used \am\ (1,1) and (2,2) observations to target high-mass star forming clumps in the Disk, shown as purple triangles. We also show data from \cite{Veena24}, who targeted the Helix stream (pink star in Figure \ref{3D}) as light pink circles. We note that \am\ (1,1) GBT observations of Sgr E exist \citep{Anderson20}, however, these are single pointing spectra and therefore no radii information is available for this dataset. We note, however, that the spectra were fit and the measured values for Sgr E range from 1.3 to 40.3 \kms.

We find that the measured velocity dispersion for the M4.7-0.8 cloud is fairly similar to values observed in the CMZ (green and blue squares), the Helix stream (pink circles), and high mass star forming regions (purple triangles) when compared with the Disk GMCs (orange diamonds). This similarity between the CMZ and Galactic Bar data points could be an indication that the turbulence observed in the CMZ is be the result of in fall to the gravitational potential. 
We note that the values measured in the Midpoint cloud are fairly similar to those observed in the Helix stream \citep{Veena24}. The Helix stream is argued to be `downstream' from the Midpoint and associated with material that has overshot the CMZ - indicating that it is likely located in Quadrant 2 (Figure \ref{3D}). We note that this region is also ``downstream" in the dust lanes from an ``interaction location" \citep{Gramze23}. Thus, the observed turbulence could be the result of a previous cloud-cloud collision. Results from \cite{Heyer09} in Figure \ref{larson} represent GMCs in the Galactic Disk. Here we see that these GMCs have significantly lower velocity dispersions, while also presenting larger physical sizes, in comparison to our data and the results of \citet{Veena24}. This is not unexpected, due to the large ($>$10pc) scale of the \citet{Heyer09} sources. However, this is useful as a contrast.
Further analysis of the structure of M4.7-0.8, with a focus on comparing cloud and clump properties to previously identified samples is beyond the scope of the current work and is planned for a follow-up investigation (Morgan et al., in prep).

The higher linewidths observed in M4.7-0.8 is also consistent with the detection of the \cyanobut\ (9-8) transition. \cite{Chevance23}, and references within, noted that clouds that are rich with complex molecules often display broader linewidths when compared with other clouds in the ISM. The detection of the \cyanobut\ (9-8) transition could be an indication that M4.7-0.8 is rich in other complex molecules, similar to other broad-line regions like the CMZ. 
Follow-up observations of additional complex molecules is currently underway with Atacama Large Millimeter/sub-millimeter Array (ALMA) to address this theory.

\subsection{Evidence of Star Formation at the Midpoint}
\label{sec:sf}

Understanding star formation in barred spiral galaxies has been gaining momentum in recent years with the onset of high-resolution, high-sensitivity telescopes like ALMA and JWST \citep[e.g.,][]{Phillips96, regan99, Sheth00,Verley07, Maeda23,  Neumann19, Neumann24b, Fraser20, Renaud15}. Bars within spiral galaxies are observed to contain abundant molecular material, which could potentially form stars \citep{Diaz21}. This region of the galaxy also has a higher stellar density and more metal rich than the disk at similar radii \citep{Neumann24a}. These extragalactic studies of barred galaxies found that primarily low-mass galaxies host star formation in their bars and are typically usually on the leading edge of the bar \citep{Fraser20}. Furthermore, these sites of star formation are regulated by shear forces, turbulence, and gas flows \citep{Renaud15, Fraser20}. In this section we explore possible signs of star formation in the cloud by focusing on Knots B (\S \ref{knotB}) and E (\S \ref{knotE}) and investigate possible stellar feedback producing the Shell (\S \ref{feedback}).

\subsubsection{Is there Star Formation in Knot B?}
\label{knotB}

\begin{figure}[t!]
\centering
\includegraphics[width=0.47\textwidth]{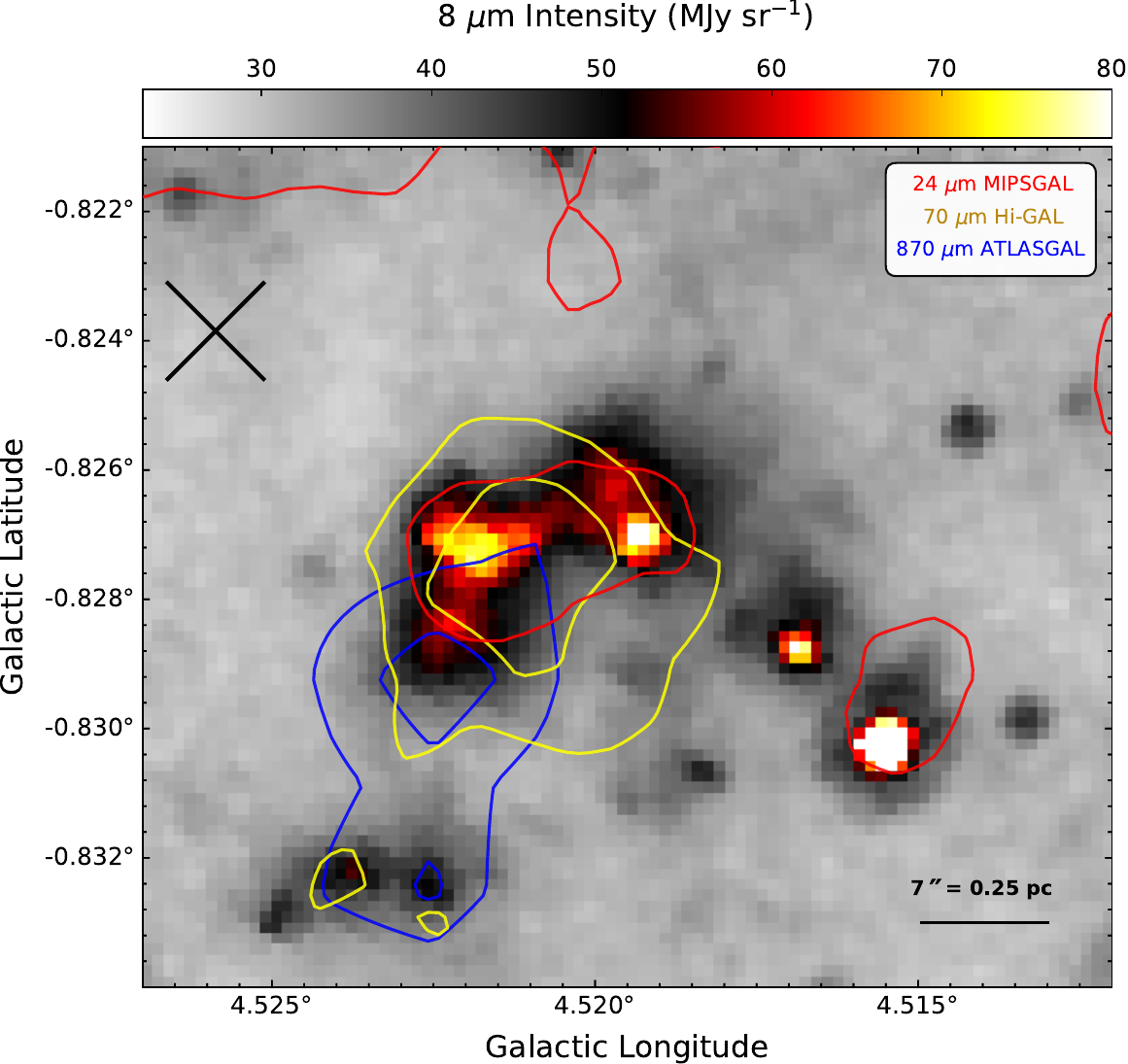}
\caption{8 \micron\ emission of the extended source located near the head of Knot B (red box in Figures \ref{atlasgal}--\ref{mid-ir}). The blue contours trace the ATLASGAL 870 \micron\ emission, from Figure \ref{atlasgal}, at 0.6 and 0.68 \jybe. The yellow contours trace the Herschel 70 \micron\ emission, from Figure \ref{herschel}, at 0.032 and 0.04 Jy pixel$^{-1}$. The red contours trace the MIPSGAL 24 \micron\ emission, from Figure \ref{mid-ir}, at 20.5 MJy sr$^{-1}$. The black $\times$ symbol shows the central location of the \am\ (3,3) maser (Section \ref{3maser}). A scale bar of the region is shown in the bottom right corner, assuming a distance of 7.0 kpc.}
\label{hii reg}
\end{figure}

Knot B is observed to contain several tracers of possible star formation, including the new detection of an \am\ (3,3) maser (Section \ref{3maser}). 
Analysis of the dust continuum emission shows compact emission southwest of the (3,3) maser, near the observed edge of the knot, in the 8, 12, 24, 70, and 870 \micron\ bands (red box in Figures \ref{atlasgal}--\ref{mid-ir}). 
Figure \ref{hii reg} shows a higher resolution image of the 8 \micron\ data in this compact emission. Overlaid are contours showing emission at 870 \micron\ (blue; Figure \ref{atlasgal}), 70 \micron\ (yellow; Figure \ref{herschel}, bottom), and 24 \micron\ (red; Figure \ref{mid-ir}, top).

The 8 \micron\ observations show this source is extended (\til0.5 pc) and contains several bright point sources. The 24 \micron\ emission is closely associated with the bright emission observed in the 8 \micron. The 70 \micron\ emission is concentrated toward the bright, extended 8 \micron\ structure near the center of the region. The 870 \micron\ emission appears to be marginally offset from this central source and slightly elongated - extending away from the main concentration of mid-IR emission. 

The detection of relatively compact emission at mid-IR (8, 12, 24, and 70 \micron) and in the sub-millimeter (870 \micron), with no compact emission detected at far-IR wavelengths (160 and 250 \micron), could indicate that this source is an \hii\ region. The detection in the 8, 12, 24 and 70 \micron\ bands is likely associated with the hot dust component. The compact emission observed in the 870 \micron\ band could be associated with the tail end of the `free-free' Bremsstrahlung emission from radio and mm wavelengths. Both of these emission mechanisms, free-free emission at sub-millimeter and hot dust at the mid-IR, are observed to be associated with \hii\ regions \citep[e.g.,][]{Wynn-Williams74}. 

Furthermore, the detection of an \am\ (3,3) maser may also support this argument of possible star formation in Knot B. \am\ (3,3) masers have been previous observed towards regions of high mass star formation \citep[e.g.,][]{Kraemer95,Zhang95}. \cite{Kraemer95} observed \am\ (3,3) masers near the ends of high-velocity molecular outflows as the results of shocks. The slight offset in the maser location could be due to the jets from the massive star formation complex, interacting with the bulk emission in the filament.

While the mid-IR and sub-millimeter observations, combined with the maser detection, support this argument for an \hii\ region, we note that the source is lacking radio counterparts. We investigated the 2--3 GHz VLASS 3.0 dataset \citep{VLASS} for radio counterparts in the region and none were found above a 3$\sigma$ level ($<$0.4 m\jybe). The GLOSTAR C-band (6 GHz) survey was also checked and no detection was found above a 3$\sigma$ level in in the B-configuration data of this survey either \citep[$<$0.3 m\jybe,][]{Yang23}. This non-detection at low radio frequencies could indicate that the turn over for the free-free emission in the radio spectrum is above this frequency range or the emission is below the noise floor in the VLASS and GLOSTAR surveys. High resolution radio and millimeter observations of this source are needed to determine 1) if this is indeed an \hii\ region by investigating the spectral energy distribution (SED) of the source, and 2) if the velocity of the source is consistent with the GBT molecular emission, thereby confirming the association with the cloud.

\subsubsection{Is Knot E a Free-floating Evaporating Gas Globule?}
\label{knotE}

Free-floating evaporating gas globules (frEGGs) are small, dense, isolated dark clouds of gas and dust that are being eroded by nearby radiation \citep[e.g.,][]{Sahai12a,Sahai12b, Wright12}. These sources tend to have a tadpole or cometary like structure, containing a bright head and a fainter extended envelope \citep[e.g.,][]{Lefloch94,Lefloch97, Sahai12b}. This fainter envelope is produced by the erosion of the globule by radiation, indicating the direction of the `tail' is orientated away from the ionizing source \citep{Bertoldi90, Wright12}. These cometary globules are argued to form under the effects of radiative-driven implosion (RDI) scenarios \citep[e.g.,][]{Bertoldi90}. 
The size of these frEGGs can be much larger (20,000--110,000 AU; 0.1--0.5 pc) than their proplyd counterparts found in Orion \citep[40–350 AU;][]{Henney99, Wright12}, as the size of the eroding source can depend on the distance to the ionizing source of radiation. 

Knot E shows many characteristics that are in agreement with traits found in frEGGs. Located at the tail end of the Filament (see Figure \ref{morphfig}), Knot E appears to be distinctly separated in both position and velocity from the neighboring Knot D and the rest of the Filament. Knot E is also observed to be relatively dense (N(H$_2$) \til\ $1.5 \times 10^{22}$ cm$^{-2}$) and compact ($d<1$ pc) suggesting the source is more similar to frEGGs than the proplyds found in Orion. Knot E also appears to be cometary in the dust continuum, containing a compact core with diffuse emission extended north of the compact source. This cometary structure is best observed in the 3-color image (Figure \ref{introfigr}), which shows a composite of the 3 Herschel bands from Figure \ref{herschel}. The orientation of the cometary tail, nearly directly north when viewed in Galactic coordinates, would suggest the source producing the radiation is located south of Knot E, and could possibly be associated with Knot D.
We note that the direction of the tail is also perpendicular to the direction of accretion, therefore the cometary morphology is not likely the result of the cloud's movement through Bar.
The extended envelope is observed to be mid-IR dark, showing extinction in the 8--24 \micron\ data while the compact emission is observed to be mid-IR bright, showing bright compact emission in the 8--24 \micron\ dataset (Figure \ref{mid-ir}). \cite{Lefloch97} saw a similar morphology in several cometary globules in IC 1848 - which showed a bright head and dark tail.

The mid- and far-IR compact emission in Knot E was previously investigated in Hi-GAL survey work conducted by \cite{Elia17} (their catalog source ID 1948; marked with a red triangle in Figures \ref{atlasgal}-\ref{mid-ir}). \cite{Elia17} calculated a mass of 11.12 \msun\ and a dust temperature of 14.59 K for the 70 \micron\ point source. We note, however, that this value of the dust temperature measured in \cite{Elia17} is significantly lower than the temperature measured in \cite{Marsh17} (20.8 K; see Figure \ref{dust-temp}, top). 
This discrepancy in the temperature is could be caused by the cometary structure of the source, which could contain a hot object embedded in a cooler envelope. Another alternative could be the inclusion of bremsstrahlung (free-free) emission in the longer wavelengths. \cite{Elia17} used 870 \micron\ emission in their dust temperature calculations, when available. Figure \ref{atlasgal} shows 870 \micron\ emission is present in this region. If, however, this emission is associated with free-free emission rather than the tail end of the Planck distribution at the longest wavelengths, this could then inflate the emission values at the longest wavelengths, resulting in a lower temperature measurement.
\cite{Marsh17} calculated their temperatures and column densities using a PPMAP technique which calculates a column density per temperature `slice'. Slice 6 (18.4015 K) and slice 7 (21.7373 K) show a structure most similar to the emission distribution shown in Figure \ref{herschel}, suggesting the dust temperature is likely within this range - similar to the values shown in Figure \ref{dust-temp}.

If Knot E is being eroded by a nearby massive star (as is necessary for a frEGG source), then one would expect to see evidence of radio emission from the winds of these massive stars and free-free emission in Knot E. As mentioned in Section \ref{knotB}, the VLASS and GLOSTAR survey were checked and no radio emission was observed in the region above a 3$\sigma$ level ($<$0.4 m\jybe). 
We note, however, that we may be sensitivity limited to detecting these massive stars. Roughly 20\% of the \cite{Lu19} sample of point sources in the Galactic Center were below this 3$\sigma$ level ($<$0.4 m\jybe), with at least one showing a positive spectral index, indicating the radio emission is likely thermal (e.g., see their source C86). Thus, without sensitive radio continuum observations \citep[][reports a sensitivity of 0.025 m\jybe\ in their low-rms fields]{Lu19}, we are unable to determine whether there are massive stars in the region.

\subsubsection{Is the Shell Evidence of Feedback?}
\label{feedback}

Feedback from evolved massive stars can influence their surrounding material by injecting energy and momentum into the nearby gas and dust. Most of this ejected material is generally modeled to be isotropic, resulting in the formation of bubbles and shells in the ISM \citep[e.g.,][]{Weaver77, Barnes23, Watkins23}.

We observe a shell-like feature in the M4.7-0.8 cloud that may be associated with feedback, labeled as the `Shell' (see Figure \ref{morphfig}). The Shell shows a hallow morphology both in the maximum intensity images (Figure \ref{morphfig}) and in the channel maps (Figure \ref{channels}). This morphology is also observed in the dust continuum as well (see Figure \ref{herschel}). Furthermore, the hollow morphology is also faintly visible in the position velocity diagram (Figure \ref{Sine-pv}). These signatures are typically characteristic of an bubble-like morphology. Along the eastern side of the shell we observe an increase in the emission. This region also corresponds to the region of the shell with the highest column density (Figure \ref{dust-temp}).

The 2--3 GHz VLASS catalog was checked for possible radio counterparts and a point-source was detected within the region denoted as the Shell: $l$=4\fdg6492524, $b$=-0\fdg8542642 (marked with a red diamond in Figure \ref{herschel}). A figure showing the VLASS point source is included in the Appendix (Figure \ref{vlass-ps}). 
A cross-examination of the location of radio point source with the far- and mid-IR datasets indicates this source is lacking an IR counterpart. 
As the nature of this radio point source is unknown, as well as the velocity of the point source, it is unclear if this source is interacting with the cloud or if this is a chance alignment. 

However, we note that the location of the radio point source is nearest to the edge of the Shell with the highest emission. If this radio feature was interacting with the Shell then we could see an increase in emission nearest to this interaction location. However, from these single survey observation, this possible interaction is speculative, at best. Therefore, high resolution follow up study of this region at radio wavelengths is necessary to determine 1) the nature of this radio point source, 2) the line-of-sight distance to the source to determine if it is in the dust lanes and 3) if this point source is interacting (or possibly creating) the Shell feature observed in the Filament.

The nature of the Shell is also unclear. We investigated several SNR catalogs and no SNRs have been discovered in the region near the midpoint. 
With no evidence found for an ionizing influence upon the observed region, 
the Green SNR catalog \citep{Green19}\footnote{Green D. A., 2022, `A Catalogue of Galactic Supernova Remnants (2022 December version)', Cavendish Laboratory, Cambridge, United Kingdom (available at \hyperlink{http://www.mrao.cam.ac.uk/surveys/snrs/}{http://www.mrao.cam.ac.uk/surveys/snrs/}).}
the Chandra SNR Catalog\footnote{\hyperlink{https://hea-www.harvard.edu/ChandraSNR/}{https://hea-www.harvard.edu/ChandraSNR/}} and the SNRCat \citep{Ferrand12}\footnote{\hyperlink{http://snrcat.physics.umanitoba.ca/}{http://snrcat.physics.umanitoba.ca/}} were searched for previous observations which might support a SNR-driven scenario. Additionally, the ATNF pulsar catalog\footnote{\hyperlink{https://www.atnf.csiro.au/research/pulsar/psrcat/}{https://www.atnf.csiro.au/research/pulsar/psrcat/}} was also searched for any coincident pulsars which might indicate the presence of a prior supernova in the region. All searches returned no likely candidates that could be associated with a driving dynamical force in this region.
Higher resolution observations of the Shell are needed to disentangle the kinematics at this location and calculate momentum and energy of the possible expansion. These calculations would then give insight on the energy needed to power the expansion and the possible mechanisms that could be driving the expansion.

\section{conclusion}
\label{con}

We report the detection of a previously unknown giant molecular cloud (GMC) located at the midpoint of the near-side Galactic Bar Dust Lanes (M4.7-0.8).
In this publication we present 25 GHz radio observations of dense gas 
that is associated with material accreting into the Galactic Center. In these observations, taken with the Green Bank Telescope (GBT), we targeted the \am\ (1,1)--(4,4), and the HC$_5$N (9--8) transitions. These single dish observations observed the main `Nexus' and `Filament' of the GMC (0\fdg5 $\times$ 0\fdg25 area) with 31\arcsec\ angular resolution (\til1 pc at a distance of 7.0 kpc to the cloud). A comparison with IR survey data shows a far-IR dust component that is well correlated with the \am\ emission (Section \ref{sec:dust}). In addition to this dense gas detection, we also report the following results:

\begin{itemize}
    \item \textbf{3D Location in the Dust Lane}: Based on the observed Galactic Longitude ($l \simeq$ 5\degree) and velocity ($v \sim$ 200 \kms) we argue that M4.7-0.8 is located on the Near-side Dust Lane (Section \ref{sect:3D}). 
    We compare the location of M4.7-0.8 to the \cite{sormani19a} toy model of the Dust lanes and argue the cloud is located at a distance of 7.0 kpc (Figure \ref{3D}). The negative Galactic Latitude of the cloud ($b$=-0\fdg85) also indicates that it is 
    about 93.5 pc below the Galactic midplane, using the geometrical model presented in \cite{Goodman14}. 

    \item \textbf{Properties of the GMC}: The M4.7-0.8 GMC has a length of roughly 60 pc in Galactic Longitude and a vertical extent of \til20 pc in Galactic Latitude (assuming a distance of 7.0 kpc). Utilizing the \cite{Marsh17} PPMAP survey data (shown in Figure \ref{dust-temp}), we measure high column densities (\til$10^{22}$ \cms) and cold dust temperatures (\til20 K) in the GMC (Section \ref{dust-prop}). Using the column density data, and the \texttt{PPMAP-MASS jupyter} notebook, we calculate a cloud mass of 1.6$\times$10$^5$ \msun\ (Section \ref{GMC}). Based on the size, mass, and column density of M4.7-0.8, all of which are typical of the properties of Galactic GMCs, we argue that M4.7-0.8 is a GMC.
    We also make a comparison of the cloud's linewidths (reported in Table \ref{spec-table}) to clouds in the Disk and CMZ, using the Larson relationship, and find relatively broad linewidths in M4.7-0.8 (Section \ref{linewdith}). These broader linewidths are similar to values reported in CMZ clouds, indicating that M4.7-0.8 has elevated turbulence compared to Disk clouds and GMCs. 
    
    \item \textbf{New \am\ (3,3) Maser Detection}: We detect a previously unknown \am\ (3,3) maser in our dataset (Section \ref{3maser}). This new \am\ (3,3) maser is located in Knot B (see black $\times$ symbol in Figure \ref{morphfig}). The spectral analysis of this maser shows it has a central velocity of 193.16$\pm$0.01 \kms\ and a narrow linewidth of 0.37$\pm$0.01 \kms.

    \item \textbf{New Sites of Star Formation}: We observe two new sites of possible star formation in the cloud: Knots B and E. 
    Knot B contains the newly detect \am\ (3,3) maser - a common tracer of shocks from outflows. This region also contains extended 8 \micron\ emission (Figure \ref{hii reg}) that also has 24, 70, and 870 \micron\ emission - known tracers of star formation (Section \ref{knotB}). This region shows a brightening in the 870 \micron\ emission, when compared with the extended structure and the 250 \micron\ emission, indicating that we may be detecting free-free emission from a possible \hii\ region. However, sensitive radio observations are needed to confirm this. 
    The other source, Knot E, displays a cometary-like structure in the dust continuum that is extended at far-IR wavelengths and compact in mid/near-IR wavelengths. Knot E is dense, showing extinction in the 8 \micron\ Spitzer data (Figure \ref{mid-ir}, bottom) and a high column density in the \cite{Marsh17} PPMAP data (Figure \ref{dust-temp}, bottom). This region also shows a relatively high dust temperature (\til21 K) compared to the cooler envelope. We further investigate Knot E as a possible free-floating evaporating gas globule (frEGG; see Section \ref{knotE}). 

    \item \textbf{Evidence of Feedback in the GMC}: The M4.7-0.8 cloud also contains what appears to be a shell-like structure that we label as the Shell 
    (see Figure \ref{morphfig}). This structure contains a brighter rim in the \am\ emission, with a cavity towards its center (Section \ref{morph}). A radio continuum point source is observed within the Shell, but its relation to the Shell is uncertain without additional observations (Section \ref{feedback}).

\end{itemize}

     
\section{Acknowledgments}

The collaboration would like to thank the GBT operators for their assistance with these observations. 
NB would also like to thank Dr. Ryan Boyden (UVA) for his helpful discussion on frEGGS. 
MCS acknowledges financial support from the European Research Council under the ERC Starting Grant ``GalFlow'' (grant 101116226) and from Fondazione Cariplo under the grant ERC attrattivit\`{a} n. 2023-3014. The authors would also like to thank the anonymous referee for their help in strengthening this publication.

This research has made use of the CIRADA cutout service at URL cutouts.cirada.ca, operated by the Canadian Initiative for Radio Astronomy Data Analysis (CIRADA). CIRADA is funded by a grant from the Canada Foundation for Innovation 2017 Innovation Fund (Project 35999), as well as by the Provinces of Ontario, British Columbia, Alberta, Manitoba and Quebec, in collaboration with the National Research Council of Canada, the US National Radio Astronomy Observatory and Australia’s Commonwealth Scientific and Industrial Research Organization.

This work is based in part on observations made with the Spitzer Space Telescope, which is operated by the Jet Propulsion Laboratory, California Institute of Technology under a contract with NASA.
Herschel is an ESA space observatory with science instruments provided by European-led Principal Investigator consortia and with important participation from NASA.
The Green Bank Observatory is a facility of the National Science Foundation operated under cooperative agreement by Associated Universities, Inc.
This publication is based on data acquired with the Atacama Pathfinder Experiment (APEX) under the ATLASGAL large programme \citep{Schuller09}. APEX is a collaboration between the Max-Planck-Institut fur Radioastronomie, the European Southern Observatory, and the Onsala Space Observatory. 

\software{\texttt{CASA} \citep{2011ascl.soft07013I}, \texttt{pyspeckit} \citep{pyspeckit}, spectral-cube, astropy, pvextractor}

\appendix

\counterwithin{figure}{section}
\counterwithin{table}{section}

\section{\am\ (5,5) Observations}
\label{55 sec}

\begin{figure}[t!]
\centering
\includegraphics[width=0.490\textwidth]{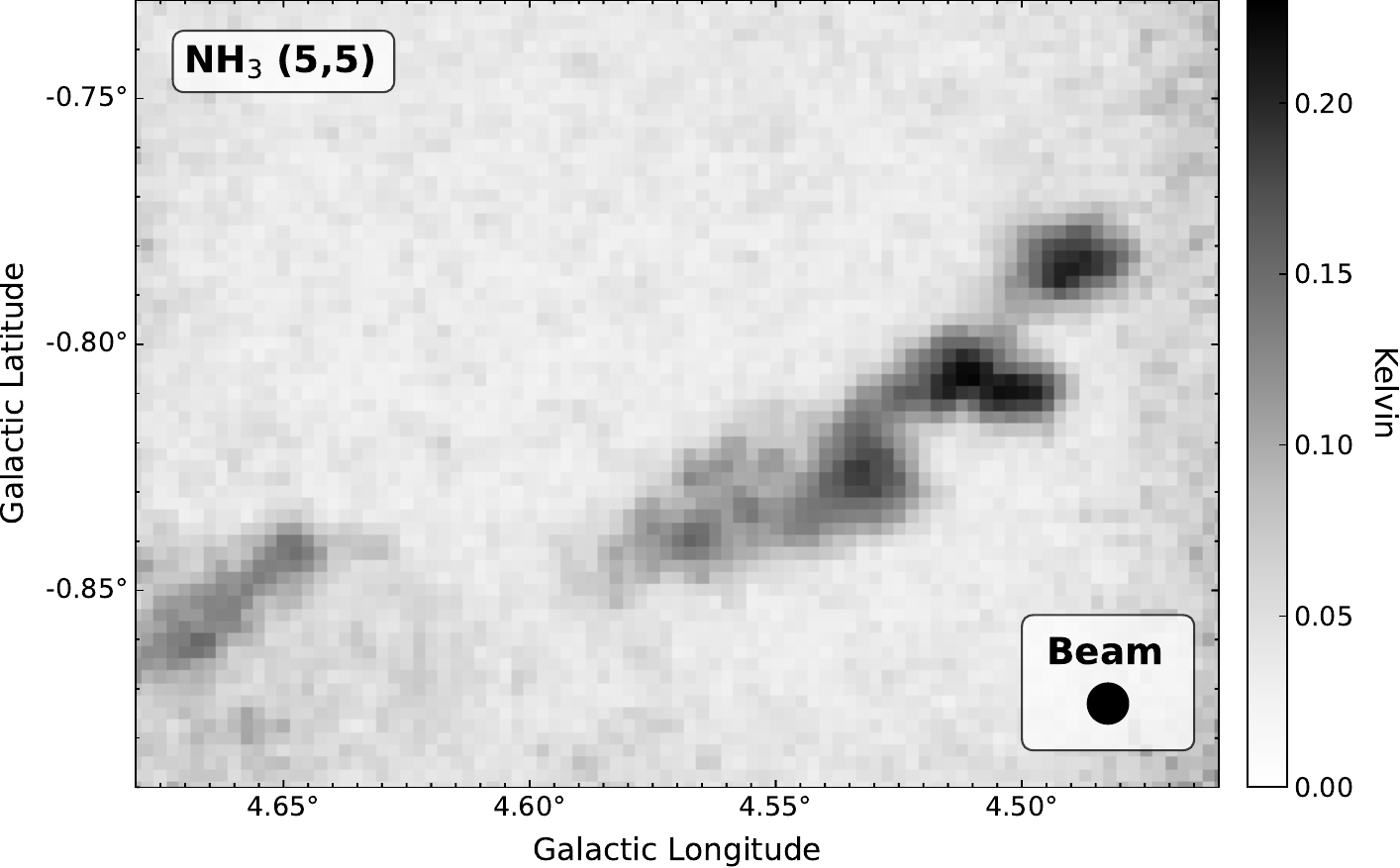}
\caption{Peak antenna temperature for the \am\ (5,5) in the Filament. The beam size is shown in the bottom right corner.}
\label{fig:Am55}
\end{figure}

In our DDT observations of the Filament, we included a spectral window centered on the (5,5) line as a test to determine whether higher transitions of \am\ were present in the cloud. This spectral window replaced the \HCsevenN\ (21--20) transition which was observed to be a non-detection in all previous observations. While we don't include this transition in our analysis, due to the incomplete coverage of the cloud, we show the detection here for future analysis of the Filament.

Figure \ref{fig:Am55} shows the maximum intensity emission in the \am\ (5,5) transition - similar to the plots shown in Figures \ref{morphfig} and \ref{morphfig2}. The \am\ (5,5) para- line emission closely follows the (4,4) para- line emission. Knots A-D are clearly visible, with Knot C being the brightest emission in this field. Similar to the (4,4) line, the knots are are more elongated than what is observed in the lower \am\ transitions, and Knot E is not detected in the the (5,5) data cube.

\section{Additional Kinematic Analysis}
\label{sec:B}

\subsection{Moment 1 Map of the Filament}

\begin{figure}
\centering
\includegraphics[width=0.48\textwidth]{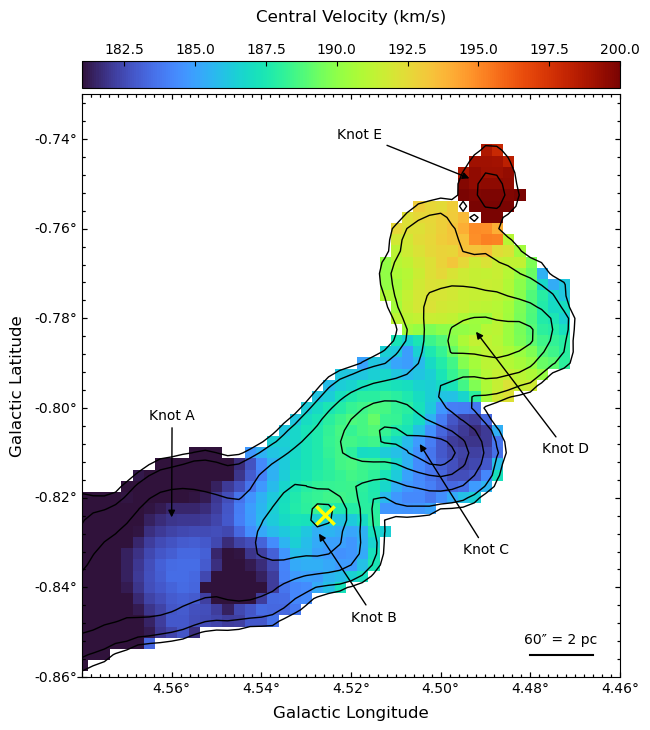}
\caption{Central velocity (Moment 1) of the \am\ (3,3) emission in the Filament Knot's A--E. This data, previously shown in Figure \ref{kin-fig}, is constrained both spatially and spectral to the region around the Filament. The location of the \am\ (3,3) maser, discussed in Section \ref{3maser}, is marked with a yellow `$\times$' symbol. }
\label{pillars}
\end{figure}

Figure \ref{pillars} shows the moment 1 map for the \am\ (3,3) emission in the Filament, constrained to the velocities in this region of the cloud (181--200 \kms). The velocity gradients in the knots are slightly more prominent in this constrained velocity range. 

\subsection{Position-Velocity Diagram}
\label{pv-section}

\begin{figure}
\centering
\includegraphics[width=0.47\textwidth]{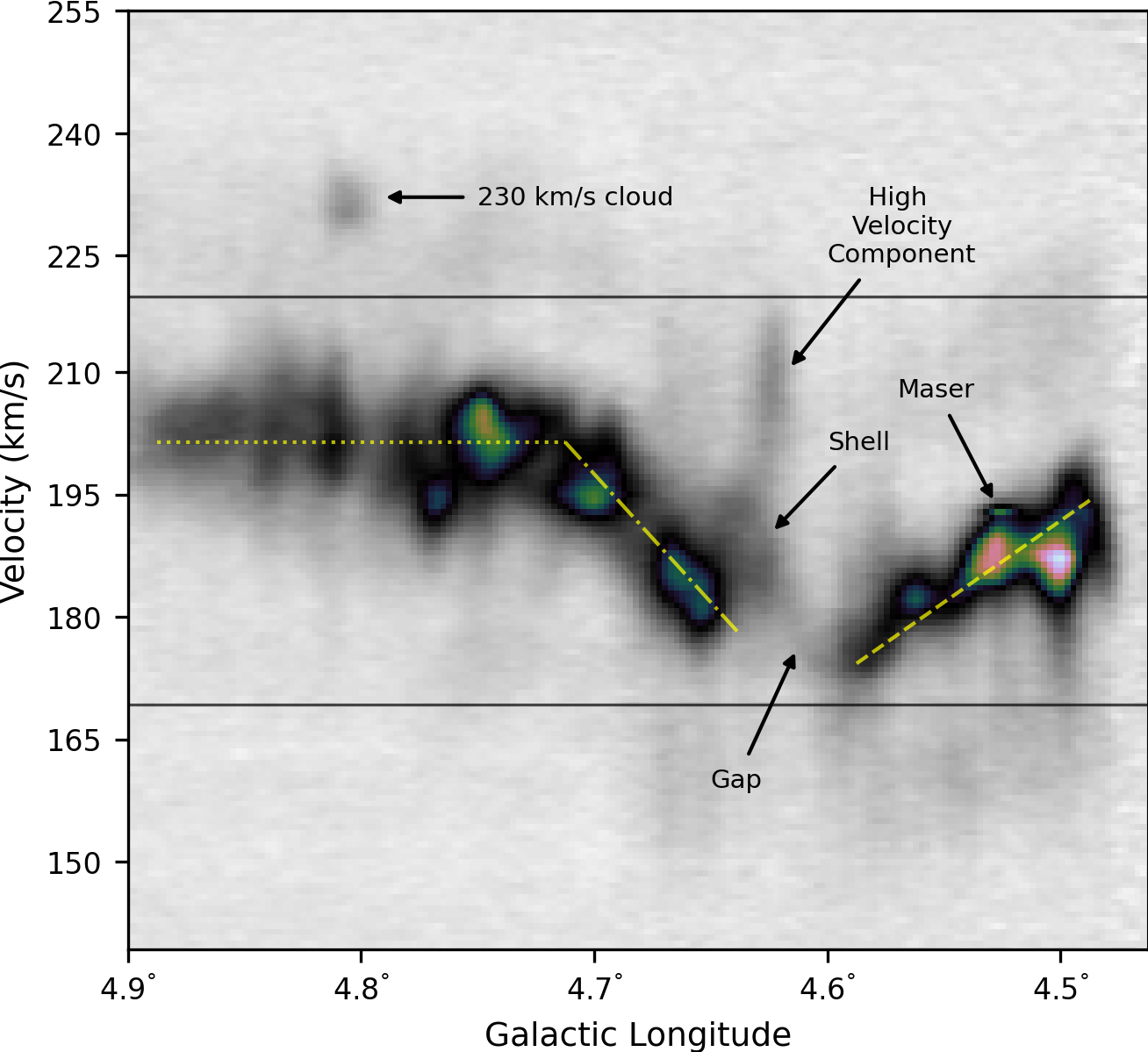}
\caption{Position-velocity diagram of the \am\ (3,3) emission in M4.7-0.8 cloud from $l$=4\fdg9 to $l$=4\fdg46, centered on $b$=-0\fdg81 with a width of 0\fdg1625. The horizontal black solid lines (at 170 and 220 \kms) show the velocity range covered in the channel maps in Figure \ref{channels}. There is a narrow velocity feature at $l$=4\fdg52 that we are interpreting as maser emission. We discuss this feature in detail in Section \ref{3maser}. The yellow dashed and dotted lines show the velocity gradients. 
}
\label{Sine-pv}
\end{figure}

Figure \ref{Sine-pv} shows a position-velocity diagram for the M4.7-0.8 molecular cloud. This plot was made in python using \texttt{pvextractor},\footnote{The \texttt{pvextractor} python program is available online at \href{https://github.com/radio-astro-tools/pvextractor}{https://github.com/radio-astro-tools/pvextractor}.} with a slice width of 0\fdg1625, centered at $b$=-0\fdg81, and covering a Galactic Longitude from $l$=4\fdg9 to 4\fdg46. Most of the emission in the cloud is between 170 and 220 \kms\ (solid black lines in Figure \ref{Sine-pv}). The 230 \kms\ cloud, discussed in Section \ref{sec:mom-map} and annotated in Figures \ref{kin-fig} \& \ref{Sine-pv}, is observed to be a distinct feature in position-velocity space.

\begin{figure*}[pt!]
\centering
\includegraphics[width=1.0\textwidth]{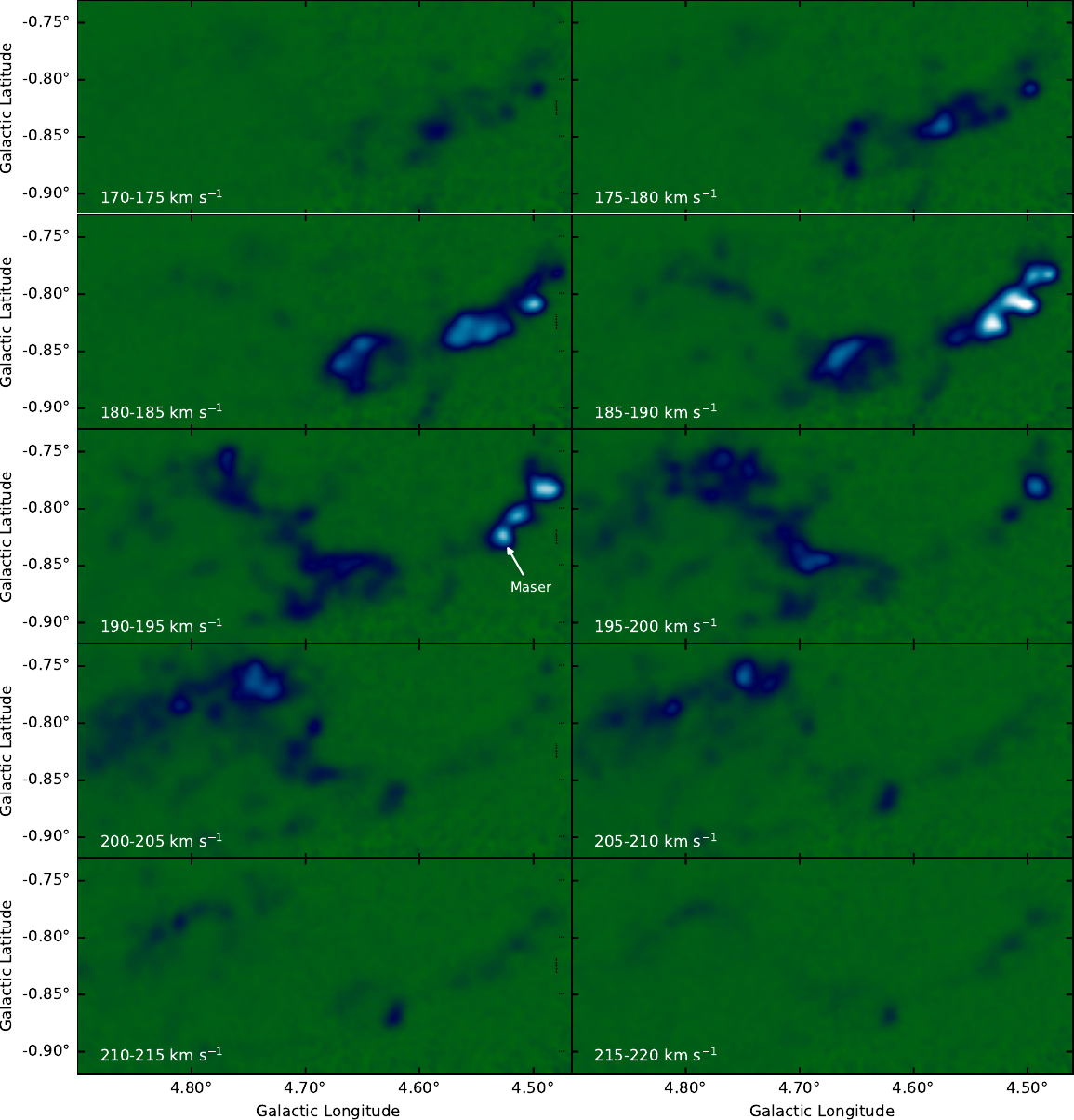}
\caption{Integrated emission (moment 0) of the \am\ (3,3) line, binned by 5 \kms, from 170 to 220 \kms. 
The \am\ (3,3) maser location is marked in the 190--195 \kms\ channel. The noise floor is more prominent in the left side of each panel due to the lower time-on-source of the DDT observations.  
}
\label{channels}
\end{figure*}

Between $l$=4\fdg9 and $l$=4\fdg7 the velocity in M4.7-0.8 is fairly constant - around 200 \kms\ (dotted yellow line). Most of this emission is associated with the region we are defining as the ``Nexus". At $l$=4\fdg7 we observe a gradual decrease in the velocity from \til200 \kms\ to \til180 \kms\ (dot-dashed line). This $\Delta v$ (\til20 \kms) and $\Delta l$ (\til0\fdg075, \til10 pc) results in a velocity gradient of \til2 \kmsp. Here, around $l$=4\fdg6--4\fdg65, we observe a gap in the structure, where the emission is relatively faint. This feature is associated with the ``Gap" region annotated in Figure \ref{morphfig} and \ref{kin-fig}. At lower Galactic longitudes we observe a second velocity gradient between 4\fdg6 and 4\fdg5, with velocity increasing from 170 \kms\ at $l$=4\fdg6 to 200 \kms\ at $l$=4\fdg5 (dashed line), resulting in a velocity gradient of \til2 \kmsp. Both of the non-zero velocity gradients are associated with the feature we are identifying as the ``Filament".

In addition to the ``Gap" between the two velocity gradients, we also observe a feature at a higher velocity: 210 \kms. This feature is associated with emission in the high velocity component (as annotated in Figure \ref{kin-fig}, top), which shows high velocities and broad linewidths in that figure (middle and bottom panels, respectively). Lastly, we also observe a narrow velocity feature around $l$=4\fdg53 and $v$=195 \kms. We have marked this feature as possible maser emission in Figure \ref{Sine-pv}, and will discuss this emission feature in more detail in Section \ref{3maser}. The hyperfine structure of the \am\ (3,3) line is also clearly visible in the Filament region of the cloud as faint emission that is $\pm$20.7 and $\pm$28.2 \kms\ from the main, bright peak. 

\subsection{Channel Maps}

Figure \ref{channels} shows channels maps of the \am\ (3,3) emission, from 170 to 220 \kms, for the M4.7-0.8 cloud. This velocity range contains all of the emission in M4.7-0.8, as observed in Figure \ref{Sine-pv}, with the exception of the 230 \kms\ cloud and a portion of the high velocity component. Each panel was made by integrated the emission over a $\Delta v$ = 5 \kms\ velocity range, as specified in the bottom left corner, essentially creating a moment 0 map for each panel (Equation \ref{M0}).

The general, large-scale velocity gradient of 0.25 \kmsp, discussed in Section \ref{sec:mom-map} is visible in this figure. Here, most of the higher velocity emission (195--210 \kms) is on the eastern side of the panels in the region associated with the Nexus. The lower velocity emission (170--195 \kms) is generally observed towards the western side of the panels in the region associated with the Filament. However, faint emission from the hyperfine lines are visible at $\pm$20.7 \kms\ from the main emission features.  
In the 175--180 and 180--185 \kms\ panels, the Shell feature is quite prominent and shows an increase in brightness along the left-most rim. 
The high velocity component, discussed in Section \ref{sec:mom-map}, is also visible in this figure and is the brightest emission in the filament in the 200--215 \kms\ panels. The spatial location of the high velocity component appears to coincide with an observed cavity visible in the 180--185 \kms\ panel.

\section{Radio Point Source}

\begin{figure}[t!]
\centering
\includegraphics[width=0.49\textwidth]{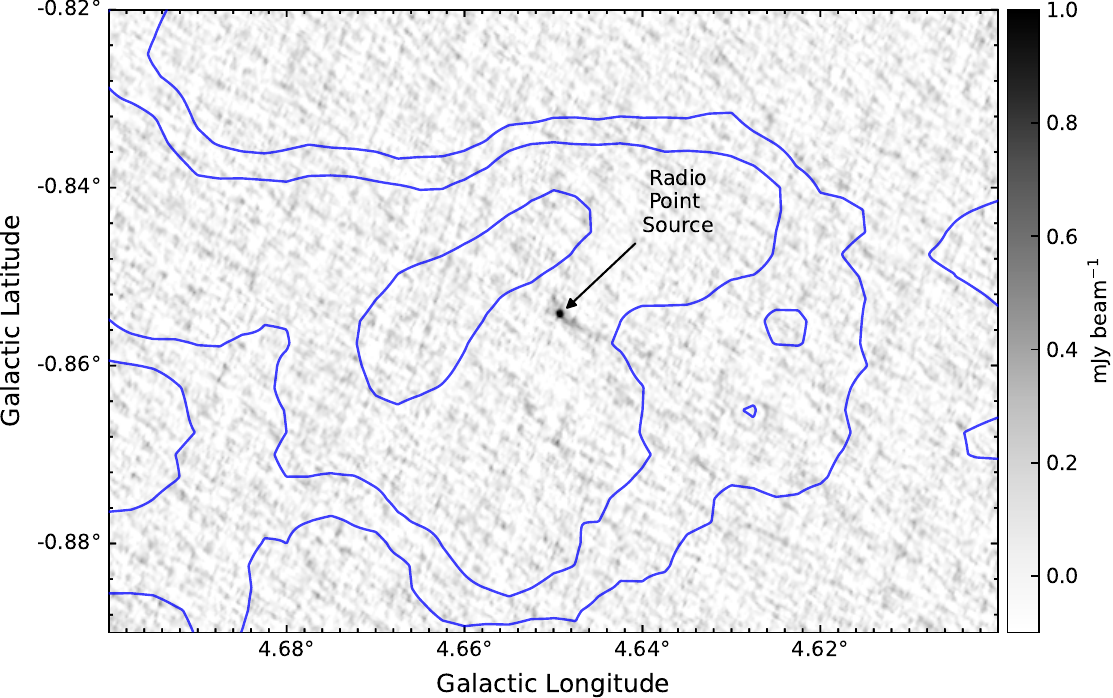}
\caption{2--4 GHz radio emission from the VLASS dataset showing the point-source detection toward the Shell \citep[2\farcs4, 120$\mu$\jybe][]{VLASS}. The blue contours show the \am\ (1,1) data at 0.15, 0.25, 0.50, 0.75 K. While not shown, this point source is also detected in the 6 GHz GLOSTAR radio survey. This unresolved point source has a flux density of 1.95 mJy, assuming the source is unresolved in the VLASS dataset. 
}
\label{vlass-ps}
\end{figure}

Figure \ref{vlass-ps} shows the radio point source detected in the vicinity of the Shell from the VLA Sky Survey (VLASS) dataset \citep{VLASS}.  This survey is an all sky survey at S-Band (2--3 GHz) with 2\arcsec\ resolution. Using this survey data, we detected a radio point source in the region denoted as the Shell (Section \ref{morph}). The radio point source is detected at 1.95 m\jybe.

\section{Three Color Dust Continuum}

\begin{figure*}[t!]
\centering
\includegraphics[width=0.75\textwidth]{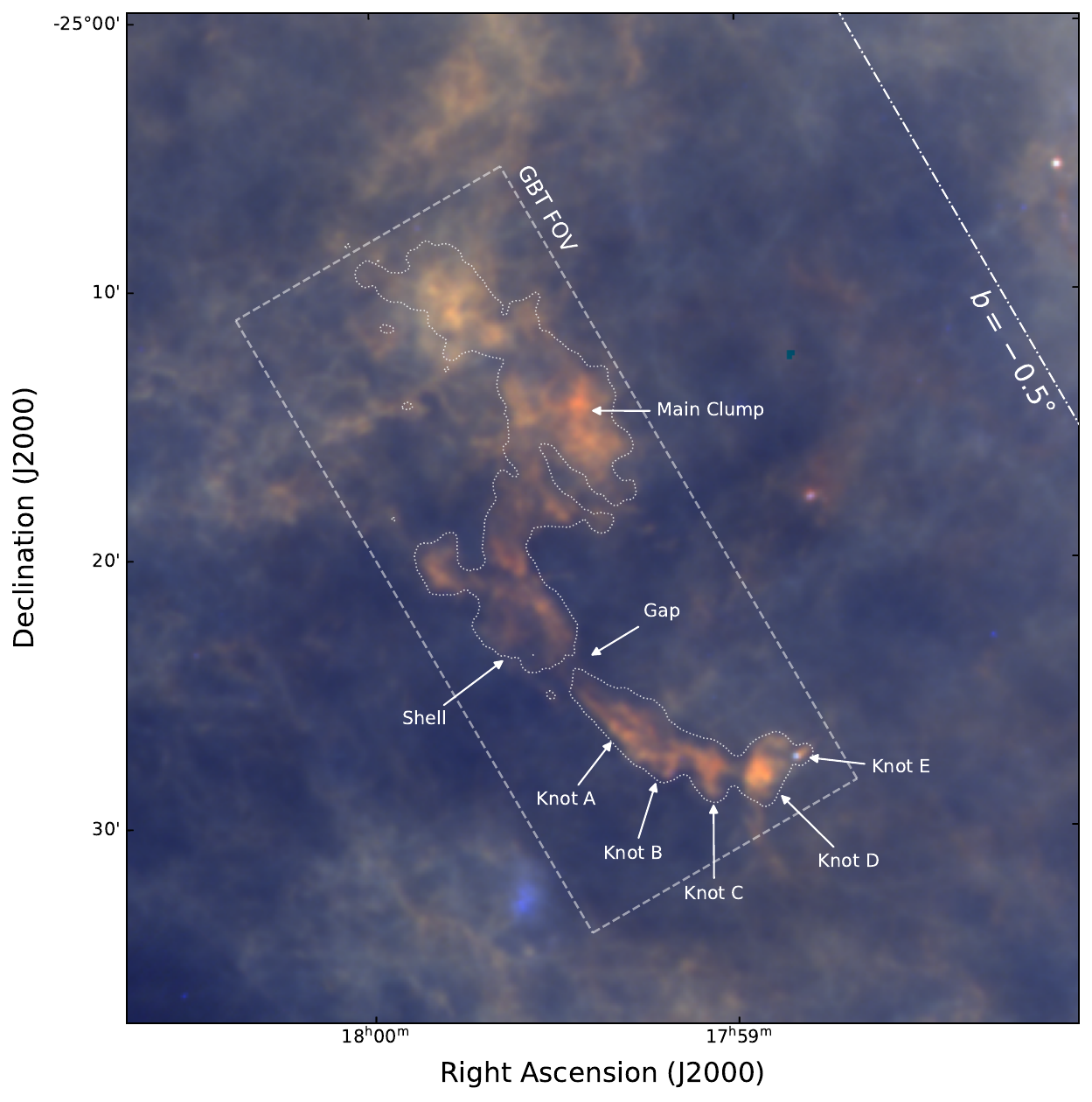}
\caption{3-color \textit{Herschel} composite image of dust at the Midpoint (M4.7-0.8) of the Galactic Bar Dust Lanes, showing PACS 70 \micron\ (Blue), PACS 160 \micron\ (Green), and SPIRE 250 \micron\ (Red). White contours show the $^{12}$CO emission associated with the Galactic Bar Dust Lanes \citep{Marshall08,sormani19a}. The dashed white box shows the field-of-view of the GBT observations (Figures \ref{morphfig}, \ref{morphfig2}, and \ref{kin-fig}). The single contour traces the \am\ (1,1) peak antenna temperature at the 15$\sigma$ level (150 mK; Figure \ref{morphfig}, top). The dot-dashed line shows $b$=-0\fdg5.}
\label{introfigr}
\end{figure*}

Figure \ref{introfigr} shows a 3-color composite image of the Herschel bands shown in Figure \ref{herschel}, where 70 \micron\ is blue, 160 \micron\ is yellow, and 250 \micron\ is red. The cometary structure of Knot E, discussed in Section \ref{knotE}, is visible in this 3-color image.

\bibliographystyle{hapj}
\bibliography{Bar.bib}

\begin{thebibliography}{92}
\expandafter\ifx\csname natexlab\endcsname\relax\def\natexlab#1{#1}\fi

\bibitem[{{Anderson} {et~al.}(2020){Anderson}, {Sormani}, {Ginsburg}, {Glover},
  {Heywood}, {Rammala}, {Schuller}, {Csengeri}, {Urquhart}, \&
  {Bronfman}}]{Anderson20}
{Anderson}, L.~D. {et~al.} 2020, \apj, 901, 51, 2008.04258

\bibitem[{{Athanassoula}(1992)}]{Athanassoula92}
{Athanassoula}, E. 1992, \mnras, 259, 345

\bibitem[{{Barnes} {et~al.}(2023){Barnes}, {Watkins}, {Meidt}, {Kreckel},
  {Sormani}, {Tre{\ss}}, {Glover}, {Bigiel}, {Chandar}, {Emsellem}, {Lee},
  {Leroy}, {Sandstrom}, {Schinnerer}, {Rosolowsky}, {Belfiore}, {Blanc},
  {Boquien}, {Brok}, {Cao}, {Chevance}, {Dale}, {Egorov}, {Eibensteiner},
  {Grasha}, {Groves}, {Hassani}, {Henshaw}, {Jeffreson}, {Jim{\'e}nez-Donaire},
  {Keller}, {Klessen}, {Koch}, {Kruijssen}, {Larson}, {Li}, {Liu}, {Lopez},
  {Murphy}, {Neumann}, {Pety}, {Pinna}, {Querejeta}, {Renaud}, {Saito},
  {Sarbadhicary}, {Sardone}, {Smith}, {Stuber}, {Sun}, {Thilker}, {Usero},
  {Whitmore}, \& {Williams}}]{Barnes23}
{Barnes}, A.~T. {et~al.} 2023, \apjl, 944, L22, 2212.00812

\bibitem[{{Battersby} {et~al.}(2020){Battersby}, {Keto}, {Walker}, {Barnes},
  {Callanan}, {Ginsburg}, {Hatchfield}, {Henshaw}, {Kauffmann}, {Kruijssen},
  {Longmore}, {Lu}, {Mills}, {Pillai}, {Zhang}, {Bally}, {Butterfield},
  {Contreras}, {Ho}, {Ott}, {Patel}, \& {Tolls}}]{CMZoom}
{Battersby}, C. {et~al.} 2020, \apjs, 249, 35, 2007.05023

\bibitem[{{Bertoldi} \& {McKee}(1990)}]{Bertoldi90}
{Bertoldi}, F., \& {McKee}, C.~F. 1990, \apj, 354, 529

\bibitem[{{Binney} {et~al.}(1991){Binney}, {Gerhard}, {Stark}, {Bally}, \&
  {Uchida}}]{Binney91}
{Binney}, J., {Gerhard}, O.~E., {Stark}, A.~A., {Bally}, J., \& {Uchida}, K.~I.
  1991, \mnras, 252, 210

\bibitem[{{Bolatto} {et~al.}(2013){Bolatto}, {Wolfire}, \& {Leroy}}]{Bolatto13}
{Bolatto}, A.~D., {Wolfire}, M., \& {Leroy}, A.~K. 2013, \araa, 51, 207,
  1301.3498

\bibitem[{{Carey} {et~al.}(2009){Carey}, {Noriega-Crespo}, {Mizuno}, {Shenoy},
  {Paladini}, {Kraemer}, {Price}, {Flagey}, {Ryan}, {Ingalls}, {Kuchar},
  {Pinheiro Gon{\c{c}}alves}, {Indebetouw}, {Billot}, {Marleau}, {Padgett},
  {Rebull}, {Bressert}, {Ali}, {Molinari}, {Martin}, {Berriman}, {Boulanger},
  {Latter}, {Miville-Deschenes}, {Shipman}, \& {Testi}}]{mipsgal}
{Carey}, S.~J. {et~al.} 2009, \pasp, 121, 76

\bibitem[{{CASA Team} {et~al.}(2022){CASA Team}, {Bean}, {Bhatnagar}, {Castro},
  {Donovan Meyer}, {Emonts}, {Garcia}, {Garwood}, {Golap}, {Gonzalez Villalba},
  {Harris}, {Hayashi}, {Hoskins}, {Hsieh}, {Jagannathan}, {Kawasaki},
  {Keimpema}, {Kettenis}, {Lopez}, {Marvil}, {Masters}, {McNichols},
  {Mehringer}, {Miel}, {Moellenbrock}, {Montesino}, {Nakazato}, {Ott}, {Petry},
  {Pokorny}, {Raba}, {Rau}, {Schiebel}, {Schweighart}, {Sekhar}, {Shimada},
  {Small}, {Steeb}, {Sugimoto}, {Suoranta}, {Tsutsumi}, {van Bemmel},
  {Verkouter}, {Wells}, {Xiong}, {Szomoru}, {Griffith}, {Glendenning}, \&
  {Kern}}]{CASA}
{CASA Team} {et~al.} 2022, \pasp, 134, 114501, 2210.02276

\bibitem[{{Chen} {et~al.}(2023){Chen}, {Sun}, {Feng}, {Zhang}, {Guo}, {Xu},
  {Su}, {Sun}, {Zhang}, {Zhou}, {Chen}, {Yan}, {Zhang}, {Fang}, \&
  {Yang}}]{Chen23}
{Chen}, X. {et~al.} 2023, \aj, 165, 16, 2211.00810

\bibitem[{{Chevance} {et~al.}(2023){Chevance}, {Krumholz}, {McLeod},
  {Ostriker}, {Rosolowsky}, \& {Sternberg}}]{Chevance23}
{Chevance}, M., {Krumholz}, M.~R., {McLeod}, A.~F., {Ostriker}, E.~C.,
  {Rosolowsky}, E.~W., \& {Sternberg}, A. 2023, in Astronomical Society of the
  Pacific Conference Series, Vol. 534, Protostars and Planets VII, ed.
  S.~{Inutsuka}, Y.~{Aikawa}, T.~{Muto}, K.~{Tomida}, \& M.~{Tamura}, 1,
  2203.09570

\bibitem[{{Chiang} {et~al.}(2024){Chiang}, {Sandstrom}, {Chastenet}, {Bolatto},
  {Koch}, {Leroy}, {Sun}, {Teng}, \& {Williams}}]{Chang24}
{Chiang}, I.-D. {et~al.} 2024, \apj, 964, 18, 2311.00407

\bibitem[{{Churchwell} {et~al.}(2009){Churchwell}, {Babler}, {Meade},
  {Whitney}, {Benjamin}, {Indebetouw}, {Cyganowski}, {Robitaille}, {Povich},
  {Watson}, \& {Bracker}}]{Churchwell09}
{Churchwell}, E. {et~al.} 2009, \pasp, 121, 213

\bibitem[{{Dame} {et~al.}(2001){Dame}, {Hartmann}, \& {Thaddeus}}]{Dame01}
{Dame}, T.~M., {Hartmann}, D., \& {Thaddeus}, P. 2001, \apj, 547, 792,
  astro-ph/0009217

\bibitem[{{de Vaucouleurs}(1964)}]{deVauc64}
{de Vaucouleurs}, G. 1964, in The Galaxy and the Magellanic Clouds, ed. F.~J.
  {Kerr}, Vol.~20, 195

\bibitem[{{D{\'\i}az-Garc{\'\i}a} {et~al.}(2021){D{\'\i}az-Garc{\'\i}a},
  {Lisenfeld}, {P{\'e}rez}, {Zurita}, {Verley}, {Combes}, {Espada}, {Leon},
  {Mart{\'\i}nez-Badenes}, {Sabater}, \& {Verdes-Montenegro}}]{Diaz21}
{D{\'\i}az-Garc{\'\i}a}, S. {et~al.} 2021, \aap, 654, A135, 2106.13099

\bibitem[{{Elia} {et~al.}(2017){Elia}, {Molinari}, {Schisano}, {Pestalozzi},
  {Pezzuto}, {Merello}, {Noriega-Crespo}, {Moore}, {Russeil}, {Mottram},
  {Paladini}, {Strafella}, {Benedettini}, {Bernard}, {Di Giorgio}, {Eden},
  {Fukui}, {Plume}, {Bally}, {Martin}, {Ragan}, {Jaffa}, {Motte}, {Olmi},
  {Schneider}, {Testi}, {Wyrowski}, {Zavagno}, {Calzoletti}, {Faustini},
  {Natoli}, {Palmeirim}, {Piacentini}, {Piazzo}, {Pilbratt}, {Polychroni},
  {Baldeschi}, {Beltr{\'a}n}, {Billot}, {Cambr{\'e}sy}, {Cesaroni},
  {Garc{\'\i}a-Lario}, {Hoare}, {Huang}, {Joncas}, {Liu}, {Maiolo}, {Marsh},
  {Maruccia}, {M{\`e}ge}, {Peretto}, {Rygl}, {Schilke}, {Thompson},
  {Traficante}, {Umana}, {Veneziani}, {Ward-Thompson}, {Whitworth}, {Arab},
  {Bandieramonte}, {Becciani}, {Brescia}, {Buemi}, {Bufano}, {Butora},
  {Cavuoti}, {Costa}, {Fiorellino}, {Hajnal}, {Hayakawa}, {Kacsuk}, {Leto}, {Li
  Causi}, {Marchili}, {Martinavarro-Armengol}, {Mercurio}, {Molinaro},
  {Riccio}, {Sano}, {Sciacca}, {Tachihara}, {Torii}, {Trigilio}, {Vitello}, \&
  {Yamamoto}}]{Elia17}
{Elia}, D. {et~al.} 2017, \mnras, 471, 100, 1706.01046

\bibitem[{{Emsellem} {et~al.}(2015){Emsellem}, {Renaud}, {Bournaud},
  {Elmegreen}, {Combes}, \& {Gabor}}]{Emsellem15}
{Emsellem}, E., {Renaud}, F., {Bournaud}, F., {Elmegreen}, B., {Combes}, F., \&
  {Gabor}, J.~M. 2015, \mnras, 446, 2468, 1410.6479

\bibitem[{{Ferrand} \& {Safi-Harb}(2012)}]{Ferrand12}
{Ferrand}, G., \& {Safi-Harb}, S. 2012, Advances in Space Research, 49, 1313,
  1202.0245

\bibitem[{{Fraser-McKelvie} {et~al.}(2020){Fraser-McKelvie},
  {Arag{\'o}n-Salamanca}, {Merrifield}, {Masters}, {Nair}, {Emsellem},
  {Kraljic}, {Krishnarao}, {Andrews}, {Drory}, \& {Neumann}}]{Fraser20}
{Fraser-McKelvie}, A. {et~al.} 2020, \mnras, 495, 4158, 2005.08987

\bibitem[{{Friesen} {et~al.}(2017){Friesen}, {Pineda}, {co-PIs}, {Rosolowsky},
  {Alves}, {Chac{\'o}n-Tanarro}, {How-Huan Chen}, {Chun-Yuan Chen}, {Di
  Francesco}, {Keown}, {Kirk}, {Punanova}, {Seo}, {Shirley}, {Ginsburg},
  {Hall}, {Offner}, {Singh}, {Arce}, {Caselli}, {Goodman}, {Martin}, {Matzner},
  {Myers}, {Redaelli}, \& {GAS Collaboration}}]{gas1}
{Friesen}, R.~K. {et~al.} 2017, \apj, 843, 63, 1704.06318

\bibitem[{{Fux}(1999)}]{fux99}
{Fux}, R. 1999, \aap, 345, 787, astro-ph/9903154

\bibitem[{{Gadotti} {et~al.}(2019){Gadotti}, {S{\'a}nchez-Bl{\'a}zquez},
  {Falc{\'o}n-Barroso}, {Husemann}, {Seidel}, {P{\'e}rez}, {de
  Lorenzo-C{\'a}ceres}, {Martinez-Valpuesta}, {Fragkoudi}, {Leung}, {van de
  Ven}, {Leaman}, {Coelho}, {Martig}, {Kim}, {Neumann}, \&
  {Querejeta}}]{timer1}
{Gadotti}, D.~A. {et~al.} 2019, \mnras, 482, 506, 1810.01425

\bibitem[{{Ginsburg} {et~al.}(2022){Ginsburg}, {Sokolov}, {de Val-Borro},
  {Rosolowsky}, {Pineda}, {Sip{\H{o}}cz}, \& {Henshaw}}]{pyspeckit}
{Ginsburg}, A., {Sokolov}, V., {de Val-Borro}, M., {Rosolowsky}, E., {Pineda},
  J.~E., {Sip{\H{o}}cz}, B.~M., \& {Henshaw}, J.~D. 2022, \aj, 163, 291,
  2205.04987

\bibitem[{{Goodman} {et~al.}(2014){Goodman}, {Alves}, {Beaumont}, {Benjamin},
  {Borkin}, {Burkert}, {Dame}, {Jackson}, {Kauffmann}, {Robitaille}, \&
  {Smith}}]{Goodman14}
{Goodman}, A.~A. {et~al.} 2014, \apj, 797, 53, 1408.0001

\bibitem[{{Gramze} {et~al.}(2023){Gramze}, {Ginsburg}, {Meier}, {Ott},
  {Shirley}, {Sormani}, \& {Svoboda}}]{Gramze23}
{Gramze}, S.~R., {Ginsburg}, A., {Meier}, D.~S., {Ott}, J., {Shirley}, Y.,
  {Sormani}, M.~C., \& {Svoboda}, B.~E. 2023, \apj, 959, 93, 2309.16403

\bibitem[{{Green}(2019)}]{Green19}
{Green}, D.~A. 2019, Journal of Astrophysics and Astronomy, 40, 36, 1907.02638

\bibitem[{{Hatchfield} {et~al.}(2021){Hatchfield}, {Sormani}, {Tress},
  {Battersby}, {Smith}, {Glover}, \& {Klessen}}]{Hatchfield21}
{Hatchfield}, H.~P., {Sormani}, M.~C., {Tress}, R.~G., {Battersby}, C.,
  {Smith}, R.~J., {Glover}, S. C.~O., \& {Klessen}, R.~S. 2021, \apj, 922, 79,
  2106.08461

\bibitem[{{Henney} \& {O'Dell}(1999)}]{Henney99}
{Henney}, W.~J., \& {O'Dell}, C.~R. 1999, \aj, 118, 2350, astro-ph/9908018

\bibitem[{{Henshaw} {et~al.}(2023){Henshaw}, {Barnes}, {Battersby}, {Ginsburg},
  {Sormani}, \& {Walker}}]{Henshaw23}
{Henshaw}, J.~D., {Barnes}, A.~T., {Battersby}, C., {Ginsburg}, A., {Sormani},
  M.~C., \& {Walker}, D.~L. 2023, in Astronomical Society of the Pacific
  Conference Series, Vol. 534, Protostars and Planets VII, ed. S.~{Inutsuka},
  Y.~{Aikawa}, T.~{Muto}, K.~{Tomida}, \& M.~{Tamura}, 83, 2203.11223

\bibitem[{{Heyer} {et~al.}(2009){Heyer}, {Krawczyk}, {Duval}, \&
  {Jackson}}]{Heyer09}
{Heyer}, M., {Krawczyk}, C., {Duval}, J., \& {Jackson}, J.~M. 2009, \apj, 699,
  1092, 0809.1397

\bibitem[{{Ho} \& {Townes}(1983)}]{Ho83}
{Ho}, P.~T.~P., \& {Townes}, C.~H. 1983, \araa, 21, 239

\bibitem[{{Hogge} {et~al.}(2018){Hogge}, {Jackson}, {Stephens}, {Whitaker},
  {Foster}, {Camarata}, {Anish Roshi}, {Di Francesco}, {Longmore}, {Loughnane},
  {Moore}, {Rathborne}, {Sanhueza}, \& {Walsh}}]{ramps1}
{Hogge}, T. {et~al.} 2018, \apjs, 237, 27, 1808.02533

\bibitem[{{International Consortium Of Scientists}(2011)}]{2011ascl.soft07013I}
{International Consortium Of Scientists}. 2011, {CASA: Common Astronomy
  Software Applications}, Astrophysics Source Code Library, 1107.013

\bibitem[{{Kauffmann} {et~al.}(2017){Kauffmann}, {Pillai}, {Zhang}, {Menten},
  {Goldsmith}, {Lu}, \& {Guzm{\'a}n}}]{Kauffmann17}
{Kauffmann}, J., {Pillai}, T., {Zhang}, Q., {Menten}, K.~M., {Goldsmith},
  P.~F., {Lu}, X., \& {Guzm{\'a}n}, A.~E. 2017, \aap, 603, A89, 1610.03499

\bibitem[{{Kennicutt} {et~al.}(2003){Kennicutt}, {Armus}, {Bendo}, {Calzetti},
  {Dale}, {Draine}, {Engelbracht}, {Gordon}, {Grauer}, {Helou}, {Hollenbach},
  {Jarrett}, {Kewley}, {Leitherer}, {Li}, {Malhotra}, {Regan}, {Rieke},
  {Rieke}, {Roussel}, {Smith}, {Thornley}, \& {Walter}}]{sings}
{Kennicutt}, Jr., R.~C. {et~al.} 2003, \pasp, 115, 928, astro-ph/0305437

\bibitem[{{Keown} {et~al.}(2019){Keown}, {Di Francesco}, {Rosolowsky}, {Singh},
  {Figura}, {Kirk}, {Anderson}, {Chen}, {Elia}, {Friesen}, {Ginsburg},
  {Marston}, {Pezzuto}, {Schisano}, {Bontemps}, {Caselli}, {Liu}, {Longmore},
  {Motte}, {Myers}, {Offner}, {Sanhueza}, {Schneider}, {Stephens}, {Urquhart},
  \& {KEYSTONE Collaboration}}]{keystone1}
{Keown}, J. {et~al.} 2019, \apj, 884, 4, 1908.10514

\bibitem[{{Kim} {et~al.}(2024){Kim}, {Gadotti}, {Querejeta}, {P{\'e}rez},
  {Zurita}, {Neumann}, {van de Ven}, {M{\'e}ndez-Abreu}, {de
  Lorenzo-C{\'a}ceres}, {S{\'a}nchez-Bl{\'a}zquez}, {Fragkoudi}, {Martins},
  {Silva-Lima}, {Kim}, \& {Park}}]{Kim24}
{Kim}, T. {et~al.} 2024, \apj, 968, 87, 2405.00107

\bibitem[{{Kraemer} \& {Jackson}(1995)}]{Kraemer95}
{Kraemer}, K.~E., \& {Jackson}, J.~M. 1995, \apjl, 439, L9

\bibitem[{{Lacy} {et~al.}(2020){Lacy}, {Baum}, {Chandler}, {Chatterjee},
  {Clarke}, {Deustua}, {English}, {Farnes}, {Gaensler}, {Gugliucci},
  {Hallinan}, {Kent}, {Kimball}, {Law}, {Lazio}, {Marvil}, {Mao}, {Medlin},
  {Mooley}, {Murphy}, {Myers}, {Osten}, {Richards}, {Rosolowsky}, {Rudnick},
  {Schinzel}, {Sivakoff}, {Sjouwerman}, {Taylor}, {White}, {Wrobel},
  {Andernach}, {Beasley}, {Berger}, {Bhatnager}, {Birkinshaw}, {Bower},
  {Brandt}, {Brown}, {Burke-Spolaor}, {Butler}, {Comerford}, {Demorest}, {Fu},
  {Giacintucci}, {Golap}, {G{\"u}th}, {Hales}, {Hiriart}, {Hodge}, {Horesh},
  {Ivezi{\'c}}, {Jarvis}, {Kamble}, {Kassim}, {Liu}, {Loinard}, {Lyons},
  {Masters}, {Mezcua}, {Moellenbrock}, {Mroczkowski}, {Nyland}, {O'Dea},
  {O'Sullivan}, {Peters}, {Radford}, {Rao}, {Robnett}, {Salcido}, {Shen},
  {Sobotka}, {Witz}, {Vaccari}, {van Weeren}, {Vargas}, {Williams}, \&
  {Yoon}}]{VLASS}
{Lacy}, M. {et~al.} 2020, \pasp, 132, 035001, 1907.01981

\bibitem[{{Larson}(1981)}]{Larson81}
{Larson}, R.~B. 1981, \mnras, 194, 809

\bibitem[{{Lefloch} \& {Lazareff}(1994)}]{Lefloch94}
{Lefloch}, B., \& {Lazareff}, B. 1994, \aap, 289, 559

\bibitem[{{Lefloch} {et~al.}(1997){Lefloch}, {Lazareff}, \&
  {Castets}}]{Lefloch97}
{Lefloch}, B., {Lazareff}, B., \& {Castets}, A. 1997, \aap, 324, 249

\bibitem[{{Leroy} {et~al.}(2021){Leroy}, {Schinnerer}, {Hughes}, {Rosolowsky},
  {Pety}, {Schruba}, {Usero}, {Blanc}, {Chevance}, {Emsellem}, {Faesi},
  {Herrera}, {Liu}, {Meidt}, {Querejeta}, {Saito}, {Sandstrom}, {Sun},
  {Williams}, {Anand}, {Barnes}, {Behrens}, {Belfiore}, {Benincasa},
  {Be{\v{s}}li{\'c}}, {Bigiel}, {Bolatto}, {den Brok}, {Cao}, {Chandar},
  {Chastenet}, {Chiang}, {Congiu}, {Dale}, {Deger}, {Eibensteiner}, {Egorov},
  {Garc{\'\i}a-Rodr{\'\i}guez}, {Glover}, {Grasha}, {Henshaw}, {Ho}, {Kepley},
  {Kim}, {Klessen}, {Kreckel}, {Koch}, {Kruijssen}, {Larson}, {Lee}, {Lopez},
  {Machado}, {Mayker}, {McElroy}, {Murphy}, {Ostriker}, {Pan}, {Pessa},
  {Puschnig}, {Razza}, {S{\'a}nchez-Bl{\'a}zquez}, {Santoro}, {Sardone},
  {Scheuermann}, {Sliwa}, {Sormani}, {Stuber}, {Thilker}, {Turner}, {Utomo},
  {Watkins}, \& {Whitmore}}]{phangs}
{Leroy}, A.~K. {et~al.} 2021, \apjs, 257, 43, 2104.07739

\bibitem[{{Li} {et~al.}(2016){Li}, {Gerhard}, {Shen}, {Portail}, \&
  {Wegg}}]{Li16}
{Li}, Z., {Gerhard}, O., {Shen}, J., {Portail}, M., \& {Wegg}, C. 2016, \apj,
  824, 13, 1603.09650

\bibitem[{{Liszt}(2006)}]{Liszt06}
{Liszt}, H.~S. 2006, \aap, 447, 533

\bibitem[{{Lu} {et~al.}(2019){Lu}, {Mills}, {Ginsburg}, {Walker}, {Barnes},
  {Butterfield}, {Henshaw}, {Battersby}, {Kruijssen}, {Longmore}, {Zhang},
  {Bally}, {Kauffmann}, {Ott}, {Rickert}, \& {Wang}}]{Lu19}
{Lu}, X. {et~al.} 2019, \apjs, 244, 35, 1909.02338

\bibitem[{{Maeda} {et~al.}(2023){Maeda}, {Egusa}, {Ohta}, {Fujimoto}, \&
  {Habe}}]{Maeda23}
{Maeda}, F., {Egusa}, F., {Ohta}, K., {Fujimoto}, Y., \& {Habe}, A. 2023, \apj,
  943, 7, 2211.15681

\bibitem[{{Mangum} \& {Wootten}(1994)}]{Mangum94}
{Mangum}, J.~G., \& {Wootten}, A. 1994, \apjl, 428, L33

\bibitem[{{Marganian} {et~al.}(2006){Marganian}, {Garwood}, {Braatz},
  {Radziwill}, \& {Maddalena}}]{gbtidl}
{Marganian}, P., {Garwood}, R.~W., {Braatz}, J.~A., {Radziwill}, N.~M., \&
  {Maddalena}, R.~J. 2006, in Astronomical Society of the Pacific Conference
  Series, Vol. 351, Astronomical Data Analysis Software and Systems XV, ed.
  C.~{Gabriel}, C.~{Arviset}, D.~{Ponz}, \& S.~{Enrique}, 512

\bibitem[{{Marsh} {et~al.}(2017){Marsh}, {Whitworth}, {Lomax}, {Ragan},
  {Becciani}, {Cambr{\'e}sy}, {Di Giorgio}, {Eden}, {Elia}, {Kacsuk},
  {Molinari}, {Palmeirim}, {Pezzuto}, {Schneider}, {Sciacca}, \&
  {Vitello}}]{Marsh17}
{Marsh}, K.~A. {et~al.} 2017, \mnras, 471, 2730, 1707.03808

\bibitem[{{Marshall} {et~al.}(2008){Marshall}, {Fux}, {Robin}, \&
  {Reyl{\'e}}}]{Marshall08}
{Marshall}, D.~J., {Fux}, R., {Robin}, A.~C., \& {Reyl{\'e}}, C. 2008, \aap,
  477, L21, 0711.2471

\bibitem[{{Molinari} {et~al.}(2016){Molinari}, {Schisano}, {Elia},
  {Pestalozzi}, {Traficante}, {Pezzuto}, {Swinyard}, {Noriega-Crespo}, {Bally},
  {Moore}, {Plume}, {Zavagno}, {di Giorgio}, {Liu}, {Pilbratt}, {Mottram},
  {Russeil}, {Piazzo}, {Veneziani}, {Benedettini}, {Calzoletti}, {Faustini},
  {Natoli}, {Piacentini}, {Merello}, {Palmese}, {Del Grande}, {Polychroni},
  {Rygl}, {Polenta}, {Barlow}, {Bernard}, {Martin}, {Testi}, {Ali},
  {Andr{\'e}}, {Beltr{\'a}n}, {Billot}, {Carey}, {Cesaroni}, {Compi{\`e}gne},
  {Eden}, {Fukui}, {Garcia-Lario}, {Hoare}, {Huang}, {Joncas}, {Lim}, {Lord},
  {Martinavarro-Armengol}, {Motte}, {Paladini}, {Paradis}, {Peretto},
  {Robitaille}, {Schilke}, {Schneider}, {Schulz}, {Sibthorpe}, {Strafella},
  {Thompson}, {Umana}, {Ward-Thompson}, \& {Wyrowski}}]{Molinari16}
{Molinari}, S. {et~al.} 2016, \aap, 591, A149, 1604.05911

\bibitem[{{Neumann} {et~al.}(2019){Neumann}, {Gadotti}, {Wisotzki}, {Husemann},
  {Busch}, {Combes}, {Croom}, {Davis}, {Gaspari}, {Krumpe}, {P{\'e}rez-Torres},
  {Scharw{\"a}chter}, {Smirnova-Pinchukova}, {Tremblay}, \&
  {Urrutia}}]{Neumann19}
{Neumann}, J. {et~al.} 2019, \aap, 627, A26, 1905.05214

\bibitem[{{Neumann} {et~al.}(2024{\natexlab{a}}){Neumann}, {Thomas},
  {Maraston}, {Gleis}, {Mao}, {Schinnerer}, \& {Stuber}}]{Neumann24a}
{Neumann}, J., {Thomas}, D., {Maraston}, C., {Gleis}, D.~R., {Mao}, C.,
  {Schinnerer}, E., \& {Stuber}, S.~K. 2024{\natexlab{a}}, \mnras, 534, 2438,
  2409.18180

\bibitem[{{Neumann} {et~al.}(2024{\natexlab{b}}){Neumann}, {Bigiel}, {Barnes},
  {Gallagher}, {Leroy}, {Usero}, {Rosolowsky}, {Be{\v{s}}li{\'c}}, {Boquien},
  {Cao}, {Chevance}, {Colombo}, {Dale}, {Eibensteiner}, {Grasha}, {Henshaw},
  {Jim{\'e}nez-Donaire}, {Meidt}, {Menon}, {Murphy}, {Pan}, {Querejeta},
  {Saito}, {Schinnerer}, {Stuber}, {Teng}, \& {Williams}}]{Neumann24b}
{Neumann}, L. {et~al.} 2024{\natexlab{b}}, \aap, 691, A121, 2406.12025

\bibitem[{{Nilipour} {et~al.}(2024){Nilipour}, {Ott}, {Meier}, {Svoboda},
  {Sormani}, {Ginsburg}, {Gramze}, {Butterfield}, \& {Klessen}}]{Nilipour24}
{Nilipour}, A. {et~al.} 2024, arXiv e-prints, arXiv:2410.09258, 2410.09258

\bibitem[{{Oka} {et~al.}(2001{\natexlab{a}}){Oka}, {Hasegawa}, {Sato},
  {Tsuboi}, \& {Miyazaki}}]{Oka01b}
{Oka}, T., {Hasegawa}, T., {Sato}, F., {Tsuboi}, M., \& {Miyazaki}, A.
  2001{\natexlab{a}}, \pasj, 53, 787

\bibitem[{{Oka} {et~al.}(2001{\natexlab{b}}){Oka}, {Hasegawa}, {Sato},
  {Tsuboi}, {Miyazaki}, \& {Sugimoto}}]{Oka01a}
{Oka}, T., {Hasegawa}, T., {Sato}, F., {Tsuboi}, M., {Miyazaki}, A., \&
  {Sugimoto}, M. 2001{\natexlab{b}}, \apj, 562, 348

\bibitem[{{Phillips}(1996)}]{Phillips96}
{Phillips}, A.~C. 1996, in Astronomical Society of the Pacific Conference
  Series, Vol.~91, IAU Colloq. 157: Barred Galaxies, ed. R.~{Buta}, D.~A.
  {Crocker}, \& B.~G. {Elmegreen}, 44

\bibitem[{{Purcell} {et~al.}(2012){Purcell}, {Longmore}, {Walsh}, {Whiting},
  {Breen}, {Britton}, {Brooks}, {Burton}, {Cunningham}, {Green},
  {Harvey-Smith}, {Hindson}, {Hoare}, {Indermuehle}, {Jones}, {Lo}, {Lowe},
  {Phillips}, {Thompson}, {Urquhart}, {Voronkov}, \& {White}}]{purcell12}
{Purcell}, C.~R. {et~al.} 2012, \mnras, 426, 1972, 1207.6159

\bibitem[{{Regan} {et~al.}(1999){Regan}, {Sheth}, \& {Vogel}}]{regan99}
{Regan}, M.~W., {Sheth}, K., \& {Vogel}, S.~N. 1999, \apj, 526, 97,
  astro-ph/9908121

\bibitem[{{Renaud} {et~al.}(2015){Renaud}, {Bournaud}, {Emsellem}, {Agertz},
  {Athanassoula}, {Combes}, {Elmegreen}, {Kraljic}, {Motte}, \&
  {Teyssier}}]{Renaud15}
{Renaud}, F. {et~al.} 2015, \mnras, 454, 3299, 1509.06567

\bibitem[{{Rodr{\'\i}guez-Fern{\'a}ndez}(2006)}]{rod06}
{Rodr{\'\i}guez-Fern{\'a}ndez}, N.~J. 2006, in Journal of Physics Conference
  Series, Vol.~54, Journal of Physics Conference Series, ed. R.~{Sch{\"o}del},
  G.~C. {Bower}, M.~P. {Muno}, S.~{Nayakshin}, \& T.~{Ott} (IOP), 35--41

\bibitem[{{Rosolowsky} {et~al.}(2003){Rosolowsky}, {Engargiola}, {Plambeck}, \&
  {Blitz}}]{ros03}
{Rosolowsky}, E., {Engargiola}, G., {Plambeck}, R., \& {Blitz}, L. 2003, \apj,
  599, 258, astro-ph/0307322

\bibitem[{{Sahai} {et~al.}(2012{\natexlab{a}}){Sahai}, {G{\"u}sten}, \&
  {Morris}}]{Sahai12b}
{Sahai}, R., {G{\"u}sten}, R., \& {Morris}, M.~R. 2012{\natexlab{a}}, \apjl,
  761, L21, 1211.0345

\bibitem[{{Sahai} {et~al.}(2012{\natexlab{b}}){Sahai}, {Morris}, \&
  {Claussen}}]{Sahai12a}
{Sahai}, R., {Morris}, M.~R., \& {Claussen}, M.~J. 2012{\natexlab{b}}, \apj,
  751, 69, 1201.5067

\bibitem[{{Schinnerer} {et~al.}(2023){Schinnerer}, {Emsellem}, {Henshaw},
  {Liu}, {Meidt}, {Querejeta}, {Renaud}, {Sormani}, {Sun}, {Egorov}, {Larson},
  {Leroy}, {Rosolowsky}, {Sandstrom}, {Williams}, {Barnes}, {Bigiel},
  {Chevance}, {Cao}, {Chandar}, {Dale}, {Eibensteiner}, {Glover}, {Grasha},
  {Hannon}, {Hassani}, {Kim}, {Klessen}, {Kruijssen}, {Murphy}, {Neumann},
  {Pan}, {Pety}, {Saito}, {Stuber}, {Tre{\ss}}, {Usero}, {Watkins}, {Whitmore},
  \& {Phangs}}]{Schinnerer23}
{Schinnerer}, E. {et~al.} 2023, \apjl, 944, L15, 2212.09168

\bibitem[{{Schuller} {et~al.}(2009){Schuller}, {Menten}, {Contreras},
  {Wyrowski}, {Schilke}, {Bronfman}, {Henning}, {Walmsley}, {Beuther},
  {Bontemps}, {Cesaroni}, {Deharveng}, {Garay}, {Herpin}, {Lefloch}, {Linz},
  {Mardones}, {Minier}, {Molinari}, {Motte}, {Nyman}, {Reveret}, {Risacher},
  {Russeil}, {Schneider}, {Testi}, {Troost}, {Vasyunina}, {Wienen}, {Zavagno},
  {Kovacs}, {Kreysa}, {Siringo}, \& {Wei{\ss}}}]{Schuller09}
{Schuller}, F. {et~al.} 2009, \aap, 504, 415, 0903.1369

\bibitem[{{Schuller} {et~al.}(2021){Schuller}, {Urquhart}, {Csengeri},
  {Colombo}, {Duarte-Cabral}, {Mattern}, {Ginsburg}, {Pettitt}, {Wyrowski},
  {Anderson}, {Azagra}, {Barnes}, {Beltran}, {Beuther}, {Billington},
  {Bronfman}, {Cesaroni}, {Dobbs}, {Eden}, {Lee}, {Medina}, {Menten}, {Moore},
  {Montenegro-Montes}, {Ragan}, {Rigby}, {Riener}, {Russeil}, {Schisano},
  {Sanchez-Monge}, {Traficante}, {Zavagno}, {Agurto}, {Bontemps}, {Finger},
  {Giannetti}, {Gonzalez}, {Hernandez}, {Henning}, {Kainulainen}, {Kauffmann},
  {Leurini}, {Lopez}, {Mac-Auliffe}, {Mazumdar}, {Molinari}, {Motte}, {Muller},
  {Nguyen-Luong}, {Parra}, {Perez-Beaupuits}, {Schilke}, {Schneider}, {Suri},
  {Testi}, {Torstensson}, {Veena}, {Venegas}, {Wang}, \& {Wienen}}]{Schuller21}
------. 2021, \mnras, 500, 3064, 2012.01527

\bibitem[{{Sheth} {et~al.}(2000){Sheth}, {Regan}, {Vogel}, \&
  {Teuben}}]{Sheth00}
{Sheth}, K., {Regan}, M.~W., {Vogel}, S.~N., \& {Teuben}, P.~J. 2000, \apj,
  532, 221, astro-ph/9911280

\bibitem[{{Shetty} {et~al.}(2012){Shetty}, {Beaumont}, {Burton}, {Kelly}, \&
  {Klessen}}]{Shetty12}
{Shetty}, R., {Beaumont}, C.~N., {Burton}, M.~G., {Kelly}, B.~C., \& {Klessen},
  R.~S. 2012, \mnras, 425, 720, 1206.5803

\bibitem[{{Sormani}(2021)}]{sormani21}
{Sormani}, M.~C. 2021, in Astronomical Society of the Pacific Conference
  Series, Vol. 528, New Horizons in Galactic Center Astronomy and Beyond, ed.
  M.~{Tsuboi} \& T.~{Oka}, 51

\bibitem[{{Sormani} \& {Barnes}(2019)}]{sormani19a}
{Sormani}, M.~C., \& {Barnes}, A.~T. 2019, \mnras, 484, 1213, 1901.00867

\bibitem[{{Sormani} {et~al.}(2018{\natexlab{a}}){Sormani}, {Sobacchi},
  {Fragkoudi}, {Ridley}, {Tre{\ss}}, {Glover}, \& {Klessen}}]{sormani18b}
{Sormani}, M.~C., {Sobacchi}, E., {Fragkoudi}, F., {Ridley}, M., {Tre{\ss}},
  R.~G., {Glover}, S. C.~O., \& {Klessen}, R.~S. 2018{\natexlab{a}}, \mnras,
  481, 2, 1805.07969

\bibitem[{{Sormani} {et~al.}(2019){Sormani}, {Tre{\ss}}, {Glover}, {Klessen},
  {Barnes}, {Battersby}, {Clark}, {Hatchfield}, \& {Smith}}]{sormani19b}
{Sormani}, M.~C. {et~al.} 2019, \mnras, 488, 4663, 1906.10129

\bibitem[{{Sormani} {et~al.}(2018{\natexlab{b}}){Sormani}, {Tre{\ss}},
  {Ridley}, {Glover}, {Klessen}, {Binney}, {Magorrian}, \&
  {Smith}}]{sormani18a}
{Sormani}, M.~C., {Tre{\ss}}, R.~G., {Ridley}, M., {Glover}, S. C.~O.,
  {Klessen}, R.~S., {Binney}, J., {Magorrian}, J., \& {Smith}, R.
  2018{\natexlab{b}}, \mnras, 475, 2383, 1707.03650

\bibitem[{{Teng} {et~al.}(2023){Teng}, {Sandstrom}, {Sun}, {Gong}, {Bolatto},
  {Chiang}, {Leroy}, {Usero}, {Glover}, {Klessen}, {Liu}, {Querejeta},
  {Schinnerer}, {Bigiel}, {Cao}, {Chevance}, {Eibensteiner}, {Grasha},
  {Israel}, {Murphy}, {Neumann}, {Pan}, {Pinna}, {Sormani}, {Smith}, {Walter},
  \& {Williams}}]{Teng23}
{Teng}, Y.-H. {et~al.} 2023, \apj, 950, 119, 2304.04732

\bibitem[{{Tress} {et~al.}(2020){Tress}, {Sormani}, {Glover}, {Klessen},
  {Battersby}, {Clark}, {Hatchfield}, \& {Smith}}]{Tress20b}
{Tress}, R.~G., {Sormani}, M.~C., {Glover}, S. C.~O., {Klessen}, R.~S.,
  {Battersby}, C.~D., {Clark}, P.~C., {Hatchfield}, H.~P., \& {Smith}, R.~J.
  2020, \mnras, 499, 4455, 2004.06724

\bibitem[{{Urquhart} {et~al.}(2015){Urquhart}, {Figura}, {Moore}, {Csengeri},
  {Lumsden}, {Pillai}, {Thompson}, {Eden}, \& {Morgan}}]{Urquhart15}
{Urquhart}, J.~S. {et~al.} 2015, \mnras, 452, 4029, 1507.02187

\bibitem[{{Veena} {et~al.}(2024){Veena}, {Kim}, {S{\'a}nchez-Monge}, {Schilke},
  {Menten}, {Fuller}, {Sormani}, {Wyrowski}, {Banda-Barrag{\'a}n}, {Riquelme},
  {Tarr{\'\i}o}, \& {de Vicente}}]{Veena24}
{Veena}, V.~S. {et~al.} 2024, \aap, 689, A121, 2407.14338

\bibitem[{{Verley} {et~al.}(2007){Verley}, {Combes}, {Verdes-Montenegro},
  {Bergond}, \& {Leon}}]{Verley07}
{Verley}, S., {Combes}, F., {Verdes-Montenegro}, L., {Bergond}, G., \& {Leon},
  S. 2007, \aap, 474, 43, 0707.4127

\bibitem[{{Wallace} {et~al.}(2022){Wallace}, {Battersby}, {Mills}, {Henshaw},
  {Sormani}, {Ginsburg}, {Barnes}, {Hatchfield}, {Glover}, \&
  {Anderson}}]{Wallace22}
{Wallace}, J. {et~al.} 2022, \apj, 939, 58, 2209.11781

\bibitem[{{Watkins} {et~al.}(2023){Watkins}, {Kreckel}, {Groves}, {Glover},
  {Whitmore}, {Leroy}, {Schinnerer}, {Meidt}, {Egorov}, {Barnes}, {Lee},
  {Bigiel}, {Boquien}, {Chandar}, {Chevance}, {Dale}, {Grasha}, {Klessen},
  {Kruijssen}, {Larson}, {Li}, {M{\'e}ndez-Delgado}, {Pessa}, {Saito},
  {Sanchez-Blazquez}, {Sarbadhicary}, {Scheuermann}, {Thilker}, \&
  {Williams}}]{Watkins23}
{Watkins}, E.~J. {et~al.} 2023, \aap, 676, A67, 2302.03699

\bibitem[{{Weaver} {et~al.}(1977){Weaver}, {McCray}, {Castor}, {Shapiro}, \&
  {Moore}}]{Weaver77}
{Weaver}, R., {McCray}, R., {Castor}, J., {Shapiro}, P., \& {Moore}, R. 1977,
  \apj, 218, 377

\bibitem[{{Wright} {et~al.}(2010){Wright}, {Eisenhardt}, {Mainzer}, {Ressler},
  {Cutri}, {Jarrett}, {Kirkpatrick}, {Padgett}, {McMillan}, {Skrutskie},
  {Stanford}, {Cohen}, {Walker}, {Mather}, {Leisawitz}, {Gautier}, {McLean},
  {Benford}, {Lonsdale}, {Blain}, {Mendez}, {Irace}, {Duval}, {Liu}, {Royer},
  {Heinrichsen}, {Howard}, {Shannon}, {Kendall}, {Walsh}, {Larsen}, {Cardon},
  {Schick}, {Schwalm}, {Abid}, {Fabinsky}, {Naes}, \& {Tsai}}]{wise}
{Wright}, E.~L. {et~al.} 2010, \aj, 140, 1868, 1008.0031

\bibitem[{{Wright} {et~al.}(2012){Wright}, {Drake}, {Drew}, {Guarcello},
  {Gutermuth}, {Hora}, \& {Kraemer}}]{Wright12}
{Wright}, N.~J., {Drake}, J.~J., {Drew}, J.~E., {Guarcello}, M.~G.,
  {Gutermuth}, R.~A., {Hora}, J.~L., \& {Kraemer}, K.~E. 2012, \apjl, 746, L21,
  1201.2404

\bibitem[{{Wynn-Williams}(1974)}]{Wynn-Williams74}
{Wynn-Williams}, C.~G. 1974, in IAU Symposium, Vol.~60, Galactic Radio
  Astronomy, ed. F.~J. {Kerr} \& S.~C. {Simonson}, 259

\bibitem[{{Yang} {et~al.}(2023){Yang}, {Dzib}, {Urquhart}, {Brunthaler},
  {Medina}, {Menten}, {Wyrowski}, {Ortiz-Le{\'o}n}, {Cotton}, {Gong}, {Dokara},
  {Rugel}, {Beuther}, {Pandian}, {Csengeri}, {Veena}, {Roy}, {Nguyen},
  {Winkel}, {Ott}, {Carrasco-Gonzalez}, {Khan}, \& {Cheema}}]{Yang23}
{Yang}, A.~Y. {et~al.} 2023, \aap, 680, A92, 2310.09777

\bibitem[{{Zhang} {et~al.}(2016){Zhang}, {Li}, {Wyrowski}, {Wang}, {Yuan},
  {Xu}, {Gong}, {Yeh}, \& {Menten}}]{Zhang16}
{Zhang}, C.-P. {et~al.} 2016, \aap, 585, A117, 1510.06114

\bibitem[{{Zhang} \& {Ho}(1995)}]{Zhang95}
{Zhang}, Q., \& {Ho}, P. T.~P. 1995, \apjl, 450, L63

\bibitem[{{Zucker} {et~al.}(2015){Zucker}, {Battersby}, \&
  {Goodman}}]{Zucker15}
{Zucker}, C., {Battersby}, C., \& {Goodman}, A. 2015, \apj, 815, 23, 1506.08807

\end{thebibliography}

\end{document}